\documentclass[acmtog]{acmart}
\AtBeginDocument{%
  }

\setcopyright{arxiv}
\copyrightyear{2025}
\acmJournal{TOG}
\acmYear{2025} \acmVolume{44} \acmNumber{4} \acmArticle{} \acmMonth{8} \acmDOI{10.1145/3730853}

\usepackage{amsmath}
\usepackage{subcaption}
\usepackage{placeins}
\usepackage{float}
\usepackage{microtype}

\usepackage{multirow}
\usepackage{booktabs}
\usepackage{pifont}
\usepackage{makecell}
\usepackage{graphicx}

\newcommand{\crmod}[1]{\textcolor{black}{#1}}
\newcommand{\crmodsecond}[1]{\textcolor{black}{#1}}

\usepackage[linesnumbered, ruled, vlined, noend]{algorithm2e}
\SetKwProg{Procedure}{Procedure}{:}{}

\usepackage{xcolor}

\definecolor{mycolor}{RGB}{153, 0, 0}
\SetKwFor{ForEach}{\textcolor{mycolor}{foreach}}{\textcolor{mycolor}{do}}{\textcolor{mycolor}{end}}%
\SetKwFor{For}{\textcolor{mycolor}{for}}{\textcolor{mycolor}{do}}{\textcolor{mycolor}{end}}%
\SetKwFor{While}{\textcolor{mycolor}{while}}{\textcolor{mycolor}{do}}{\textcolor{mycolor}{end}}%

\SetCommentSty{mycommfont}

\SetKwIF{If}{ElseIf}{Else}{\textcolor{teal}{if}}{\textcolor{teal}{then}}{\textcolor{teal}{else if}}{\textcolor{teal}{else}}{\textcolor{teal}{end if}}%

\newcommand{\ie}{\textit{i.e.}}
\newcommand{\eg}{\textit{e.g.}}

\usepackage{cleveref}

\begin{document}

\title{Claycode: Stylable and Deformable 2D Scannable Codes}

\author{Marco Maida}
\affiliation{%
  \institution{Independent Researcher}
  \city{London}
  \country{United Kingdom}}
\email{mmaidacs@gmail.com}

\author{Alberto Crescini}
\affiliation{%
  \institution{Independent Researcher}
  \city{London}
  \country{United Kingdom}}
\email{info@alberto.fun}

\author{Marco Perronet}
\affiliation{%
  \institution{Independent Researcher}
  \city{London}
  \country{United Kingdom}}
\email{perronet.marco@gmail.com}

\author{Elena Camuffo}
\affiliation{%
  \institution{Independent Researcher}
  \city{Padova}
  \country{Italy}}
\email{elenacamuffo97@gmail.com}

\NewDocumentCommand\emojieyes{}{
    \includegraphics[trim=1cm 0.9cm 2.5cm 0.7cm,clip,height=0.3cm]{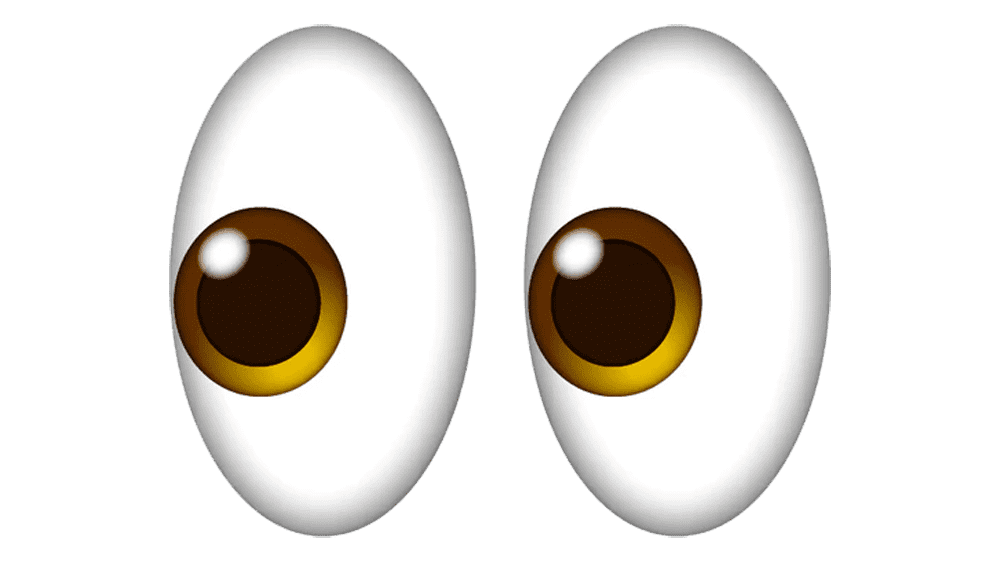}
}
\NewDocumentCommand\emojisparkles{}{
    \includegraphics[trim=0cm 0.3cm 0cm 0cm,clip,height=0.3cm]{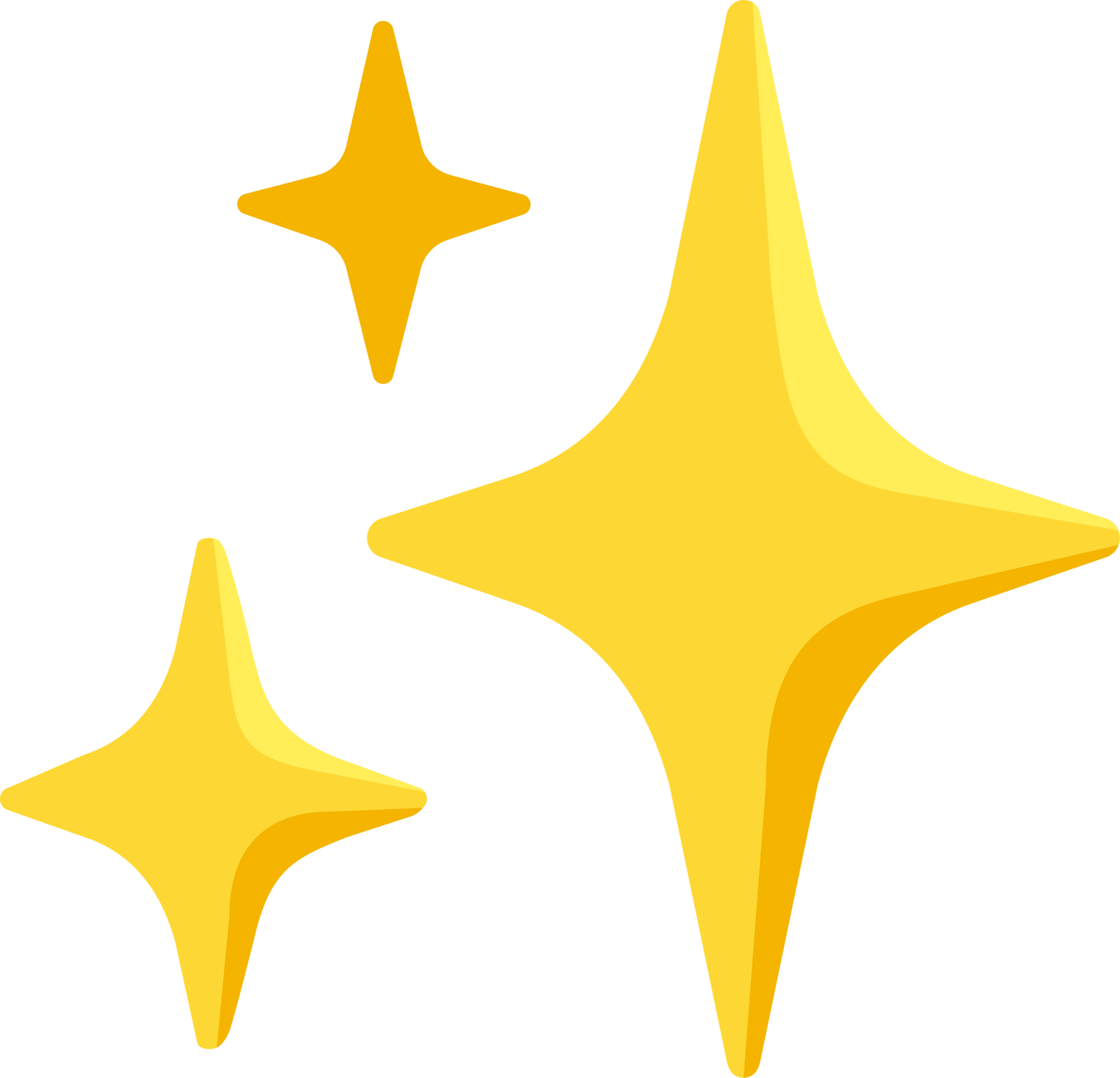}
}

\renewcommand{\shortauthors}{Maida et al.}

\begin{abstract}
  This paper introduces Claycode, a novel 2D scannable code designed for extensive stylization and deformation. Unlike traditional matrix-based codes (\eg, QR codes), Claycodes encode their message in a tree structure. During the encoding process, bits are mapped into a topology tree, which is then depicted as a nesting of color regions drawn within the boundaries of a target polygon shape. When decoding, Claycodes are extracted and interpreted in real-time from a camera stream. 
  We detail the end-to-end pipeline and show that Claycodes allow for extensive stylization without compromising their functionality. We then empirically demonstrate Claycode's high tolerance to heavy deformations, outperforming traditional 2D scannable codes in scenarios where they typically fail. 
\end{abstract}




\begin{teaserfigure}
  \centering
  \includegraphics[width=0.24\textwidth]{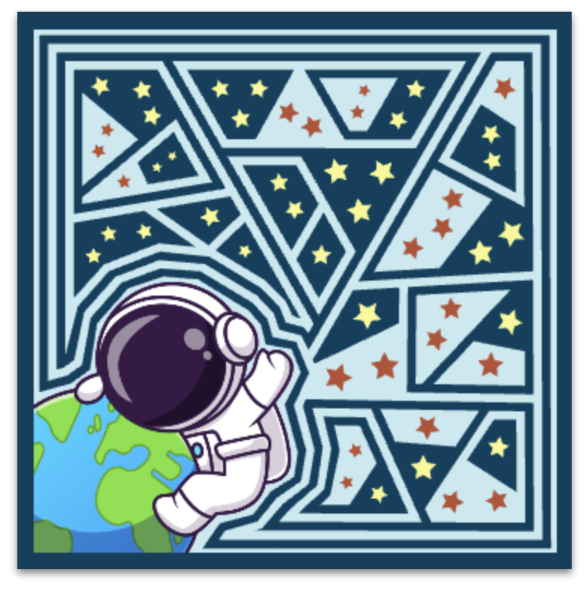}
  \hfill
  \includegraphics[width=0.22\textwidth]{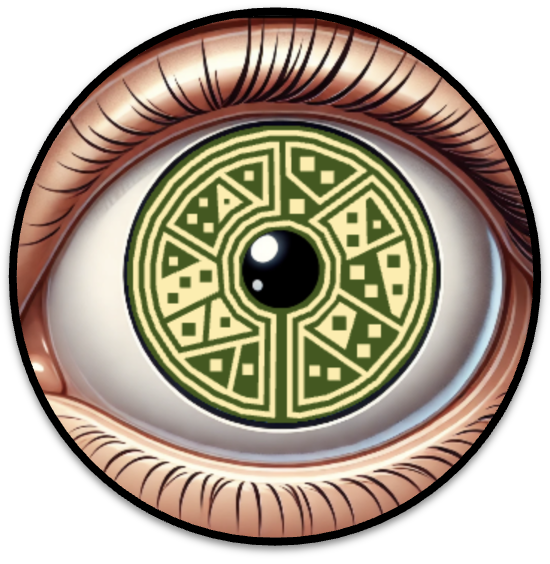}
  \hfill
  \includegraphics[width=0.23\textwidth]{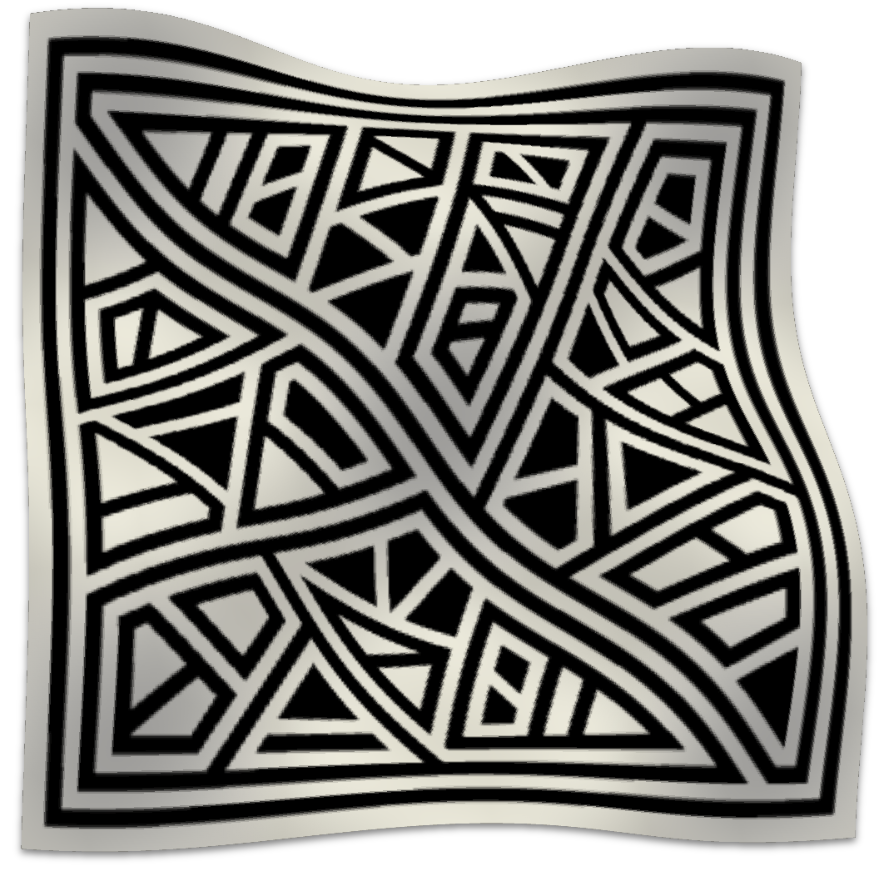}
  \hfill
  \includegraphics[width=0.28\textwidth]{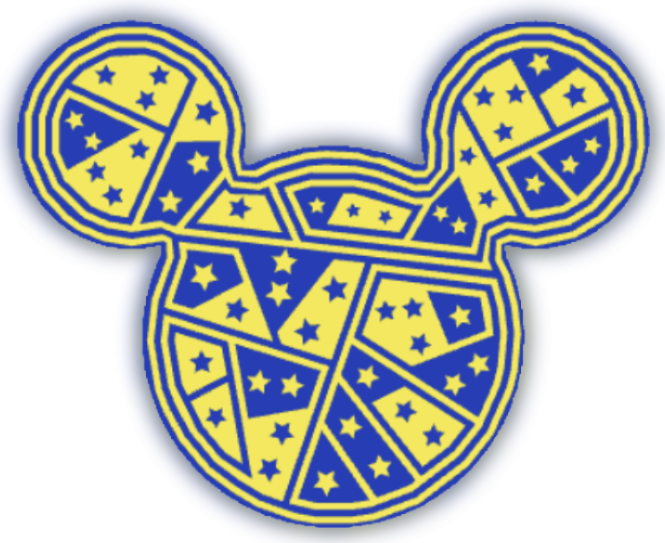}
  \caption{Examples of Claycodes using different shapes and styles. From the left: \textit{``space", ``\emojieyes", ``hello", ``magic"}.}
  \Description{}
  \label{fig:teaser}
\end{teaserfigure}

\balance

\maketitle

\begin{figure*}
    \centering
    \includegraphics[width=0.85\textwidth]{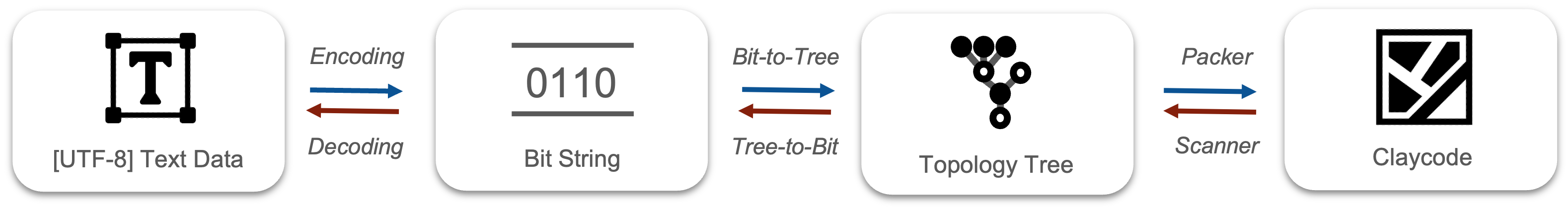} 
    \caption{The encoding and decoding pipeline of Claycodes.}
    \label{fig:pipeline}
\end{figure*}

\section{Introduction}

With the large adoption of smartphones, 2D scannable codes (such as QR codes~\cite{tiwari2016introduction}) have transcended the industrial domain and become a part of everyday life. Scannable codes are ubiquitous in ticketing, mobile payments, advertising, as gateways to digital experiences in museums and cultural institutions, and augmented reality applications~\cite{ozkaya2015factors}. While QR codes and other barcodes have proven highly efficient at pictorially encoding information, their design often lacks aesthetic appeal. Constrained by strict algorithmic requirements, they often look out of place on otherwise carefully designed posters and artworks. Over the years, the demand for visual adaptability has grown significantly, leading to custom-branded codes like Spotify Codes~\cite{spotifycodes} and App Clips~\cite{apple_hig_appclips}, businesses that provide styled QR codes to corporations and individuals~\cite{flowcode,qrcode_ai},
and codes purpose-built for the visually impaired~\cite{navilens}. 
In this work, we introduce Claycodes, two-dimensional scannable codes that can be styled and significantly deformed, making it possible to embed a message in an artwork in a visually pleasing way. The high deformability of Claycodes allows them to remain scannable on elastic surfaces, like textiles and skin tissue~\cite{wan2023implantable}, and deformations induced by a large scanning angle~\cite{qrcodes2019}. Moreover, unlike traditional scannable codes, Claycodes are not confined to a predefined shape, making them ideal for embedding within shape-constrained spaces like electrical components. 

Claycodes are the first \textit{topological codes}: they embed a message within the tree structure defined by the topology of its color regions. Several other works explored the use of topology as a mean to build visual markers~\cite{seedmarkers,dtouch0,yu2020topotag}, but Claycodes are the first fully functional topological codes able to carry an arbitrary bit payload.

This paper details the end-to-end pipeline of Claycodes (\cref{fig:pipeline}) starting from \cref{sec:bit_to_tree_encoding}, where we devise a bidirectional mapping between bit strings and trees optimized to produce readable codes. In \cref{sec:packer}, we show how to render the resulting trees 
within a target polygon, ultimately producing a Claycode. Finally, in \cref{sec:scanner} we outline the process of detecting and decoding Claycodes from a camera stream in real-time, and discuss error detection and redundancy techniques. We conclude with \cref{sec:results}, where we empirically demonstrate the resistance of Claycodes to deformations and stylization by conducting a comparative scanning test between our implementation of Claycodes, QR Codes, and Code-128 Barcodes~\cite{ISO15417-2007}.

\section{Related Work}

QR-Codes~\cite{tiwari2016introduction} receive large interest from the academic community. Earlier attempts to embed images in QR-Codes focus on deforming~\cite{chu2013halftone} or re-shuffling~\cite{cox2012qartcodes, xu2021art} the codes' modules to approximate the target image, while minimizing the error introduced.
More recent advancements, based on neural techniques, have dramatically improved the quality of image-embedded QR Codes. Notably,~\cite{8604076} pioneered the use of Neural Style Transfer to QR Codes, a method further refined by ArtCoder~\cite{su2021artcoder}, and then evolved into diffusion techniques, such as in GladCoder~\cite{Xie2024GladCoderSQ}.
A common theme in these works is the balance between visual quality and robustness, sometimes biasing the generators with saliency maps~\cite{lin2013appearance, 7112509} to the parts of the image visually considered important. All of the cited techniques generally damage the code to some degree. Other works propose codes purpose-built to be styled: PiCodes~\cite{7479568} are generated from an input image, and its pixel intensities carry the message. 
\crmod{Several other systems explore visual or structural embedding strategies that preserve, rather than compromise, the overall appearance of the carrier. Notably, StructCode~\cite{Dogan2023StructCode} encodes bits directly in the geometric features of laser-cut objects, offering an unobtrusive and aesthetically aligned approach. Related examples include FontCode~\cite{Xiao2018FontCode}, BrightMarker~\cite{Dogan2023BrightMarker}, and Imprinto~\cite{Feick2025Imprinto}.}

To the best of our knowledge, Claycodes are the first topological 2D scannable codes that can carry arbitrary bit sequences. However, the literature contains several works on topological \textit{fiducial markers}, pioneered by D-Touch~\cite{dtouch0, dtouch1} and followed by ReacTIVision~\cite{reactivision}, ARTag~\cite{ARTag}, ARTcodes~\cite{artcodes}, TopoTag~\cite{yu2020topotag}, and Seedmarkers~\cite{seedmarkers}. Fiducial markers are popular in augmented reality applications, but their visual rigidity often detracts from the immersive experience. ARTTag~\cite{higashino2016arttag}, is a recent example of stylable fiducial marker, designed around pairs of circles that can blend into an artwork. Claycode builds on the foundations laid by the aforementioned work, evolving the topological approach to 2D codes.  
\crmod{Along this direction, Jung et al.~\cite{Jung2018Methods, Jung2019Automating} propose human-designable visual markers that combine structure and stylistic freedom, supporting intentional design and fine control over appearance.}

\crmod{By contrast, a separate line of work focuses on invisibility or machine-only readability. Examples include AirCode~\cite{Li2017AirCode}, LayerCode~\cite{Maia2019LayerCode}, StegaStamp~\cite{Tancik2020StegaStamp}, and again InfraredTags~\cite{Dogan2022InfraredTags}, which rely on materials or wavelengths not visible to the human eye. These methods fall more squarely into steganography or optical tracking, and while technologically adjacent, they differ in goal: hiding or isolating the code from human perception, rather than integrating it visually as Claycode does.}

\section{Bit-Tree Encoding}\label{sec:bit_to_tree_encoding} 

We begin by introducing Claycode's \textit{bit-tree encoding}, a bidirectional mapping between the set of bit strings $\{0, 1\}^*$ and trees. Throughout this paper, we work with \textit{rooted, unlabeled, ordered} trees, \ie, directed acyclic graphs where there is a single distinguished node with no incoming edges called \textit{root}, and every other node has exactly one incoming edge -- \textit{its parent}. $T$ may refer to either a tree $T \in \mathcal{T}$ or its root node, depending on the context. Nodes sharing the same parent are referred to as \textit{siblings}. The order among siblings is significant: given a parent node $T$, we indicate its ordered children with a duplicate-free list $C(T)=[T_1, T_2, \ldots]$. If $C(T)=[]$, then $T$ is a \textit{leaf}. A node $T \neq T'$ is a \textit{descendant} of $T'$ if there exists a path from $T'$ to $T$. We call $D(T)$ the set of all descendants of $T$. Finally, two trees $T,T' \in \mathcal{T}$ are \textit{isomorphic} $(T \sim T')$ if one can be transformed into the other via a sequence of permutations of siblings. 

A \emph{bit-tree encoding} is a pair of functions $(f,g)$ such that:

\begin{align}
    &f: \{0, 1\}^* \to \mathcal{T}, \quad g: \mathcal{T} \to \{0, 1\}^*, \notag \\
    &g(f(b)) = b \quad \forall b \in \{0, 1\}^* \notag \\
    &T \sim T' \implies g(T) = g(T'), \quad \forall  T, T' \in \mathcal{T}. \label{eq:unordered-bit-tree}
\end{align}

\begin{figure}[t]
    \centering \includegraphics[width=0.8\linewidth]{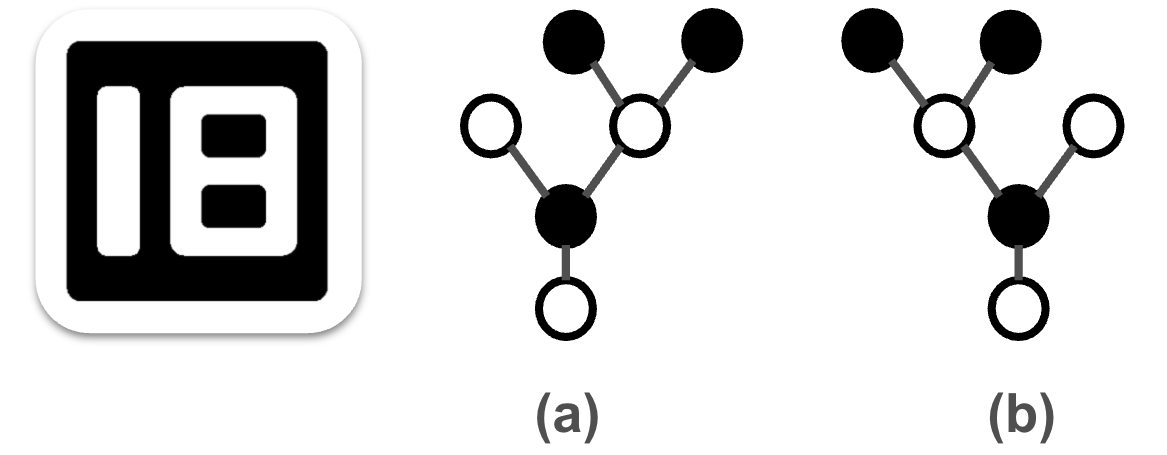} 
    \caption{Isomorphic trees map to the same Claycode.}
    \label{fig:ordering}
\end{figure}

The last line of \cref{eq:unordered-bit-tree} is needed as Claycodes have no way of visually encoding sibling's ordering (\cref{fig:ordering}). We hence define an encoding that yields the same message when permuting siblings. 
Defining a pair of functions that formally respects \cref{eq:unordered-bit-tree} is relatively straightforward. However, the bit-tree encoding largely determines the data capacity and readability of Claycodes, and as such, it is subject to a series of additional requirements.  

\begin{figure}
  \begin{subfigure}[b]{0.15\textwidth}
    \includegraphics[width=\textwidth]{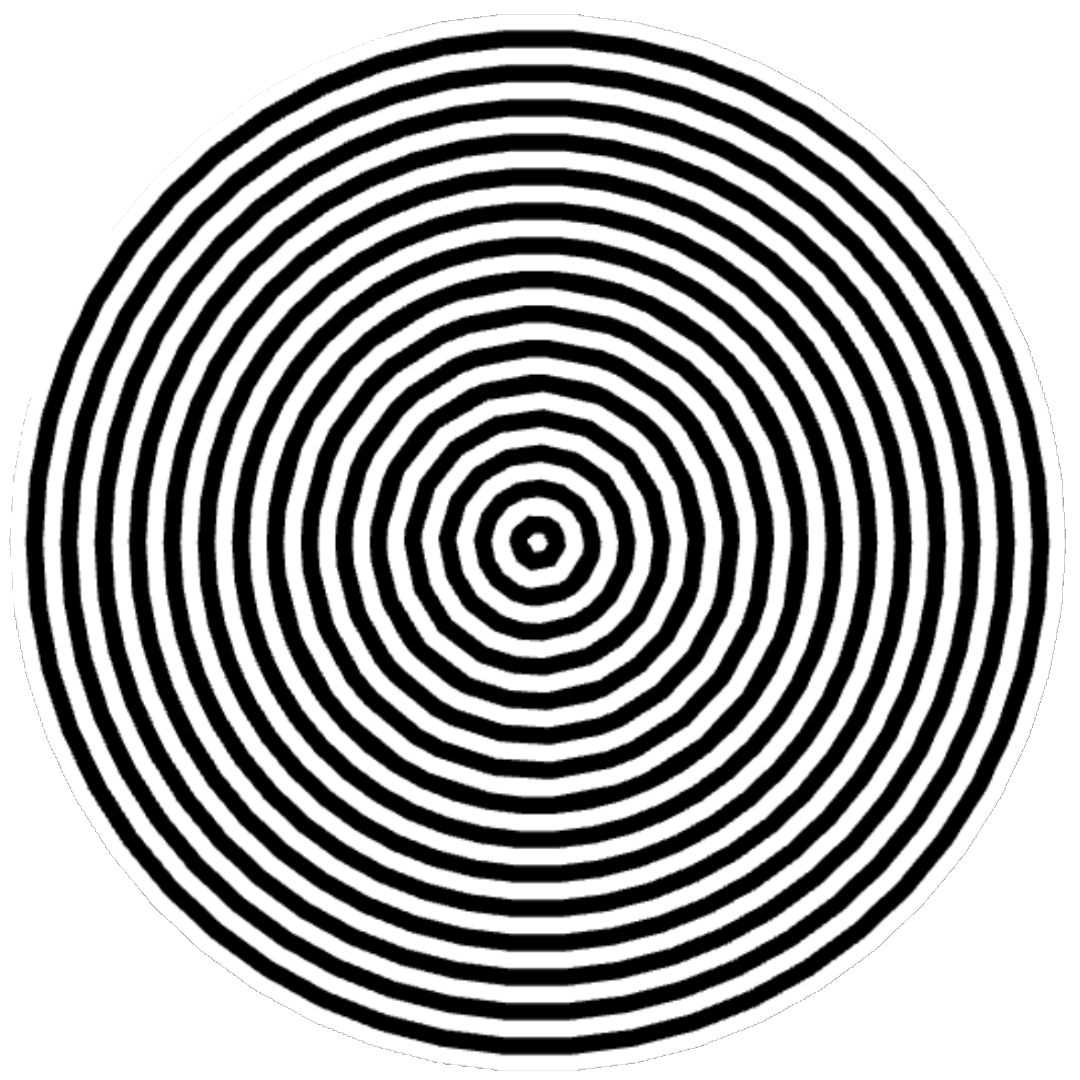}
    \caption*{(a)}
    \caption*{30 nodes, $F_p(T)$: 465}
    \label{fig:shallow-tree-a}
  \end{subfigure}
  \hfill
  \begin{subfigure}[b]{0.16\textwidth}
    \includegraphics[width=\textwidth]{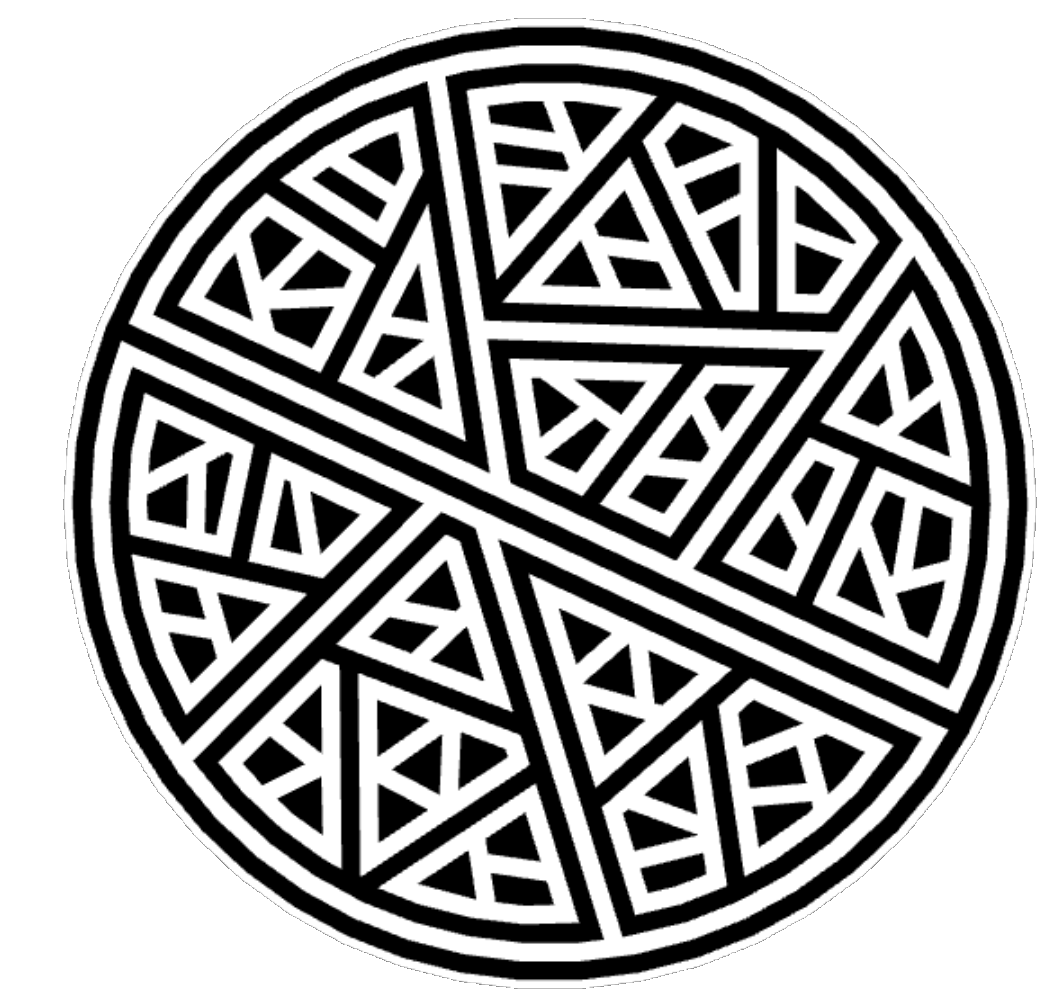}
    \caption*{(b)}
    \caption*{100 nodes, $F_p(T)$: 454}
    \label{fig:shallow-tree-b}
  \end{subfigure}
  \hfill
  \begin{subfigure}[b]{0.15\textwidth}
    \includegraphics[width=\textwidth]{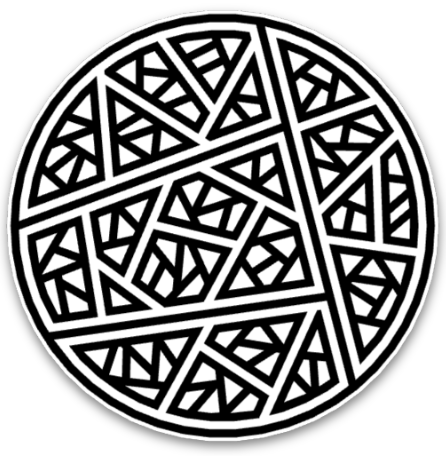}
    \caption*{(c)}
    \caption*{150 nodes, $F_p(T)$: 436}
    \label{fig:shallow-tree-c}
  \end{subfigure}
  \caption{Shallow trees are more space-efficient.}
  \label{fig:shallow-trees}
\end{figure}

\begin{figure}
    \includegraphics[width=0.46\textwidth]{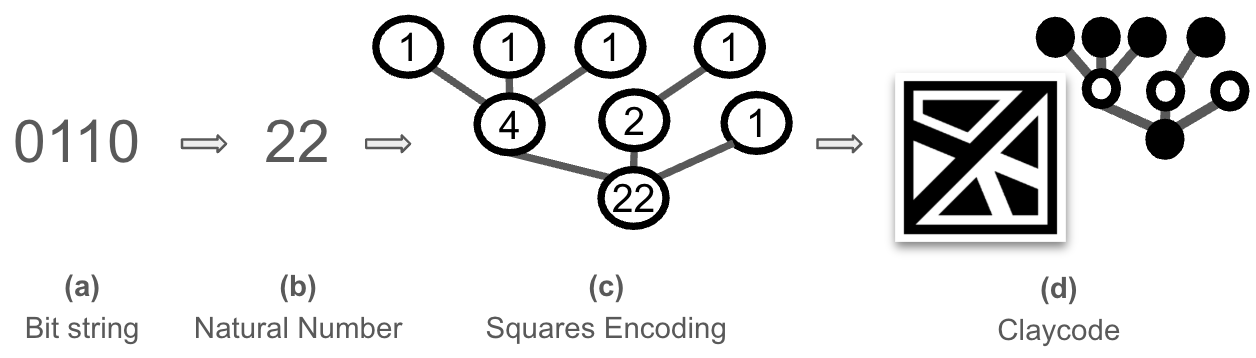} 
    \caption{The Squares bit-tree encoding.}
    \label{fig:squares}
\end{figure}

\subsection{Bit-Tree Requirements For Space-Efficient Claycodes} 

An important quality of any encoding is to minimize the size of the encoded message within its target medium. 
Claycodes pose a challenge in defining what ``size'' is: despite the number of nodes in the topology tree correlates with the space needed to draw the code, it is not a comprehensive metric by itself. As exemplified in \cref{fig:shallow-trees}, when a topology tree is rendered into a Claycode, each node surrounds all its descendants, which means that each parent node requires more space than its children. Highly nested topologies are thus significantly less space-efficient compared to shallower, more planar configurations. To capture the aforementioned aspects into a metric, we define the \textit{footprint} of a node as:
\begin{equation} \label{eq:footprint}
F(T) = 1 + |D(T)|
\end{equation}
By extension, we further call the \textit{total footprint} of a node the sum of its own footprint plus the one of all its descendants, \textit{i.e.},
$F_t(T) = F(T) + \sum_{T_d \in D(T)} F(T_d)$. 
The total footprint scales quadratically with the depth of the tree, and linearly with its width. We selected this formulation because of its simplicity and its strong -- empirically observed -- correlation with the visual complexity of the generated Claycode across different input shapes.

With the footprint metric in place, we define the requirements for Claycode's bit-tree encoding.
Given a bit string $b$, the first requirement (\textbf{R1}) of an effective bit-tree encoding $(f,g)$ is to minimize the total footprint of the generated tree $F_t(f(b))$. Additionally, $F_t(f(b))$ must strongly correlate with the length of $b$, mapping longer bit strings to trees with larger footprint (\textbf{R2}).
Further, many bit-tree encodings suffer from significant size inconsistencies: depending on the specific combination of ones and zeroes in the input bit string $b$, $F_t(f(b))$ widely varies. Such variability is problematic in Claycodes, as it creates instances of short input messages requiring disproportionately large amounts of space to be represented. Therefore, $F_t(f(b))$ must have low (positive or negative) correlation with the percentage of ones in the input message (\textbf{R3}). Finally, to the best of our knowledge, existing tree ranking and generation methods are either based on successor functions \cite{beyer1980constant,NAKANORooted,effantin2004generation,li1997generation}, requiring to compute $n-1$ predecessor trees to output the $n$th one, or prime factorization \cite{abe1994tree, cappello1988new, skliar2020one}. Scannable codes typically contain short strips of text, leading to an expected payload of hundreds of bits. Even with a short message of $100$ bits, there are $2^{100}$ possible message combinations, rendering both enumeration and prime factorization-based techniques unfeasible. A bit-tree encoding must be computable on commercial hardware within milliseconds for input messages of hundreds of bits (\textbf{R4}). Guided by (\textbf{R1-R4}), we devise a bit-tree encoding inspired by \cite{abe1994tree}.

\subsection{The Squares Encoding}

First, we modify the problem to work on natural numbers instead of bit strings. We define $f'$ and $g'$~so~that:

\begin{align} \label{eq:nat-bits}
    &f': \mathbb{N} \to \mathcal{T}, \quad g': \mathcal{T} \to \mathbb{N}, \notag \\
    &f(b) = f'(nat(b)) \quad\quad g(T) = bits(g'(T)) 
\end{align}

The functions $(nat,bits)$ define a bijection between natural numbers and bit strings, which can be easily derived by interpreting the string as a binary number. We define $(nat,bits)$ in \cref{sec:appendix-bijection}.
The inner functions $f'$ and $g'$ are detailed in \cref{alg:bit-tree}, while an illustrative example is provided in \cref{fig:squares}. The encoding function starts from a node $T$ and the input number $n$, and decomposes $n$ into a sum of squares $1 + n_1^2 + n_2^2 + \ldots, n_k^2$. 
While $n$ is guaranteed to have at least one, and often several squares decompositions \cite{pollack2018finding}, in our implementation we adopted a greedy strategy that iteratively selects the largest square in the number, \ie, $\lfloor \sqrt{n} \rfloor$. This operation can be easily implemented on arbitrary-precision integers (required by \textbf{(R4)}) via a binary search.
Next, $k$ siblings $[T_1, T_2, \ldots, T_k]$ are generated (one for each member $n_i$ of the decomposition of $n$), and assigned as children of $T$. Finally, the procedure is recursively invoked for the pairs $(T_1, n_1), \ldots, (T_k,n_k)$. Whenever $1$ is given as input number, the algorithm returns. 
Due to the commutativity of addition, reordering siblings does not change the encoded number, thus respecting \cref{eq:unordered-bit-tree}. Since each number of the decomposition $n_i$ is guaranteed to be strictly smaller than $n$, the procedure always terminates. 

\begin{algorithm}[t]
\small
\caption{The Squares Bit-Tree Encoding}
\label{alg:bit-tree}
\SetAlgoLined
\DontPrintSemicolon
\SetKwFunction{SquareDecomposition}{SQUARE\_DECOMPOSITION}
\SetKwFunction{NatToTree}{NAT\_TO\_TREE}
\SetKwFunction{TreeToNat}{TREE\_TO\_NAT}
\SetKwInOut{Input}{Input}\SetKwInOut{Output}{Output}

\Procedure{\SquareDecomposition{$n$}}{
    $n \gets n - 1; \enspace ns \gets []$\;
    \While{$n \neq 0$}{
        Append $\lfloor \sqrt{n} \rfloor$ to $ns$\;
        $n \gets n - \lfloor \sqrt{n} \rfloor^2$\;
    }
    \Return{$ns$}\;
}

\Procedure(\tcp*[f]{Implements $f'$ of \cref{eq:nat-bits}}){\NatToTree{$T$, $n$}}{
    $[n_1, n_2, \ldots, n_k] \gets$ \SquareDecomposition{$n$}\;
    $[T_1, T_2, \ldots, T_k] \gets$ generate $k$ new nodes\;
    \ForEach{$T_i \in [T_1, T_2, \ldots, T_k]$}{
        \NatToTree{$T_i, n_i$}\;
    }
    $C(T) \gets [T_1, T_2, \ldots, T_k]$\tcp*{Update children of $T$}
}

\Procedure(\tcp*[f]{Implements $g'$ of \cref{eq:nat-bits}}){\TreeToNat{$T$}}{
    $[n_1, n_2, \ldots, n_k] \gets [$\TreeToNat{$T_i$} for $T_i \in C(T)]$\;
    \Return{$1+\sum_{n_i \in [n_1,\ldots,n_k]} n_i^2$}\;
}

\end{algorithm}

\begin{figure}
    \includegraphics[width=0.9\linewidth]{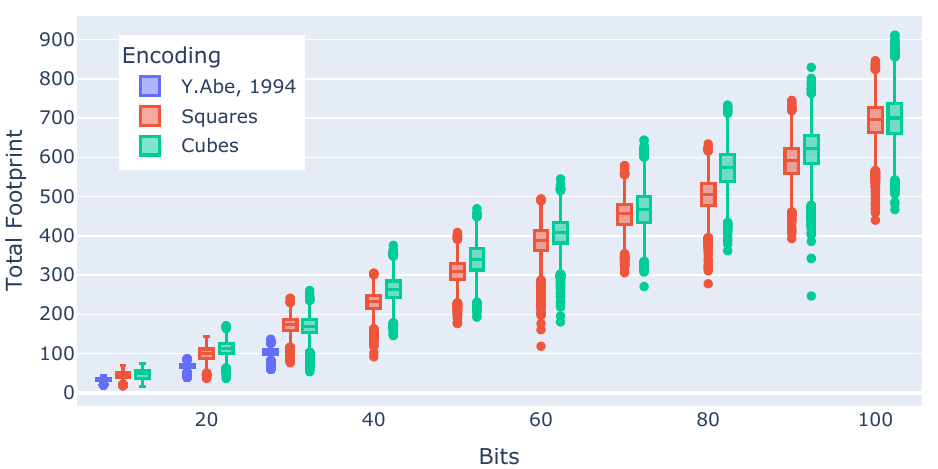} 
    \caption{The total footprint of the Squares, Cubes, and Abe's encoding.}
    \label{fig:footprint-vs-bits}
\end{figure}

\paragraph{Evaluation against {(R1-R4)}.}
We measured the footprint performance of three bit-tree encodings in \cref{fig:footprint-vs-bits}. For each bit string length, we generated 9002 samples, of which 6302 by varying the probability of ones from 0 to 1, and 2700 samples with ascending lag-1 autocorrelation (\ie, the probability that adjacent bits are the same) from $0.1$ to $0.9$. The generated dataset hence contains a mixture of fully random strings and strings with long consecutive sequences of zeros and ones, in light of \textbf{(R3)}. 
We found Abe's encoding to best satisfy \textbf{(R1)}, \textbf{(R2)}, and \textbf{(R3)}, and hence adopted it as baseline. However, Abe's bijection is based on prime factorization, thus becoming unfeasible for strings above $40$ bits, and violating \textbf{(R4)}. Squares has a median footprint $1.34$ to $1.67$ times larger than Abe's \cite{abe1994tree} encoding, with a standard deviation $1.75$ to $2.28$ times higher. While the increase is significant, Squares easily scales to thousands of bits, satisfying \textbf{(R4)}. Finally, \cref{fig:footprint-vs-bits} includes \textit{Cubes}, a variation of Squares where the input number is decomposed into a sum of cubes, showcasing it as an example result from various tests we conducted. We evaluated a total of $15$ encodings, but excluded most of them from the paper due to space constraints, and ultimately selected Squares for its best overall performance.

\begin{figure}
    \includegraphics[width=0.45\textwidth]{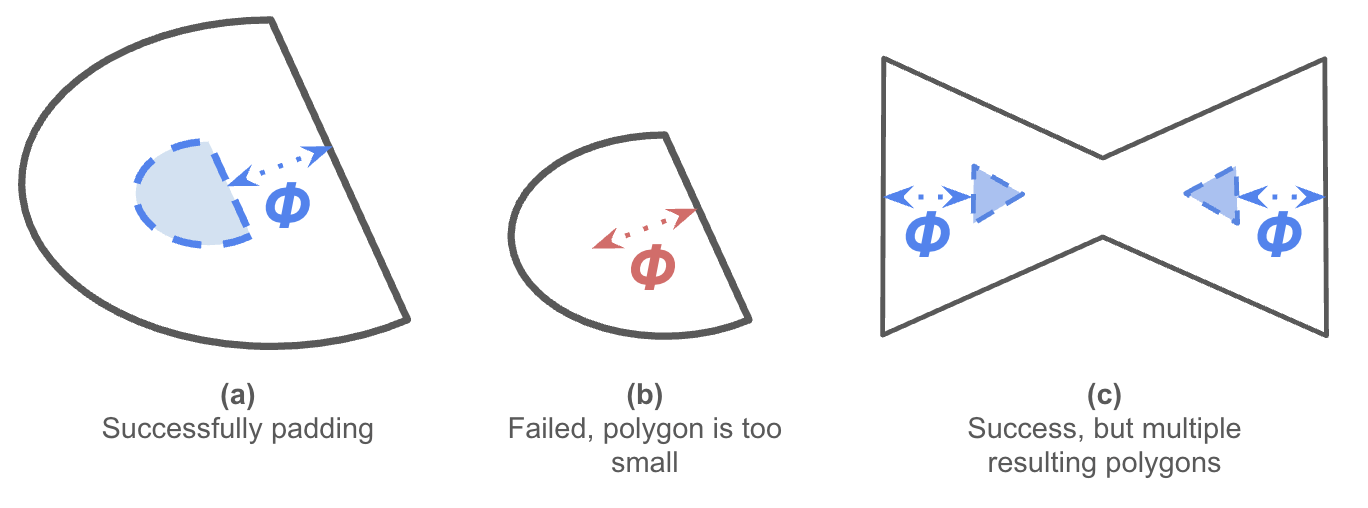} 
    \caption{The three possible outcomes of padding.}
    \label{fig:padding}
\end{figure}

\section[generation]{Packer}\label{sec:packer}

\begin{figure*}
  \centering
  \begin{subfigure}[b]{0.22\linewidth}
    \includegraphics[width=\textwidth]{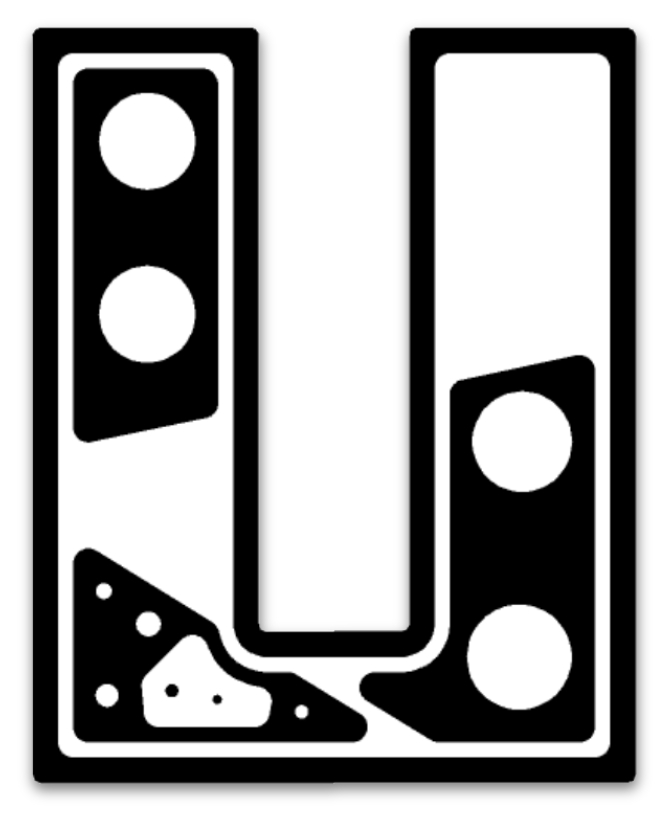}
    \caption{Seedmarkers, 2021}
    \label{fig:concave-a}
  \end{subfigure}
  \begin{subfigure}[b]{0.22\linewidth}
    \includegraphics[width=\textwidth]{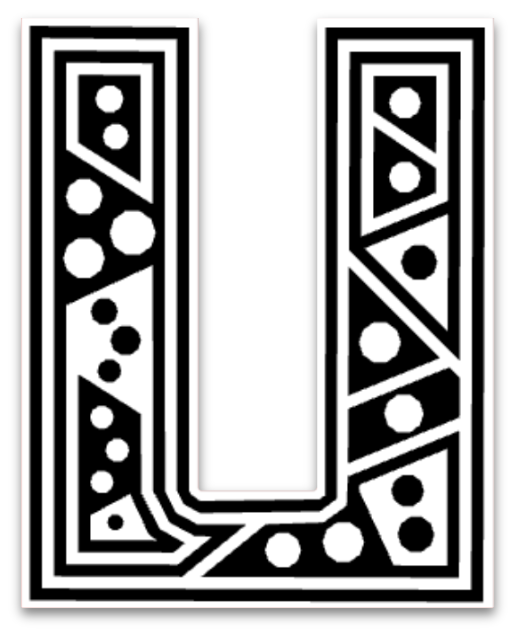}
    \caption{Claycode, \textit{``U"}}
    \label{fig:concave-b}
  \end{subfigure}
  \begin{subfigure}[b]{0.22\linewidth}
    \includegraphics[width=\textwidth]{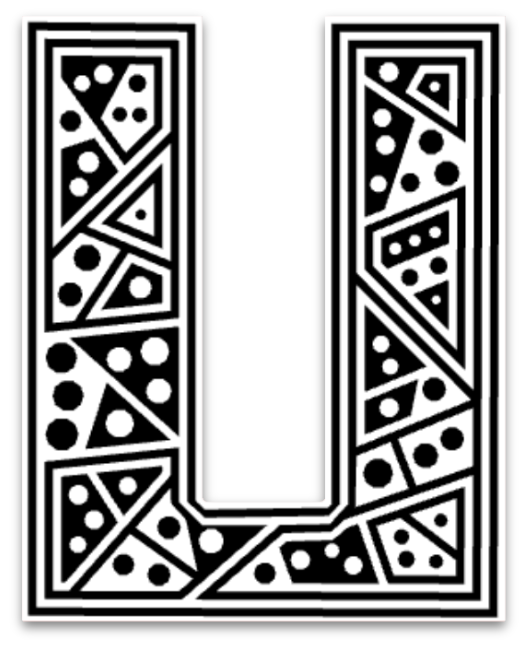}
    \caption{Claycode, \textit{``U and I"}}
    \label{fig:concave-c}
  \end{subfigure}
  \begin{subfigure}[b]{0.22\linewidth}
    \includegraphics[width=\textwidth]{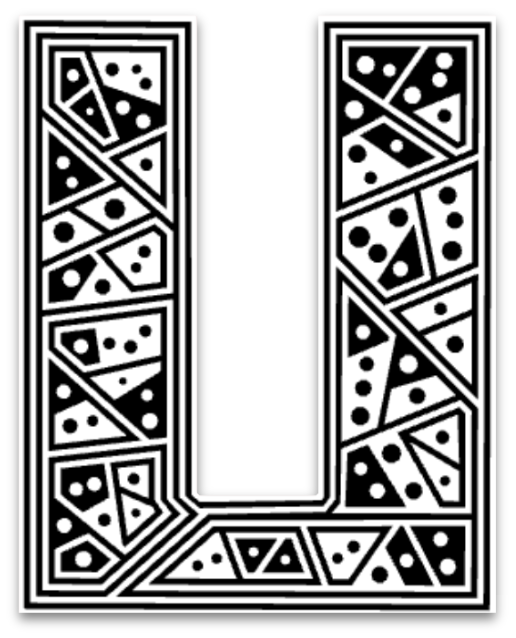}
    \caption{Claycode, 
    \textit{``U got this!"}}
    \label{fig:concave-d}
  \end{subfigure}
  \caption{Claycode's packer when compared to Seedmarkers'.}
  \label{fig:concave}
\end{figure*}

The final step of Claycode's encoding pipeline (\cref{fig:pipeline}) is the \textit{packer}, the component that renders a given topology tree within an input polygon. The packer is, in one form or another, present in the work of existing topological markers (\eg, \cite{dtouch0,reactivision,yu2020topotag}). However, Claycodes are not primarily intended to be fiducial markers, but rather to carry bits of payload. They must thus scale well to complex topologies (hundreds of nodes) and support concave shapes. A notable approach in existing literature is Seedmarkers \cite{seedmarkers}, where packing is solved by recursively computing a weighted Voronoi diagram. Although the Seedmarker's approach has the potential to work for Claycode's use case, it is mainly designed for pose estimation (\eg, leaves are always circles), instead of focusing on maximizing the scannability of complex topologies. Moreover, it fails on some highly concave shapes (\cref{fig:concave}). In the remainder of this section, we detail Claycode's packer, which works on convex and non-convex polygons without self-intersections or holes, and produces scannable Claycodes with thousands of nodes in seconds.

\subsection{The Packer Algorithm}
\begin{figure}
  \centering
  \begin{subfigure}[b]{0.2\linewidth}
    \includegraphics[width=\textwidth]{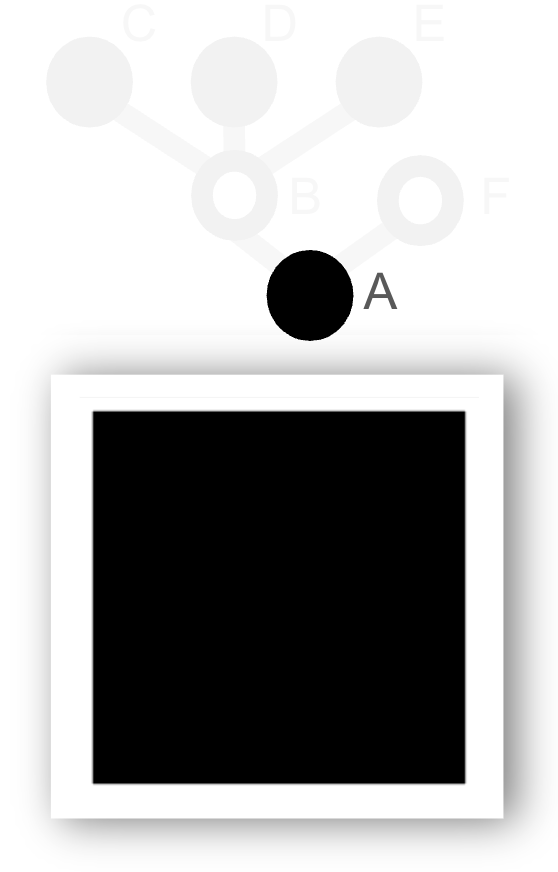}
    \caption{}
    \label{fig:packer-a}
  \end{subfigure}
  \hfill
  \begin{subfigure}[b]{0.2\linewidth}
    \includegraphics[width=\textwidth]{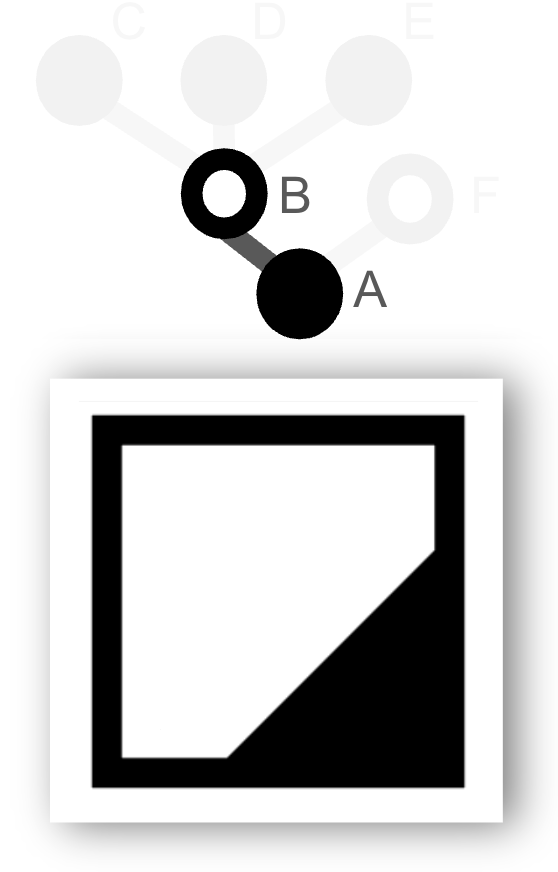}
    \caption{}
    \label{fig:packer-b}
  \end{subfigure}
  \hfill
  \begin{subfigure}[b]{0.2\linewidth}
    \includegraphics[width=\textwidth]{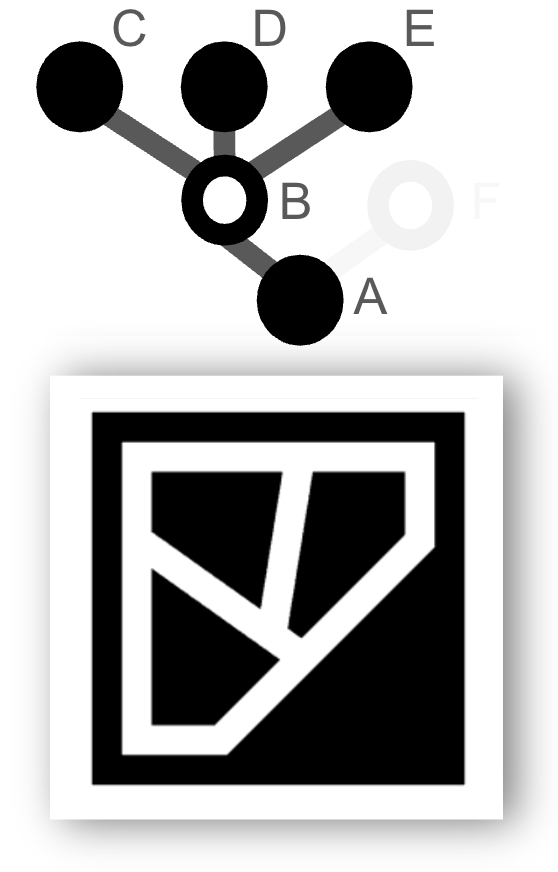}
    \caption{}
    \label{fig:packer-c}
  \end{subfigure}
  \hfill
  \begin{subfigure}[b]{0.2\linewidth}
    \includegraphics[width=\textwidth]{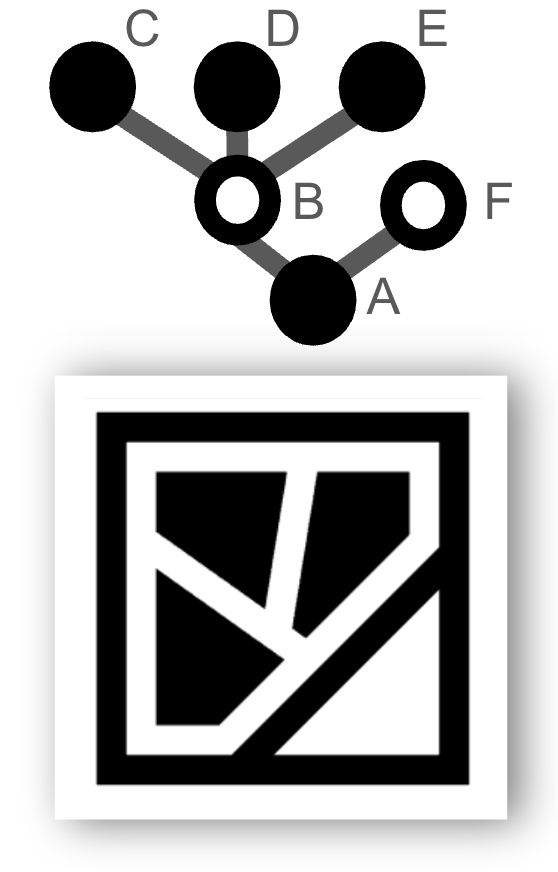}
    \caption{}
    \label{fig:packer-d}
  \end{subfigure}
  \caption{\textit{Packing} a tree, whose nodes have been highlighted for illustrative purposes, into a square polygon.}
  \Description{}
  \label{fig:packer}
\end{figure}

The packer, detailed in \cref{alg:packer}, takes in input a tree $T$, a polygon $P$, and a padding constant $\phi$. The algorithm implements a depth-first visit of $T$, where, at each invocation of the function, a new color region (\ie, node of the topology tree) is drawn. The padding constant $\phi$ defines the desired minimum thickness of each region. 
Each packing step applies two \textit{padding} and one \textit{partitioning} operations to $P$. Padding refers to contracting the boundaries of a polygon, creating a new polygon that is uniformly offset inward from the original. Partitioning means dividing the polygon into non-overlapping sub-polygons. In the following, we aim for the reader to build a visual intuition of the algorithm, and later discuss the padding and partitioning operations. \Cref{fig:packer} breaks down the packing process for a simple tree. Starting with the square input shape and the root node $A$, a new region is drawn (\cref{fig:packer-a}) after padding the initial polygon. Then, after the second round of padding, the resulting polygon is partitioned, allocating some space for $B$ and $F$. The packer assigns more of the available space to $B$, since it has a larger footprint (recall \cref{eq:footprint}). The procedure is then recursively invoked, resulting in \cref{fig:packer-b}. Similarly, the polygon associated to $B$ is then partitioned into three polygons, to make space of $C$, $D$, and $E$. \Cref{fig:packer-c} is the result of the next three iterations of the algorithm. $C$, $D$, and $E$ share a similar amount of space, as they are all leaves. Finally, the packer is invoked for $F$ (\cref{fig:packer-d}), resulting in a completed Claycode.

Note that if $\phi$ is too large the padding operation can fail, resulting in the failure of the entire procedure. Conversely, partitioning always produces a valid set of polygons. \Cref{alg:packer} heavily depends on the padding and partitioning steps, which we detail next.

\subsection{Padding Polygons} We adopt a standard padding implementation (\cref{fig:padding}) based on \cite{chen2005polygon}. Despite padding has three possible outcomes, we simplify the -- rarely observed -- scenario of \cref{fig:padding}c by selecting and returning the polygon with the largest area. %

\Cref{alg:packer} receives in input a padding constant $\phi$. While $\phi$ can be tuned to achieve different styles (\eg, a lower $\phi$ yields larger leaves but thinner color regions for intermediate nodes), we are generally interested in maximizing $\phi$ to increase the code's readability. To this end, a packing procedure typically involves multiple runs of \cref{alg:packer}, where a large $\phi$ is initially selected, and then gradually lowered until packing succeeds -- which is more and more likely to happen as $\phi$ approaches zero. It is important to note that, throughout \cref{alg:packer}, $\phi$ is kept constant, as opposed to dynamically changing whenever the padding operation fails. Varying $\phi$ is counterproductive, as it produces non-uniformly readable codes.

\begin{algorithm}[t]
\small
\caption{Packing Algorithm: attempt to draw a tree within a given input polygon.}
\label{alg:packer}
\KwIn{A tree node $T$, a polygon $P$, a padding constant $\phi > 0$}
\KwOut{$\text{SUCCESS}$ if the tree was fully drawn, $\text{FAIL}$ otherwise}
\DontPrintSemicolon
\SetKwFunction{PACK}{PACK}
\SetKwFunction{PAD}{PAD}
\SetKwFunction{PARTITION}{PARTITION}

\Procedure{\PACK{$T$, $P$, $\phi$}}{
    $P' \gets$ \PAD{$P$, $\phi/2$} \tcp*{Apply first padding}
    \lIf{$P'$ is not valid}{\Return $\text{FAIL}$ \tcp*[f]{$P$ was too small.}}
    
    Draw {$P'$}

    $P'' \gets$ \PAD{$P'$, $\phi/2$} \tcp*{Apply second padding}
    \lIf{$P''$ is not valid}{\Return $\text{FAIL}$ \tcp*[f]{$P'$ was too small.}}

    $[P''_1, \dots, P''_k] \gets$ \PARTITION{$P''$, $C(T)$}

    \ForEach{$T_i \in C(T)$}{
        \lIf{\PACK{$T_i$, $P''_i$, $\phi$} failed}{
            \Return $\text{FAIL}$
        }
    }
    \Return $\text{SUCCESS}$
}
\end{algorithm}

Finally, each step of \cref{alg:packer} involves two padding operations, which are crucial for two reasons. First, relying on a single padding operation would create uneven thicknesses in intermediate regions: if adjacent polygons were padded independently by $\phi$, the resulting distance between them would be of $2\phi$, creating inconsistencies. Second, the additional padding round of \cref{alg:packer} is applied also in the case of leaf nodes. While $P''$ is not used in that case, the procedure still terminates if padding fails. This effectively enforces a minimum thickness of $\phi$ for leaf regions.

\subsection{Partitioning polygons} Partitioning takes as input a polygon $P$ and a list of sibling nodes $[T_1, T_2, \ldots, T_k]$, and returns a list of polygons $[P_1, P_2, \ldots, P_k]$ such that $\bigcup_{i=1}^k P_i = P \land \bigcap_{i=1}^k P_i = \emptyset$. We call $\mathcal{P}(P)$ the set of all possible partitions of $P$.
Finding the optimal partitioning is a polygon decomposition problem \cite{MARKKEIL2000491} where we aim to distribute the polygon's area to the sibling nodes in a way that increases scannability. We capture the packer's requirements with two principles: 

\begin{enumerate}
    \item \textbf{Area Proportionality:} The area of each sub-polygon $A(P_i)$ should be proportional to the footprint of each sibling $F(T_i)$. We define the optimal area occupied by the polygon $P_i$ associated with a sibling $T_i$ as:
    \begin{equation}
        A^*(P_i) = A(P) \cdot \frac{F(T_i)}{\sum_{j=1}^k F(T_j)}
    \end{equation}
    \item \textbf{Circularity Maximization:} To improve readability, Polygons must avoid irregularities or elongations. We evaluate the quality of a polygon with its \textit{circularity} \cite{cox1927method}, which peaks at $1$ for a perfect circle and approaches zero for polygons with a disproportionately large perimeter relative to their area. The circularity $R$ of a polygon $P$ is defined as: 
    \begin{equation}
        R(P) = 4 \pi \frac{A(P)}{L(P)^2}
    \end{equation}
    where $L(P)$ is the perimeter of $P$.
\end{enumerate}

\begin{figure}
    \centering
\includegraphics[width=0.95\linewidth]{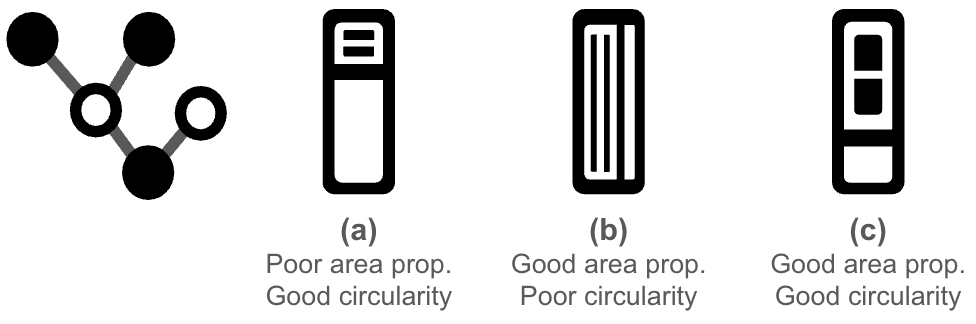} 
    \caption{Three instances of packing. The topology on the left is fit in a tall and narrow rectangle, showing the importance of both circularity and area distribution to maximize the readability of the code.}
    \label{fig:partitioning}
\end{figure}

A visual example of these two principles at work is given in \cref{fig:partitioning}. Naturally, most times circularity and area proportionality conflict with each other, so they must be weighted into a single function, resulting in the following minimization problem: 

\begin{equation}
\min_{[P_1, \ldots, P_k]\in \mathcal{P}} \sum_{i=1}^k \alpha \left| A(P_i) - A^*(P_i) \right| + (1-\alpha)\left(1-R(P_i)\right).
\label{eq:packermin}
\end{equation}

As with $\phi$, the constant $\alpha$ can be tuned to achieve different aesthetics, but in our implementation we fixed it to $0.6$, thus biasing the optimization towards a better area proportionality.
Instead of directly searching for a solution of \cref{eq:packermin}, we first solve the binary instance of the problem (\textit{\ie}, involving only two siblings), and then lift the solution to the general case. A simplified formula for the binary case (derived in \cref{sec:appendix-packer-proof}) is given by:

\begin{equation}
\label{eq:packermin-binary}
\min_{[P_1, P_2]\in \mathcal{P}} \alpha\left|A(P_1) - w_1A(P)\right| + (1-\alpha)\left(1 - \frac{R(P_1)+R(P_2)}{2}\right)
\end{equation}

where $w_1$ is the fraction of the total area that we intend $P_1$ to occupy, \textit{\ie}, $\frac{F(T_1)}{F(T_1)+F(T_2)}$. %
A solution to \cref{eq:packermin-binary} is found using a straightforward random search (\cref{alg:polygon_cut}). In our solution, we prune the search space by only considering partitions that can be obtained through straight-line cuts. A \textit{cut} of a polygon $P$ is a binary partition of $P$ formed by a segment whose endpoints lie on the perimeter of $P$. In our implementation, we upper-bounded the search to $400$ solutions, and stopped early in case of convergence. %

Using \cref{alg:polygon_cut} as a building block, we define the \FuncSty{PARTITION} procedure in \cref{alg:partitioning}. Each iteration of the algorithm processes a single sibling $T_i$, weighting its footprint against the combined footprint of all its rightmost siblings $T_{i+1}, \ldots, T_k$. The subsequent invocation of \cref{alg:polygon_cut} produces two new polygons: $P_i$, which is assigned to $T_i$, and $P_r$, dedicated to all the rightmost siblings of $T_i$. Repeating this process $k-1$ times results in a full partition.

\begin{figure}
    \includegraphics[width=0.44\textwidth]{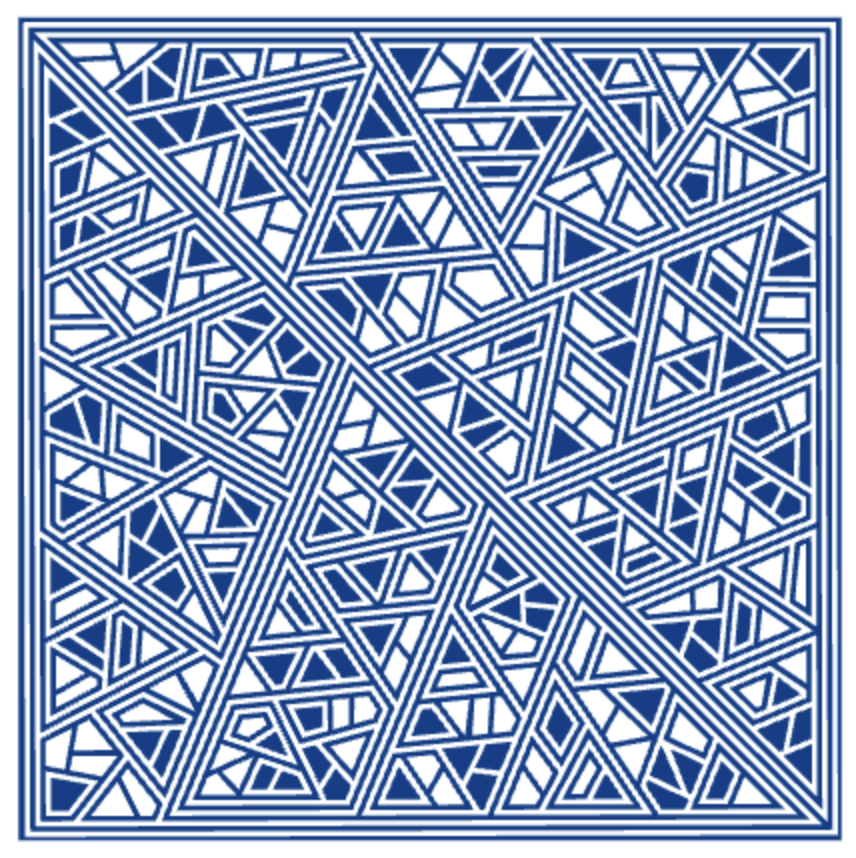} 
    \caption{\textit{``This Claycode packs more than 500 nodes \emojisparkles"}}
    \label{fig:many-nodes}
\end{figure}

\begin{figure*}
    \centering
    \includegraphics[width=0.9\linewidth]{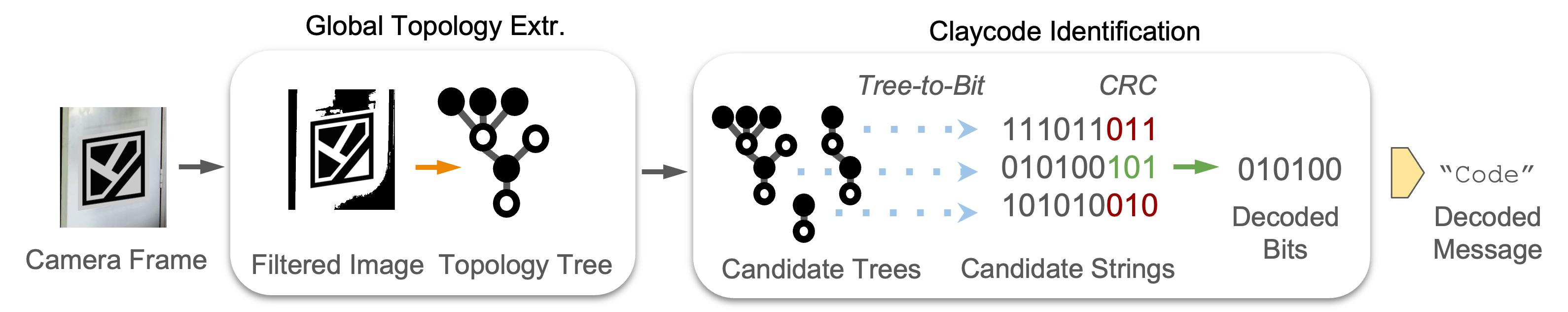}
    \caption{The pipeline followed by the \textit{scanner}.}
    \label{fig:scanner}
\end{figure*}

\section[generation]{Scanner}\label{sec:scanner}
To complete the Claycode's pipeline we discuss the \textit{scanner}, a module that performs the inverse of the encoding pipeline (\cref{fig:pipeline}), retrieving the original message from a visual representation of a Claycode. 

\begin{algorithm}
\small
\caption{Searching for a cut $(P_1, P_2)$ of $P$ such that the area of $P_1$ is close to $A(P) \cdot w_1$ and both $P_1$ and $P_2$ have good circularity.}
\label{alg:polygon_cut}
\SetAlgoLined
\SetKwFunction{PolygonCut}{POLYGON\_CUT}
\DontPrintSemicolon
\KwIn{A polygon $P$ and a percentage $w_1 \in (0,1)$}
\KwOut{Two polygons $P_1$ and $P_2$}
\Procedure{\PolygonCut{$P$, $w_1$}} {
    $best\_error \gets \infty$\;
    $best\_cut \gets \emptyset$\;
    \While{not converged}{
        $P_1, P_2 \gets$ select a random cut of $P$\;
        $\epsilon_a \gets |A(P_1) - w_1 \cdot A(P)|$ \tcp*{Area prop. error}
        $\epsilon_c \gets 1 - \frac{R(P_1) + R(P_2)}{2}$ \tcp*{Circularity error}
        $error \gets \alpha \cdot \epsilon_a + (1-\alpha) \cdot \epsilon_c$\;
        \If{$error < best\_error$}{
            $best\_cut, best\_error \gets (P_1, P_2), error$\;
        }
    }
    \Return{$best\_cut$}\;
}
\end{algorithm}

\begin{algorithm}
\small
\caption{Partitioning a polygon into $k$ sub-polygons}
\label{alg:partitioning}
\KwIn{A list of sibling nodes $[T_1, \dots, T_k]$ and a polygon $P$}
\KwOut{$k$ sub-polygons, each relative to a sibling}
\SetKwFunction{Partition}{PARTITION}
\DontPrintSemicolon
\Procedure{\Partition{$[T_1, \dots, T_k]$}} {
    $partitions \gets []; \enspace P_r \gets P$ \tcp*{$P_r$ starts as the whole $P$}
    
    \For{$T_i \in [T_1, \dots, T_k]$}{
        \eIf{$i = n$}{
            Append $P_r$ to $partitions$ \tcp*{Final polygon}
        }{
            $weight \gets \frac{F(T_i)}{\sum_{j=i}^k F(T_j)}$ \tcp*{Area fraction of $T_i$}
            $P_i, P_r \gets$ \texttt{POLYGON\_CUT}($P_{r}, weight$)
            
            Append $P_i$ to $partitions$
        }
    }
    \Return $partitions$
}
\end{algorithm}

This module is designed to run on a smartphone, using the camera to frame the code and extract the information contained. Unlike most 2D codes, %
Claycodes do not have a predefined shape or visual anchor, such as the finding patterns of QR codes~\cite{tiwari2016introduction}. The Claycode's scanner cannot hence rely on the classic paradigm of first isolating the subset of the input image containing the code and then focusing on its pixels for decoding. 
Our method, illustrated in \cref{fig:scanner}, begins instead by transforming the input image into a \textit{global topology tree}, without explicitly searching for a Claycode. We call this phase \textit{global topology extraction}. The underlying assumption is that, if a Claycode is framed within the input image, then its topology must be a subtree of the global topology tree. Once the global topology tree is extracted, the input image is discarded, and the scanner analyses the topology to extract all the contained messages. We call this phase \textit{Claycode identification}.

\begin{figure*}[ht]
    \centering
    \begin{subfigure}[b]{0.48\linewidth}
    \centering
        \includegraphics[height=0.65\linewidth]{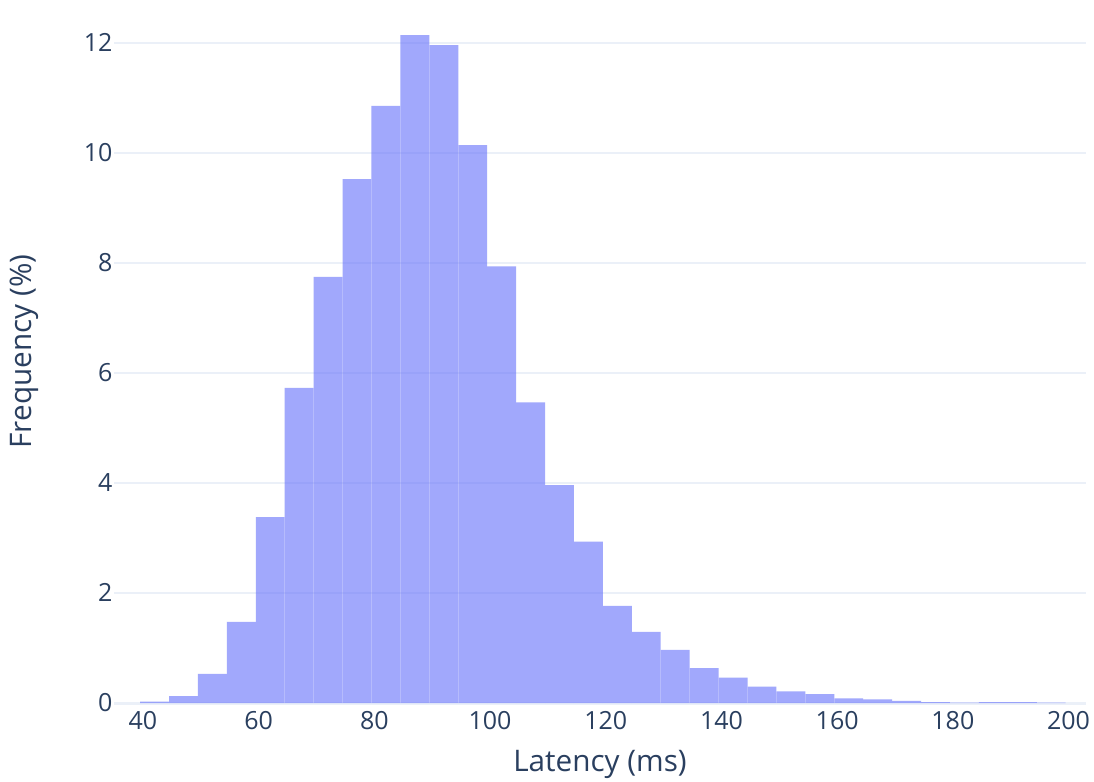}
        \caption{Latency distribution of unsuccessful scanning attempts ($10,355$ samples)\crmodsecond{.}}
        \label{fig:latencies}
    \end{subfigure}
    \hfill
    \begin{subfigure}[b]{0.48\linewidth}
    \centering
        \includegraphics[height=0.65\linewidth]{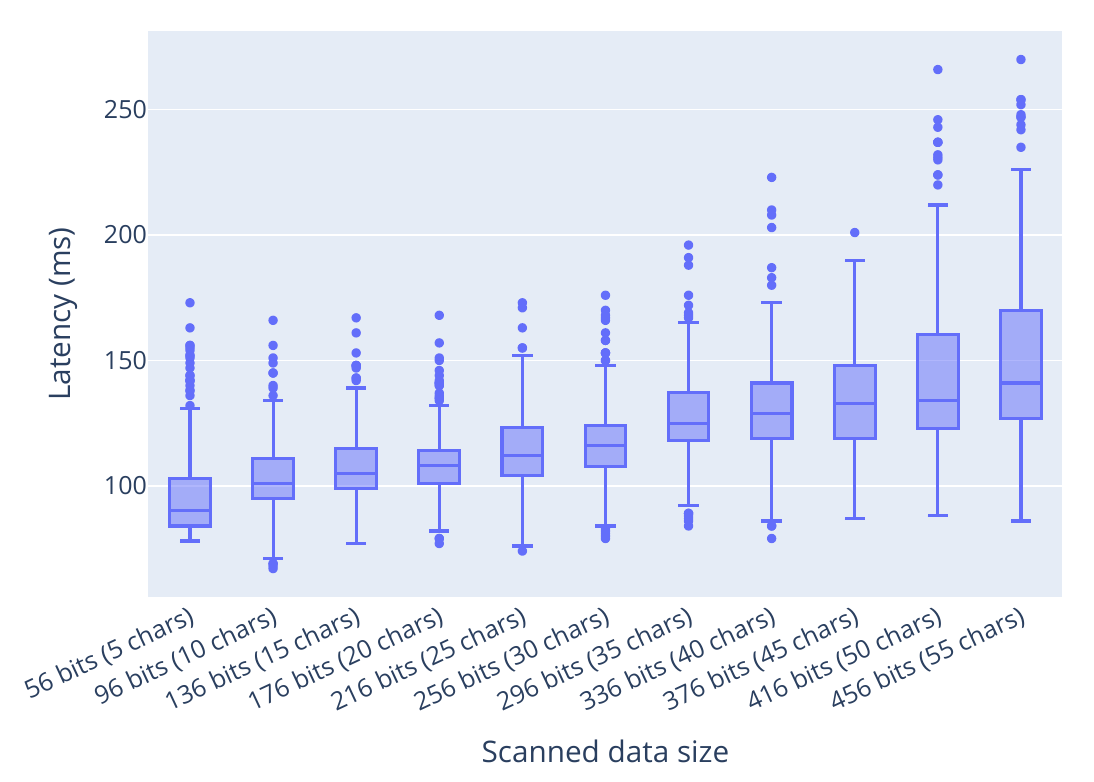}
        \caption{\crmod{Latency of successful scans with growing code size in bits ($3,410$ samples)}\crmodsecond{.}}
        \label{fig:latencies-wrt-code-size}
    \end{subfigure}
    \caption{\crmod{Scanner pipeline end-to-end latency analysis.}}
    \label{fig:combined-latencies}
\end{figure*}

\begin{figure}
    \centering
    \includegraphics[width=0.9\linewidth]{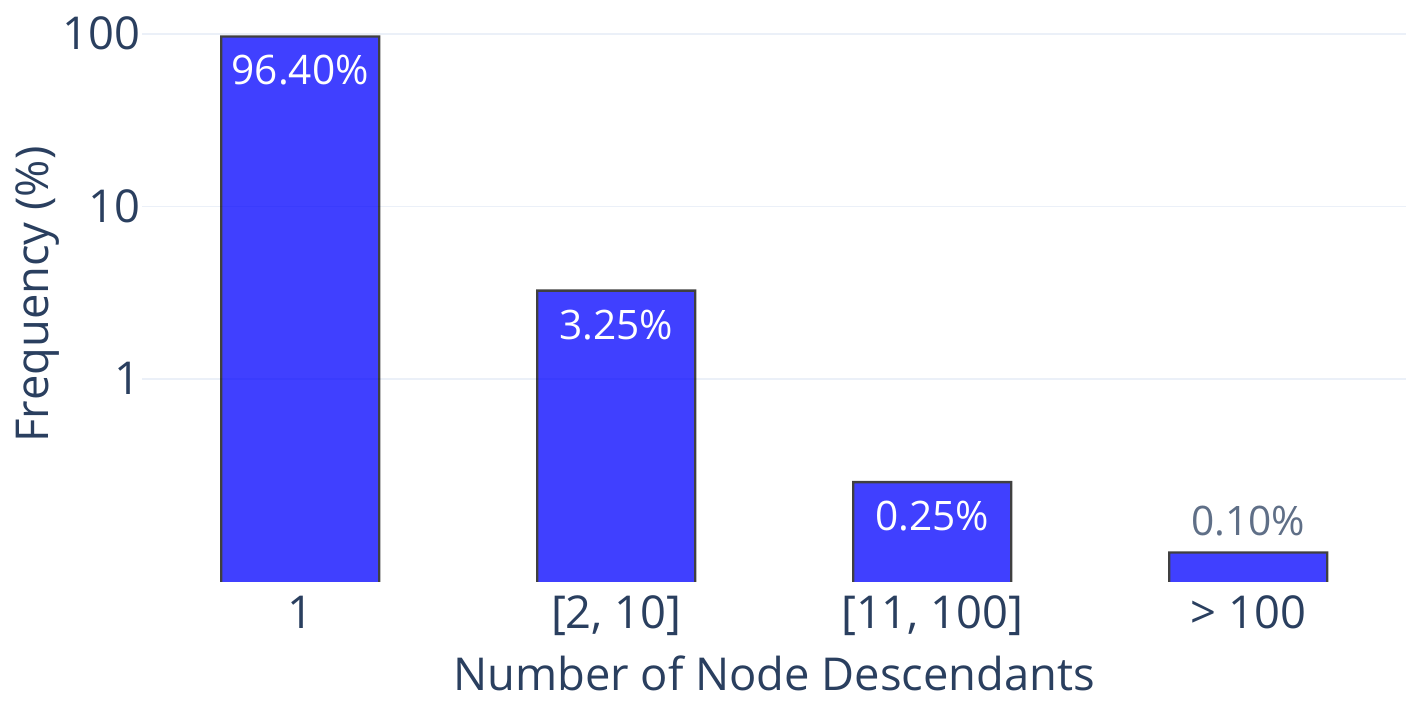} 
    \caption{Number of descendants for each node in $1,039$ global topology trees (six million nodes) derived from real-world images.}
    \label{fig:descendants}
\end{figure}

\subsection{Global Topology Extraction}
The procedure begins with a camera frame, which is filtered and then processed to build its topological structure.
The process is designed to prepare an image for contour detection by enhancing important features while reducing irrelevant details. 

\begin{figure}
    \includegraphics[width=0.37\textwidth]{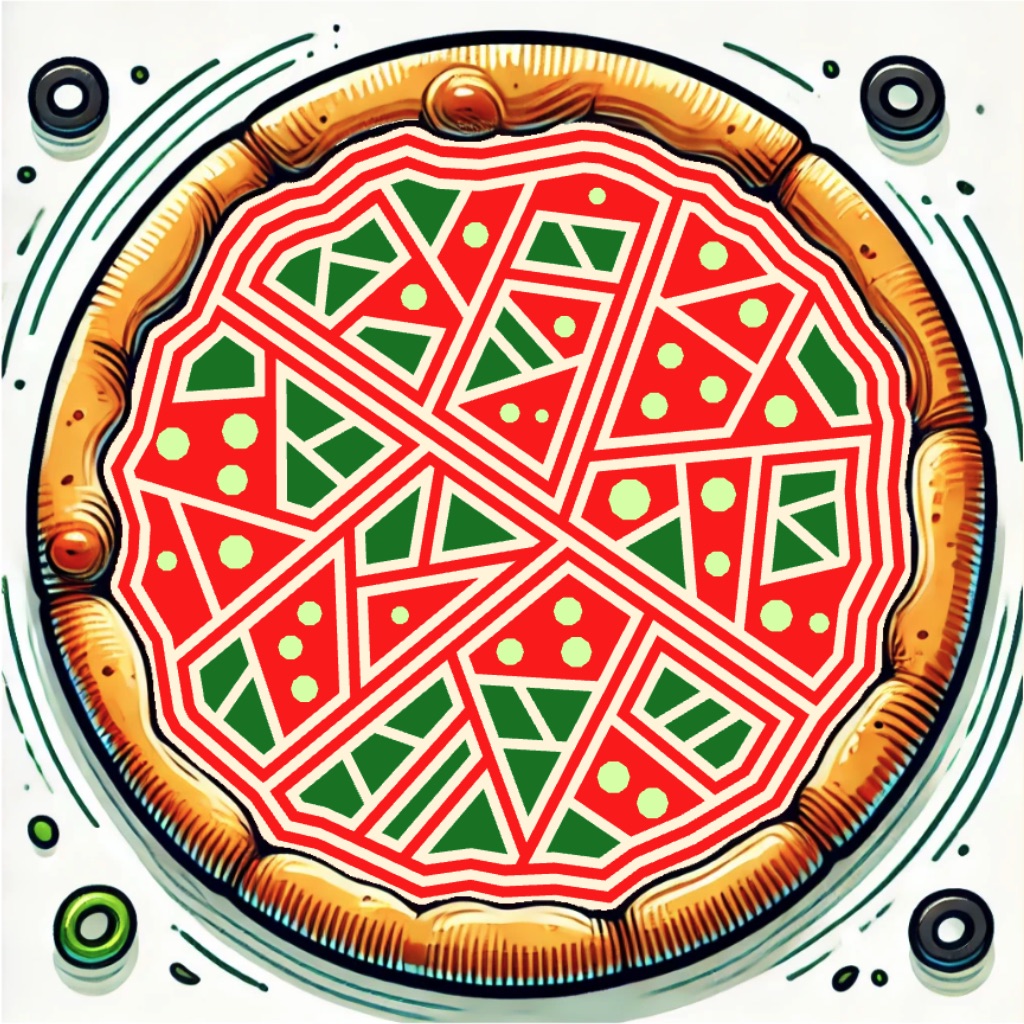} 
    \caption{\textit{``Pizza!"}}
    \label{fig:pizza}
\end{figure}

\begin{figure*}
    \centering
    \begin{subfigure}[b]{0.9\linewidth}
        \includegraphics[width=\textwidth]{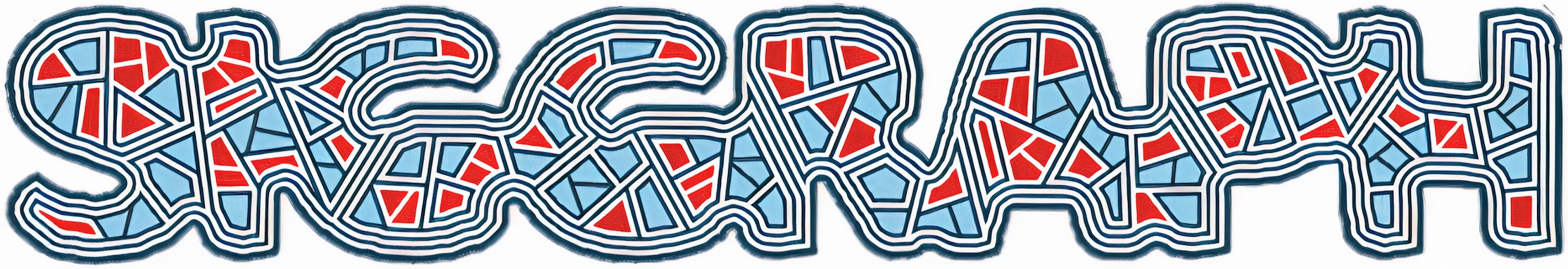}
    \end{subfigure}
    \caption{\textit{``SIGGRAPH"}}
    \label{fig:siggraph}
\end{figure*}

Let $I_{\text{RGB}}(x,y)$ be the input RGB image, with $x, y$ as their pixel coordinates. 
First, the image is converted to grayscale $I_{\text{gray}}(x,y)$, simplifying the input to intensity variations. Then, to remove noise while preserving edges, a bilateral filter \cite{bilateral} is applied, producing $I_{\text{bil}}(x,y)$, which smooths the image while retaining boundary details. The filtered image is then binarized through adaptive thresholding:
\begin{equation}
B(x, y) =
\begin{cases}
1, & \text{if } I_{\text{bil}}(x, y) > \tau(x, y), \\
0, & \text{otherwise}.
\end{cases}
\end{equation}
Here $\tau(x, y) = \frac{1}{n} \sum_{(i, j) \in N(x, y)} I_{\text{bil}}(i, j) - K$ is the local threshold value calculated for the pixel $(x, y)$  based on the intensity values in its neighborhood $N(x, y)$. The size of $N(x, y)$ is defined by the block size, which is typically a small odd integer depending on the size of the input image; $n$ is the square block size and $K$ is a constant subtracted from the mean to fine-tune the threshold.
$B(x, y)$ serves as an input for the Suzuki-Abe algorithm~\cite{Suzuki1985TopologicalSA,opencv_library} for contour extraction, where the boundaries of connected regions are identified and organized into a hierarchy (algorithm is detailed in \cref{sec:appendix-contours}), resulting in the global topology tree of the image $T_{f}$.
Note that the thresholding operation does not prevent the use of colors (\cref{fig:pizza,fig:siggraph,fig:many-nodes}), but neighboring regions must be sufficiently contrasted, in order to ensure that the boundary is detected.    

\subsection{Claycode Identification}
From this point forward, the scanner exclusively operates on $T_f$, identifying and decoding all Claycodes in the scene. On our test smartphone, $|D(T_f)|$ typically contains thousands of nodes. While some may represent the root of a Claycode, most are noise. To accelerate processing, we first construct a subset of candidate roots using a simple heuristic:

\begin{equation}
{\mathbf T} = \{T \in D(T_{f}) : |D(T)| > n \},
\end{equation}

In our implementation, $n=10$. The heuristic guarantees that no valid Claycode is discarded -- no Claycode is so simple as to contain only ten nodes -- and averagely prunes more than $~99\%$ of the nodes, since real-world topologies are predominantly flat (\cref{fig:descendants}).

Each identified potential Claycode $T \in {\mathbf T}$ is then transformed into a bit string using the function $g(T)$ described in \cref{sec:bit_to_tree_encoding}:
\begin{equation}
{\mathbf{B}} = \{ g(T) \mid T \in \mathbf{T}\}.
\end{equation}

At this point, we obtain a set of potential message candidates, containing a mixture of messages (one for each Claycode in the scene) and noise. To filter the valid messages, we designed the encoding pipeline to append a 16-bit Cyclic Redundancy Code (CRC), denoted as $b_{\text{CRC}}$, to each message $b_{\text{msg}}$. In our implementation, we adopted the same polynomial of the \texttt{CRC-15/CAN} protocol \ie, $0x4599$~\cite{BoschCAN}. As a final step, all candidate bit strings $b \in \mathbf{B}$ are validated using the CRC, resulting in the final list of decoded Claycode payloads:
\begin{equation}
\mathbf{B}^{*} = \{ b_{\text{msg}} \mid b = b_{\text{msg}} : b_{\text{CRC}}, \, b \in \mathbf{B}, \, \text{CRC}(b_{\text{msg}}) = b_{\text{CRC}} \}
\end{equation}

We found the CRC to be extremely effective avoiding false positives, especially in the case of damaged Claycodes. In \cref{sec:results} we have used our implementation to scan thousands of Claycodes without experiencing a single false positive. This is a significant achievement, as false positives are a key challenge in topological codes, as exemplified by SeedMarkers~\cite{seedmarkers} and D-Touch~\cite{costanza2003region}.

\paragraph{Redundancy scheme.} The CRC functions solely as an error detection mechanism and does not enable the recovery of partially corrupted messages. Traditional error correction schemes, such as~\cite{reed1960polynomial}, are not viable: a single error in the topology tree usually propagates through the tree-to-bit function into several bit flips and often even alters the length of the message. 
To provide some resistance to occlusion damage, we implement a simple redundancy scheme based on the fact that Claycode's scanner natively processes multiple candidate messages simultaneously. We modify the bit-tree encoding function of \cref{sec:bit_to_tree_encoding} to generate a new tree $T_R$ such that $C(T_R) = [f(b),f(b)]$. The approach can be extended to any number of copies, and critically improves the resistance of the codes, at the price of the code's data capacity. We refer to the number of copies in the Claycode as \textit{redundancy level} $R$, normally set to $1$ (\ie, only one copy of $f(b)$ is present).

\subsection{\crmod{Scanning Latency}}\label{subsec:scanning_latency}

\crmodsecond{To assess the performance of the scanner application, we measured the end-to-end latency for each scan attempt on our test device, a Google Pixel 8 Pro (12GB RAM, running Android 15). Images were captured using the primary back camera at a resolution of $1920 \times 1920$.} Although there is no hard upper limit on the duration of the scanning pipeline, we established a target average latency of $100ms$ under typical conditions. This target allows for $10$ scanning attempts per second on average, which aligns with the performance of existing scanning technologies~\cite{Yang_2018}.

\crmodsecond{
\paragraph{Experiment $1$ -- Latency of unsuccessful scans.}} We fixed the phone to a tripod and ran the scanner for eight minutes while it was pointed at a screen. The displayed footage featured scenes of varying complexity—such as portrait close-ups, landscapes, and abstract patterns—to fully exercise the pipeline. To incorporate real-world footage, we also handheld the phone and recorded for another eight minutes while walking indoors and outdoors. Notably, no Claycode was scanned during this experiment.
As shown in \cref{fig:latencies}, we achieved an average latency of $90.2ms$, with a standard deviation of $18ms$ and $10,335$ samples in the dataset. The minimum and maximum latency are $42ms$ and $196ms$, respectively.

\begin{figure*}
  \centering
  \begin{subfigure}[b]{0.33\linewidth}
    \includegraphics[trim=0.1cm 0.1cm 0.2cm 0.3cm, clip,width=\textwidth]{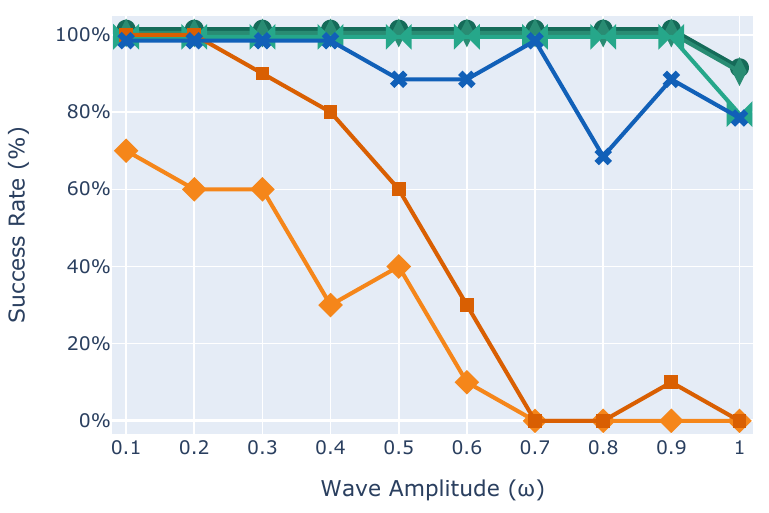}
    \caption{Exp. 1: deformation}
    \label{fig:evaluation-deformation}
  \end{subfigure}
  \hfill
  \begin{subfigure}[b]{0.33\linewidth}
    \includegraphics[trim=0.1cm 0.1cm 0.2cm 0.3cm, clip,width=\textwidth]{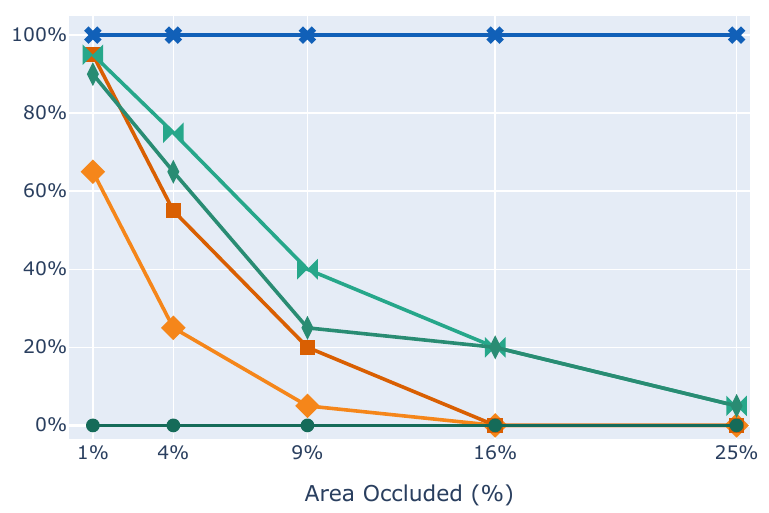}
    \caption{Exp. 2: occlusion}
    \label{fig:evaluation-occlusion}
  \end{subfigure}
  \hfill
  \begin{subfigure}[b]{0.33\linewidth}
    \includegraphics[trim=0.1cm 0.1cm 0.2cm 0.3cm, clip,width=\textwidth]{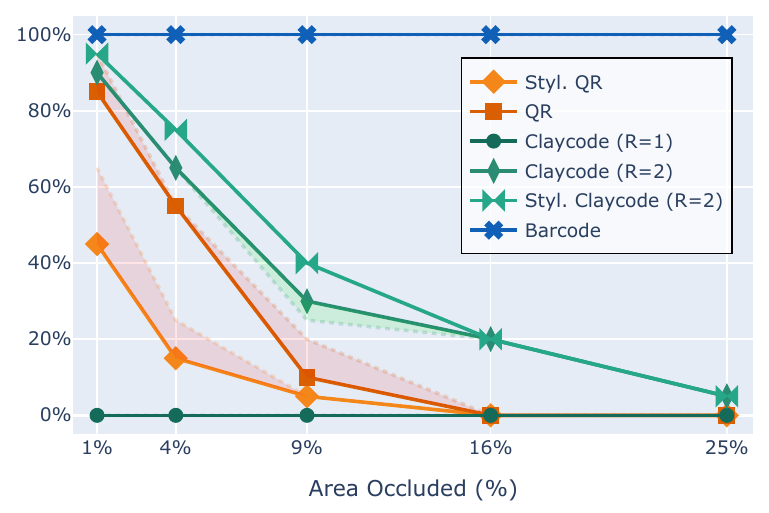}
    \caption{Exp. 3: occlusion + deformation}
    \label{fig:evaluation-occlusion-deformation}
  \end{subfigure}

    \begin{subfigure}[b]{0.138\textwidth}
    \includegraphics[width=\textwidth]{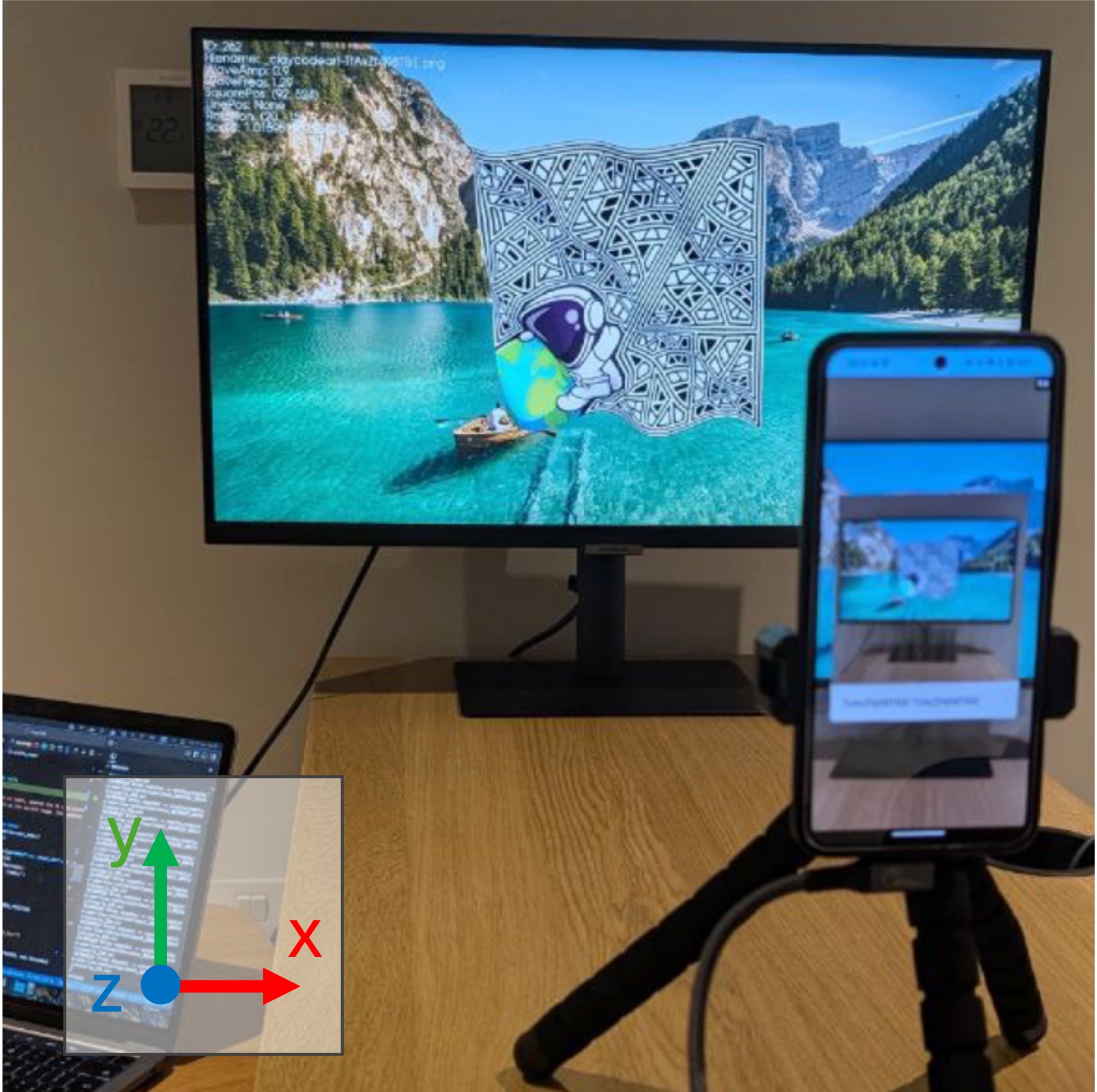}
    \caption{The test setup}
    \label{fig:eval-setup}
    \end{subfigure}
    \begin{subfigure}[b]{0.138\textwidth}
    \includegraphics[width=\textwidth]{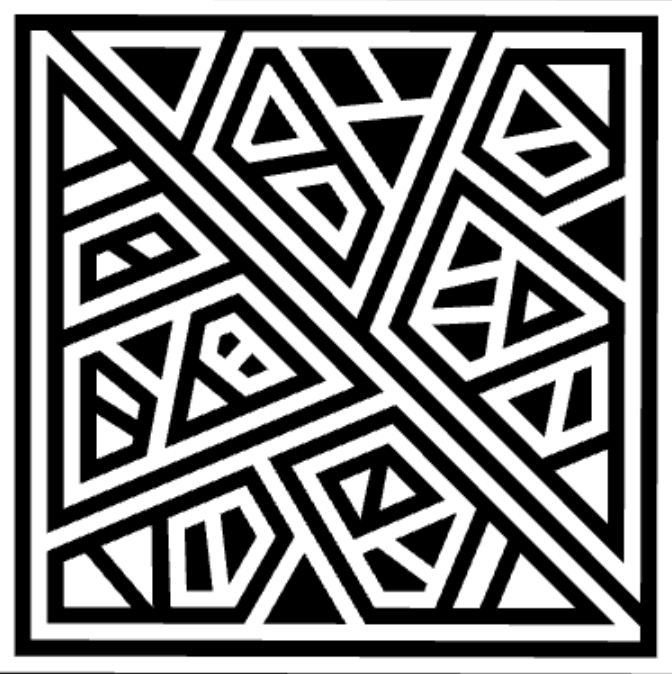}
    \caption{Claycode ($R=1$)}
    \end{subfigure}
    \hfill
    \begin{subfigure}[b]{0.138\textwidth}
    \includegraphics[width=\textwidth]{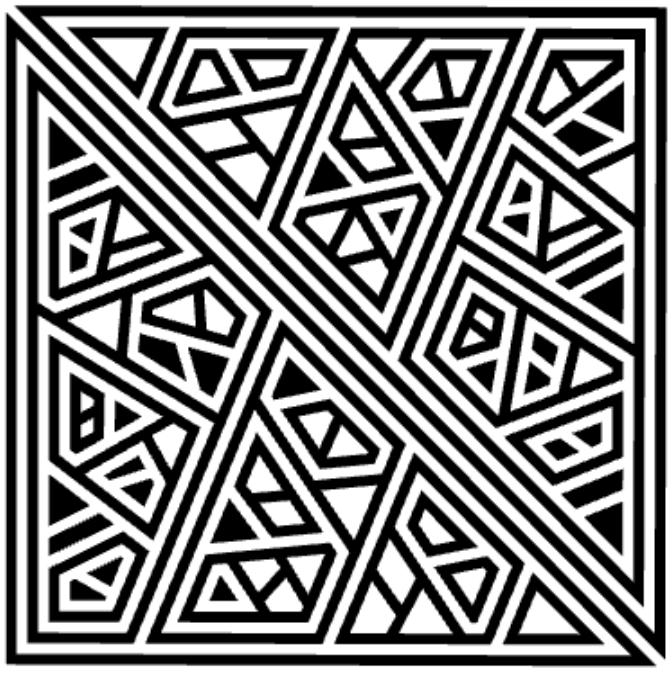}
    \caption{Claycode ($R=2$)}
    \end{subfigure}
    \hfill
    \begin{subfigure}[b]{0.138\textwidth}
    \includegraphics[width=\textwidth]{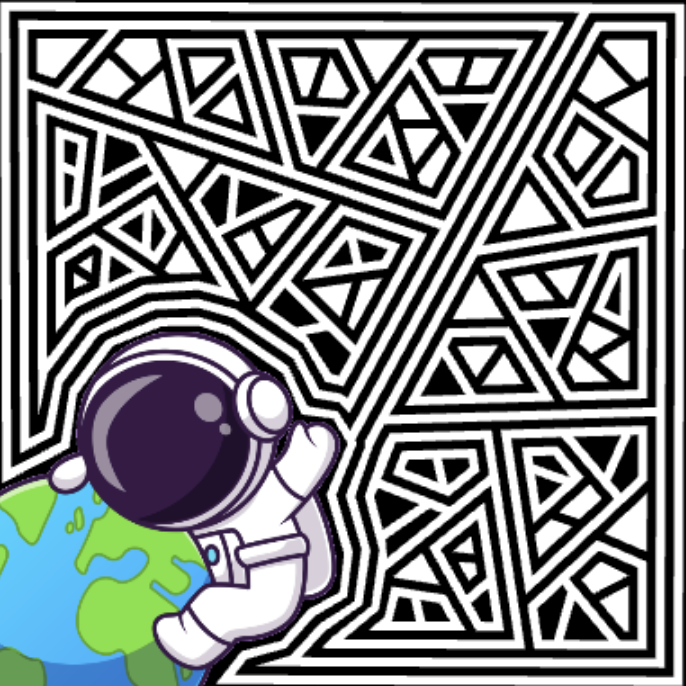}
    \caption{Styl. Clayc. ($R=2$)}
    \end{subfigure}
    \hfill
    \begin{subfigure}[b]{0.138\textwidth}
    \includegraphics[width=\textwidth]{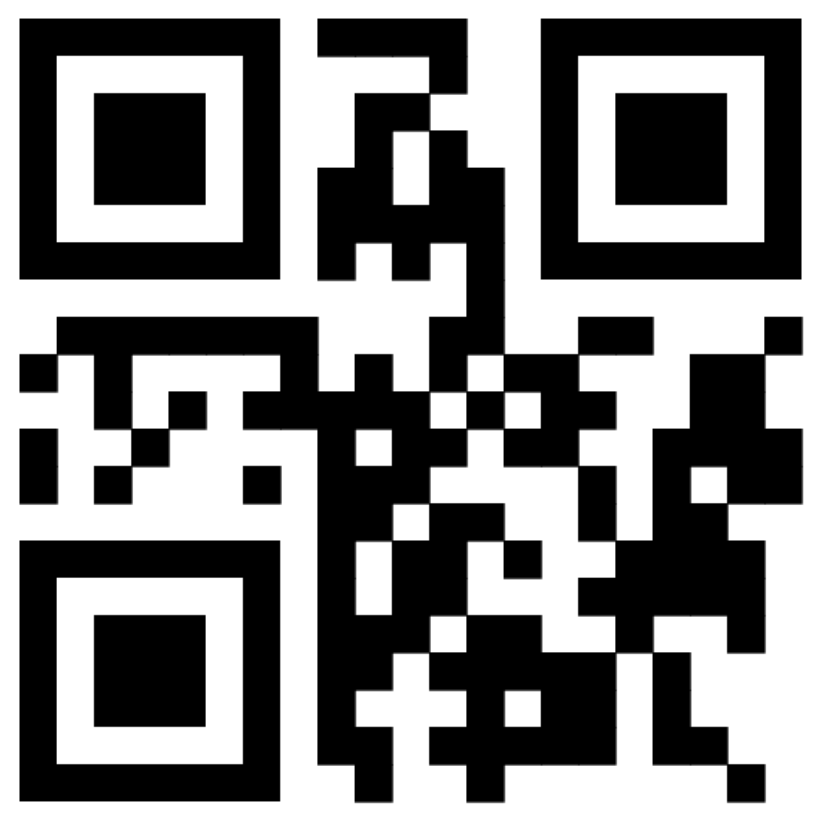}
    \caption{QR Code}
    \end{subfigure} 
    \hfill
    \begin{subfigure}[b]{0.138\textwidth}
    \includegraphics[width=\textwidth]{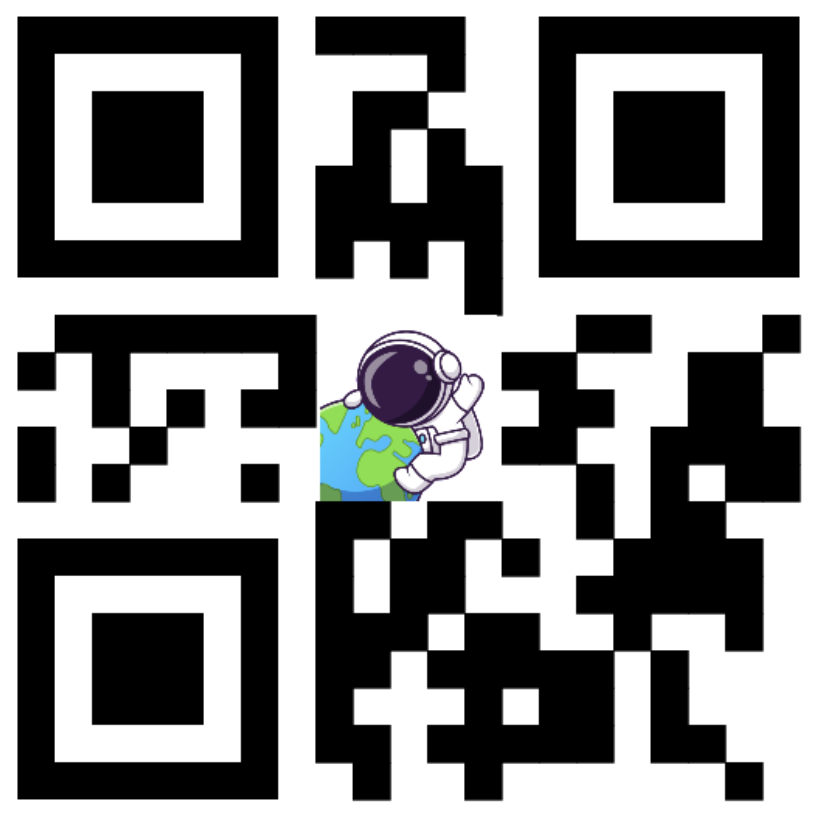}
    \caption{Stylized QR Code}
    \end{subfigure}
    \hfill
    \begin{subfigure}[b]{0.138\textwidth}
    \includegraphics[width=\textwidth]{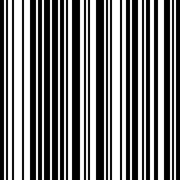}
    \caption{Barcode}
    \end{subfigure}
    \hfill

    \begin{subfigure}[b]{0.163\textwidth}
    \includegraphics[width=\textwidth]{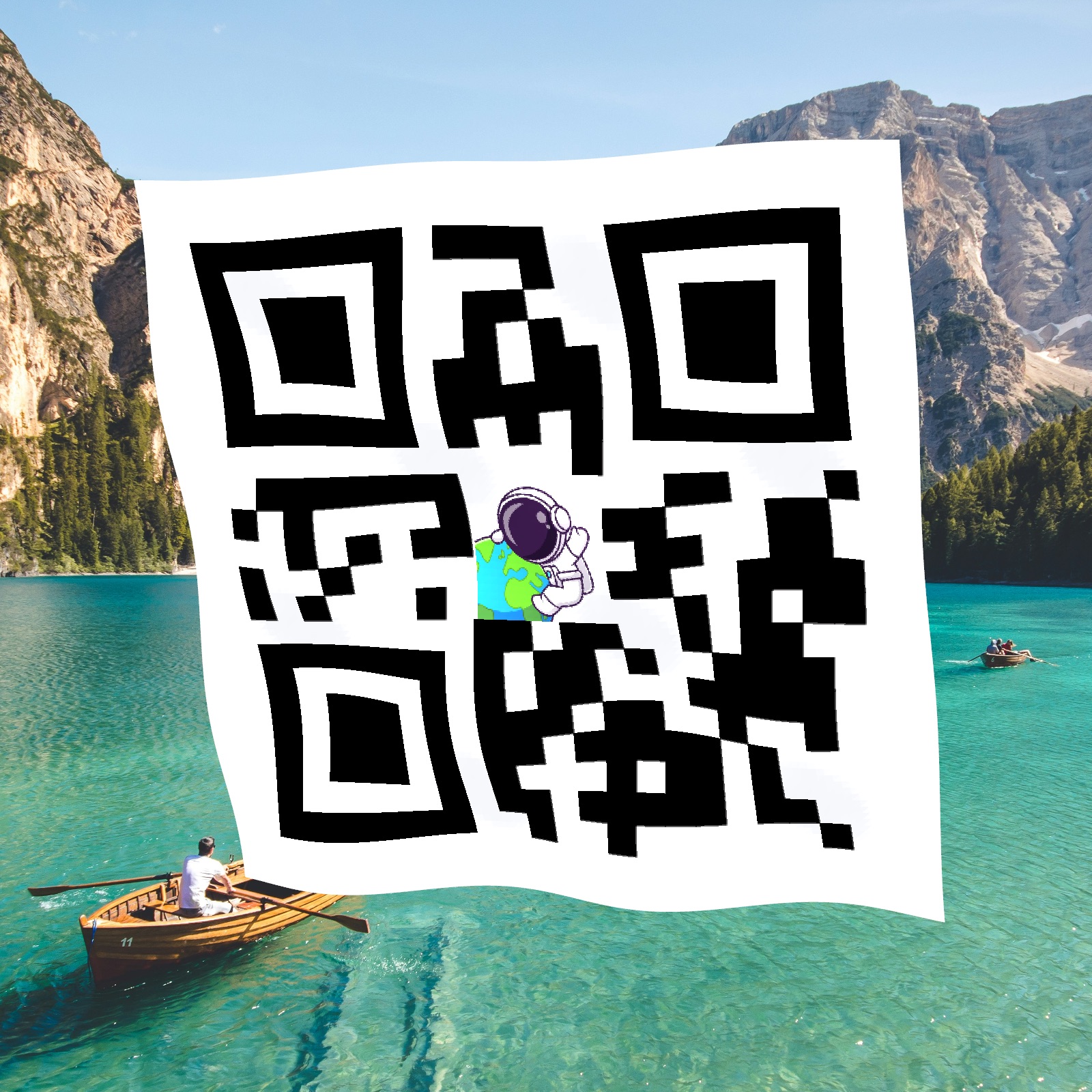}
    \caption{$\omega=0.3, \psi=0$}
    \label{fig:eval-app-distortion-0.3-occlusion-0}
    \end{subfigure}  
    \hfill
    \begin{subfigure}[b]{0.163\textwidth}
    \includegraphics[width=\textwidth]{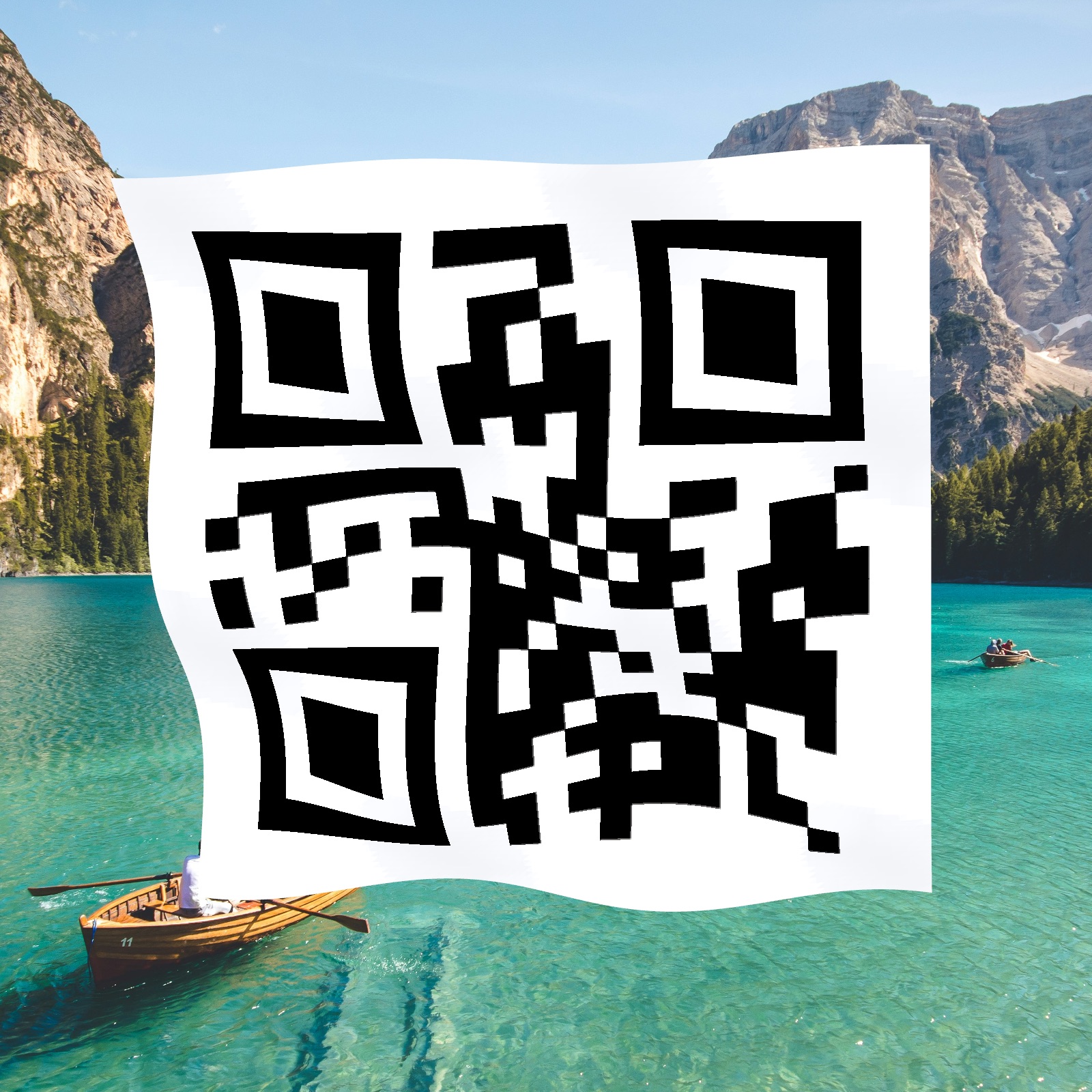}
    \caption{$\omega=0.7, \psi=0$}
    \label{fig:eval-app-distortion-0.7-occlusion-0}
    \end{subfigure}
    \hfill
    \begin{subfigure}[b]{0.163\textwidth}
    \includegraphics[width=\textwidth]{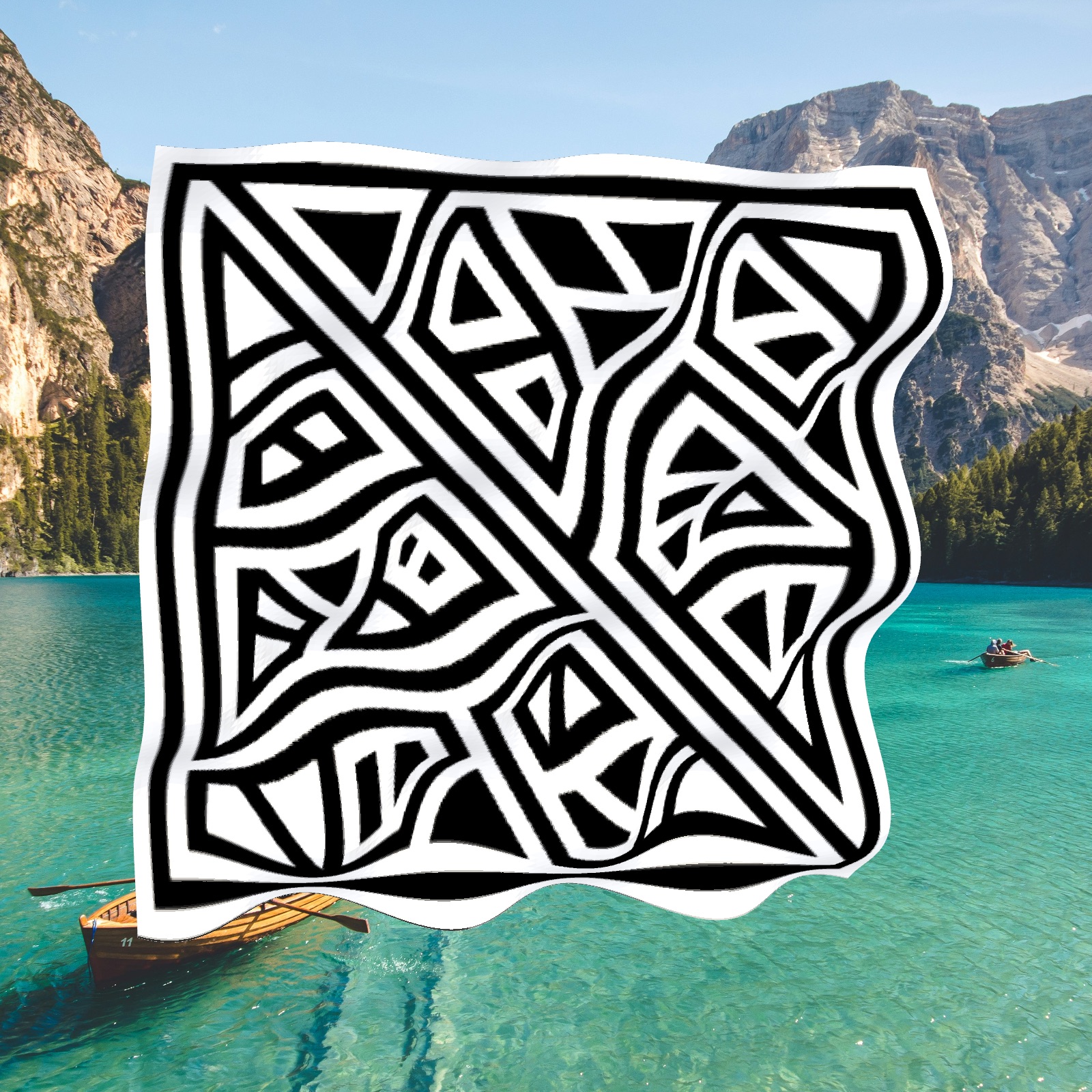}
    \caption{$\omega=1, \psi=0$}
    \label{fig:eval-app-distortion-1-occlusion-0}
    \end{subfigure}
    \hfill
    \begin{subfigure}[b]{0.163\textwidth}
    \includegraphics[width=\textwidth]{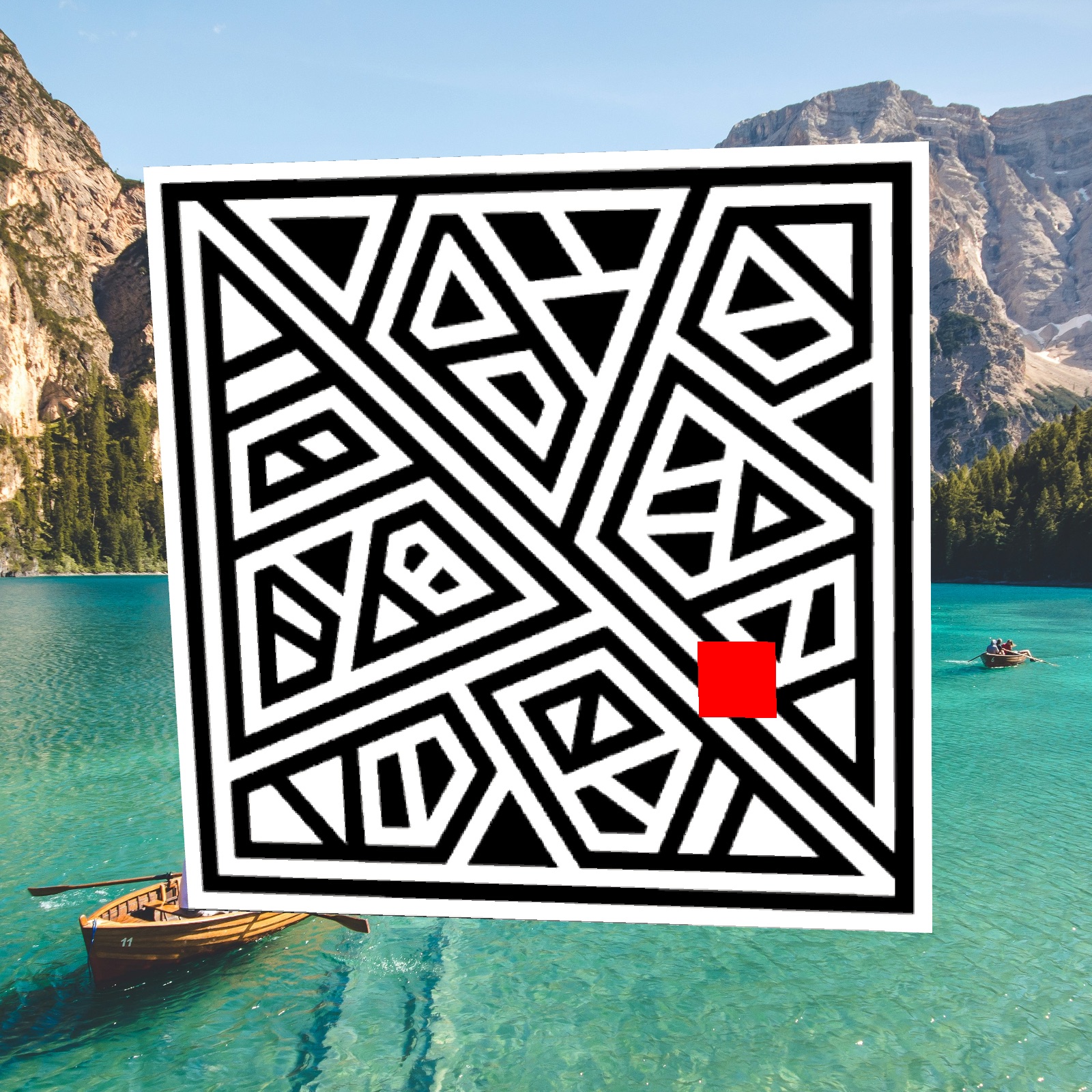}
    \caption{$\omega=0, \psi=0.01$}
    \label{fig:eval-app-distortion-0-occlusion-0.01}
    \end{subfigure}
    \hfill
    \begin{subfigure}[b]{0.163\textwidth}
    \includegraphics[width=\textwidth]{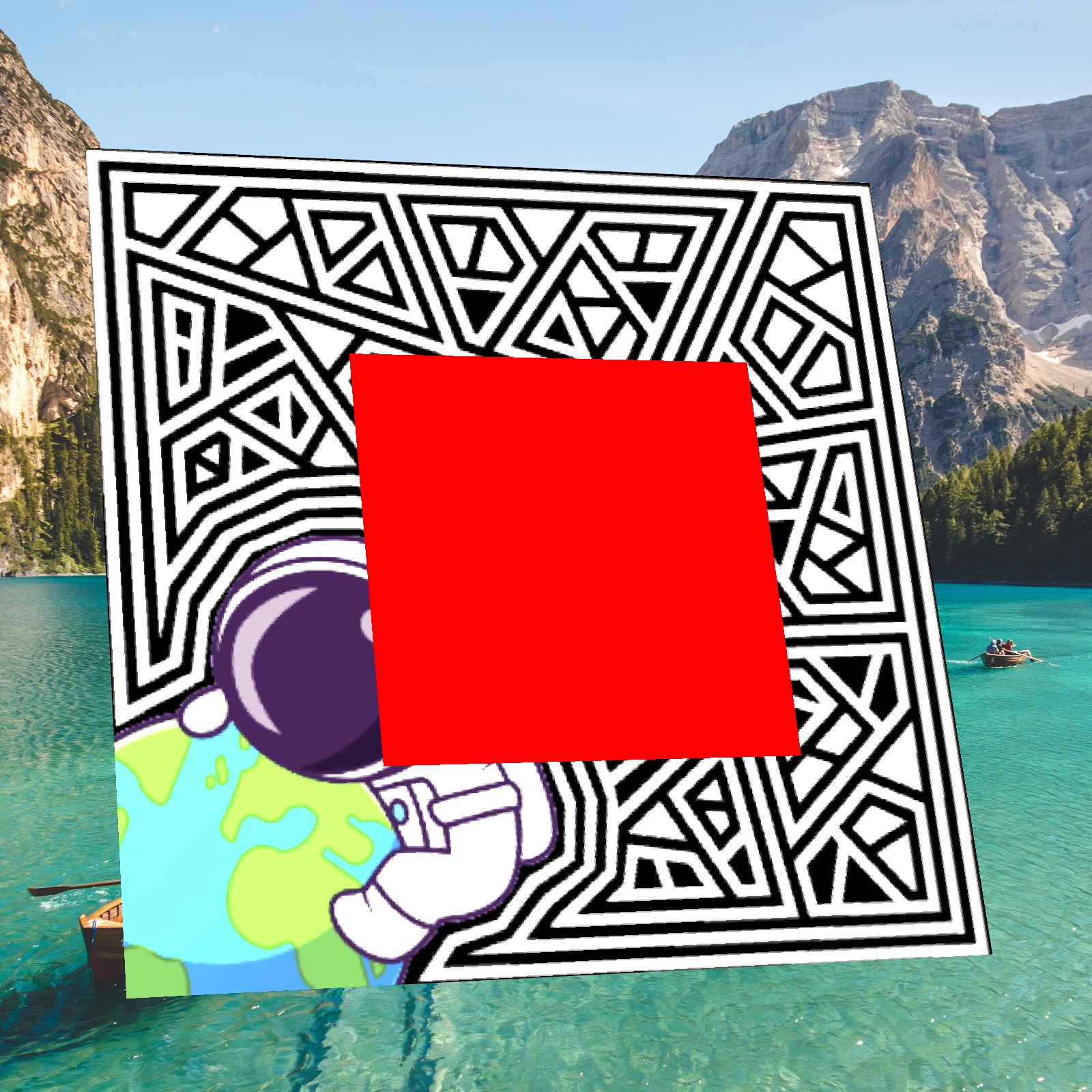}
    \caption{$\omega=0, \psi=0.25$}
    \label{fig:eval-app-distortion-0-occlusion-0.25}
    \end{subfigure}
    \hfill
    \begin{subfigure}[b]{0.163\textwidth}
    \includegraphics[width=\textwidth]{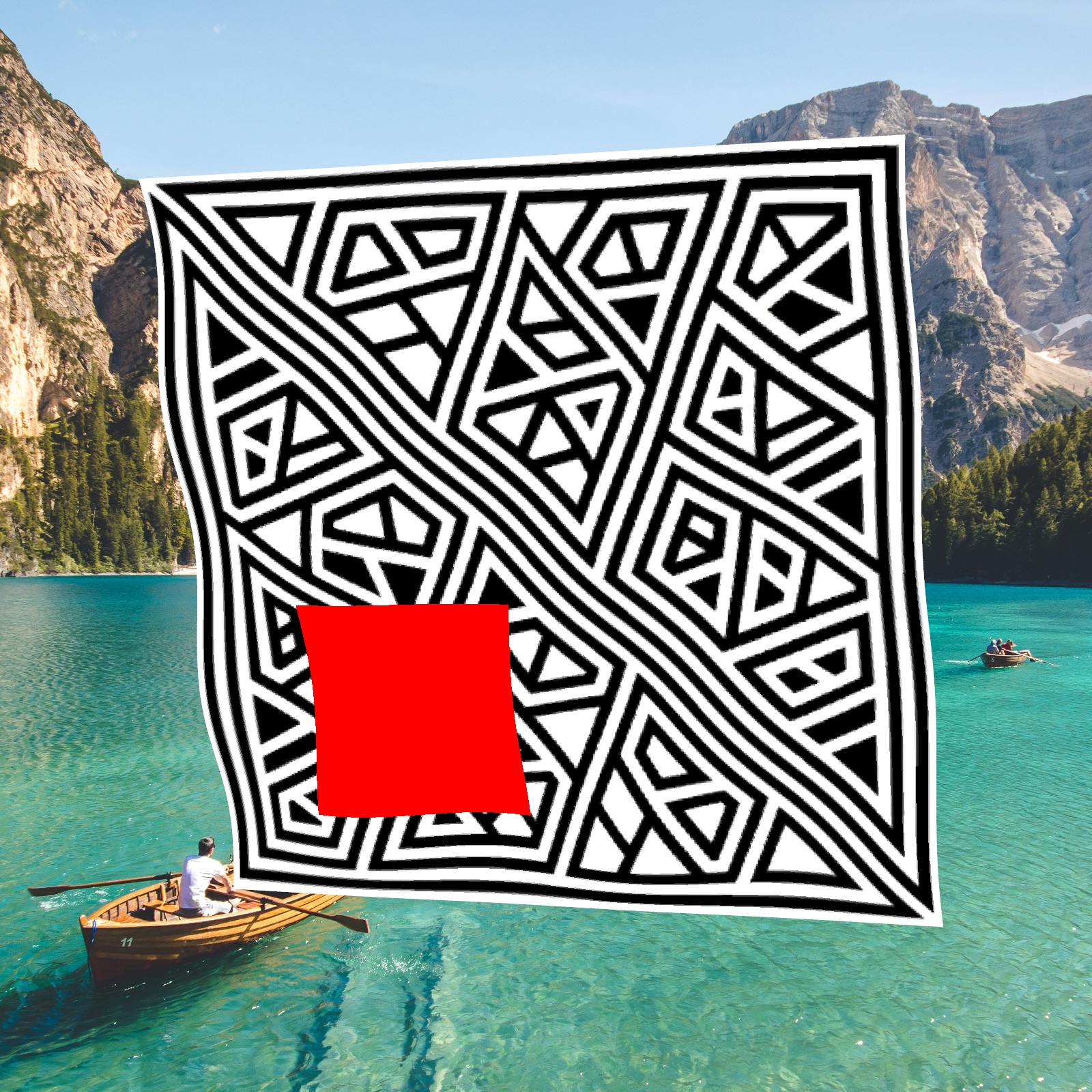}
    \caption{$\omega=0.2, \psi=0.09$}
    \label{fig:eval-app-distortion-0.2-occlusion-0.09}
    \end{subfigure}

 \caption{The comparative evaluation between Claycodes, QR Codes, and Code128-Barcodes presented in \cref{sec:results}.}
  \label{fig:eval-plots}
\end{figure*}

\paragraph{Experiment $2$ -- Latency of successful scans.} The phone was fixed in place and pointed at a screen displaying black-and-white Claycodes, each encoding a randomly-generated alphanumeric message. We varied the encoded data sizes from $56$ to $456$ bits, with a step size of $40$ bits ($5$ characters). For each data size, we generated $310$ Claycodes with random alphanumeric content, resulting in a total of $3,410$ samples. Each code was displayed for $5$ seconds.
As illustrated in \cref{fig:latencies-wrt-code-size}, the average scan latency increased approximately linearly with code size, ranging from $97\text{ms}$ (for $56$ bits) to $148\text{ms}$ (for $456$ bits), with observed extremes between $67\text{ms}$ and $270\text{ms}$. When compared to \cref{fig:latencies}, it is evident that latency is affected by the additional processing involved in detecting and decoding Claycodes. Nevertheless, this latency corresponds only to the final iteration of the scanning loop—which runs approximately $10$ times per second—and thus has a limited impact on the overall performance.

\section{Comparative Evaluation} \label{sec:results}
Scannable codes, whether printed or displayed on screens, are often subject to environmental wear and adverse scan conditions, such as the quality of camera hardware, lens distortion, uneven surfaces and scanning angle. To assess Claycode's robustness, we conducted an empirical study to measure its performance in a controlled environment compared to other popular 2D scannable codes.

\subsection{Experiment Setup}
We compared $K=6$ different types of codes: Claycodes (redundancy level $R=1$), Claycodes ($R=2$), stylized Claycodes ($R=2$), QR Codes (\textit{high} error correction level), stylized QR Codes (\textit{high} error correction level) and Code128-Barcodes. Barcodes are 1D codes and are included as a baseline. We generated $M=10$ alphanumeric random strings and produced a code for each code type, resulting in $K\cdot M=60$ different codes. To ensure reproducibility in a controlled environment, while performing the evaluation on representative hardware, we fixed an Android phone (Google Pixel 8 Pro\crmodsecond{, as described in \cref{subsec:scanning_latency}}) on a tripod, at a fixed distance from a 24-inch screen, as shown in \cref{fig:eval-setup}. The codes were rendered one by one in a 3D environment in the form of a textured plane mesh, modified to simulate different kinds of tampering and scanning conditions. 
We varied a diverse set of parameters. First, the \textbf{scanning angles} $\phi_x,\phi_y \in [-20^{\circ},-10^{\circ}] \cup [+10^{\circ}, +20^{\circ}]$ were simulated by rotating the mesh on the $x$ and $y$ axes. We never set the scanning angle to zero, as it is rarely the case in real-world scans. The \textbf{deformation damage} was simulated by introducing a wave-like oscillation in the plane mesh, in the form of an offset to the $z$ (depth) component of the mesh, in such a way that $z(x, y) = \omega \cdot \sin(\nu_x \cdot x) \cdot \cos(\nu_y \cdot y)$, where the frequencies $(\nu_x, \nu_y) \in [1,2]$ were picked at random, while the wave amplitude $\omega \in [0,1]$ was adopted as the measure of deformation. To simulate \textbf{occlusion damage}, we overlaid a red square on the mesh texture, as shown in \cref{fig:eval-app-distortion-0-occlusion-0.01,,fig:eval-app-distortion-0-occlusion-0.25,,fig:eval-app-distortion-0.2-occlusion-0.09}. The square's size, defined as a percentage of the total area $\psi \in [0,1]$, was adopted as a measure of occlusion, while its location $(\rho_x, \rho_y)$ was randomly selected ensuring the entirety of the square overlaps the texture. We call each combination of parameters $(\phi_x,\phi_y, \omega, \nu_x, \nu_y, \psi, \rho_x, \rho_y)$ a \textit{scenario}. 
In each experiment, we generated a set of scenarios and then evaluated them against the dataset of $K \cdot M$ codes. Each test case was considered either a success (the code was scanned) or a failure (the code was not scanned, or it was incorrectly scanned -- which has only happened with barcodes). Before considering a test case failed, we allowed some seconds for the scanner to recognize the code, and played an animation that adjusted the mesh size by smoothly increasing it to $1.3$ times its original size, then reducing it to $0.7$ times its size, and finally returning it to its initial dimensions. We decided to do so as varying the zoom is compatible with most real-life scenarios, and because we found barcodes to be particularly sensitive to zooming conditions. We scanned Claycodes using an Android implementation of the scanner described in \cref{sec:scanner}, while for QR Codes and barcodes we used the commercially-available app developed by~\cite{qr_code_scanner_gamma_play}, which to-date is top-rated and counts hundreds of millions of downloads.

\subsection{Results}
We performed three experiments, first considering deformations and occlusions as mutually exclusive features (experiments $1$ and $2$), and finally combining them in the last batch of our experiment.

\paragraph{Experiment $1$ -- deformations.}
In our first experiment, we studied the impact of deformations on scannability, in the absence of occlusion. For each of the $K\cdot M = 60$ codes, we randomly generated $10$ different scenarios by varying $\omega$ from $0.1$ to $1$ with a step size of $0.1$, for a total of $600$ scans. We fixed $\psi=0$ (codes were never occluded), and randomly sampled the remaining parameters. As shown in \cref{fig:evaluation-deformation}, Claycodes succeeded in nearly all cases until $\omega=1$ (\cref{fig:eval-app-distortion-1-occlusion-0}), where they started to show some minimal degradation. Comparatively, QR Code's performance started dropping at $\omega=0.3$ (\cref{fig:eval-app-distortion-0.3-occlusion-0}), and failed in almost all scenarios with $\omega\geq 0.7$ (\cref{fig:eval-app-distortion-0.7-occlusion-0}). Stylized QR Codes performed the worst, starting with a $70\%$ success rate for $\omega=0.1$, and converging to zero for $\omega\geq 0.7$.

\paragraph{Experiment $2$ -- occlusions.}
In this experiment, we focused on occlusion damage in the absence of deformation. For each of the $K\cdot M = 60$ codes, we randomly generated $10$ different scenarios by varying $\psi$ across the set $\{0.01, 0.04, 0.09, 0.16, 0.25\}$, with two samples for each value of $\psi$, and a total of $600$ scans. We fixed $\omega=0$ to eliminate any deformation, and all other parameters were randomly sampled. As shown in \cref{fig:evaluation-occlusion}, Claycodes with redundancy level $R=1$ do not tolerate any kind of occlusion (\cref{fig:eval-app-distortion-0-occlusion-0.01}), but almost always perform better than QR codes for $R=2$. Interestingly, stylized Claycodes outperform non-stylized ones. This is because the overlaid square sometimes partially or completely occludes the art, as opposed to the payload -- the same is true for QR codes. 

\paragraph{Experiment $3$ -- deformations and occlusions.}
Finally, we repeated experiment $2$ on the same dataset of scenarios, but fixed to $\omega=0.2$. We chose this value (an example is provided in \cref{fig:eval-app-distortion-0.2-occlusion-0.09}) because $\omega=0.2$ had no effect on the scannability of non-stylized QR Codes (\cref{fig:evaluation-occlusion}). As shown in \cref{fig:evaluation-occlusion-deformation}, the impact of the added deformation is not present in Claycodes. Conversely, the added deformation further reduces the success rate of QR Codes, showing how the deformation and occlusion damage compounds.

\newcommand{\gk}{\textcolor{green!70!black}{\ding{51}}}    %
\newcommand{\yk}{\textcolor{yellow!70!orange}{\ding{118}}} %
\newcommand{\rk}{\textcolor{red}{\ding{55}}}               %

\crmod{
\section{Evaluation Across Physical Media and Lighting Scenarios} \label{sec:media}
}

\begin{figure*}
  \centering

    \begin{subfigure}[b]{0.078\textwidth}
    \includegraphics[width=\textwidth]{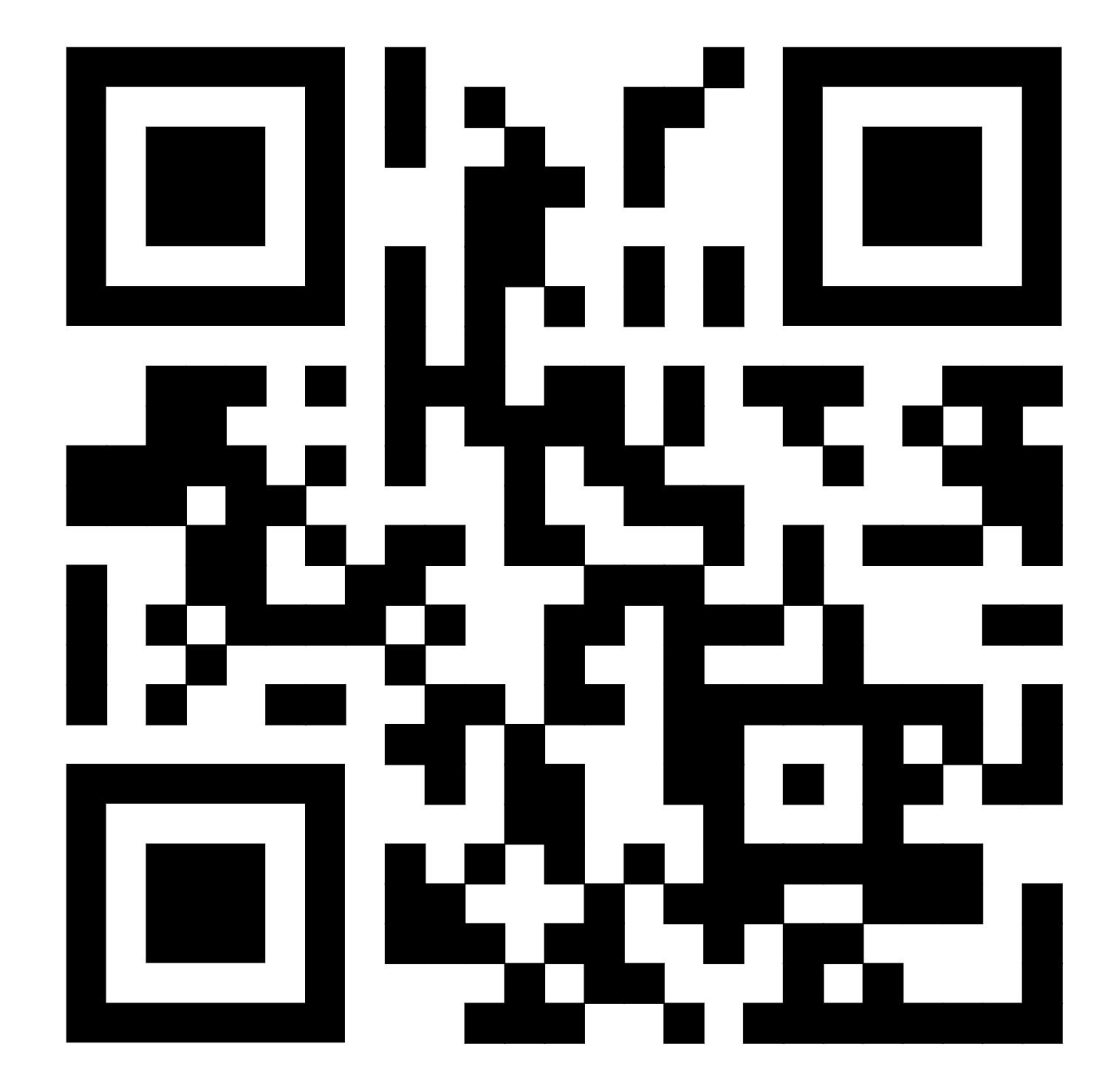}
    \caption*{\footnotesize\textit{``baseline''}}
    \end{subfigure}  
    \hfill
    \begin{subfigure}[b]{0.078\textwidth}
    \includegraphics[width=\textwidth]{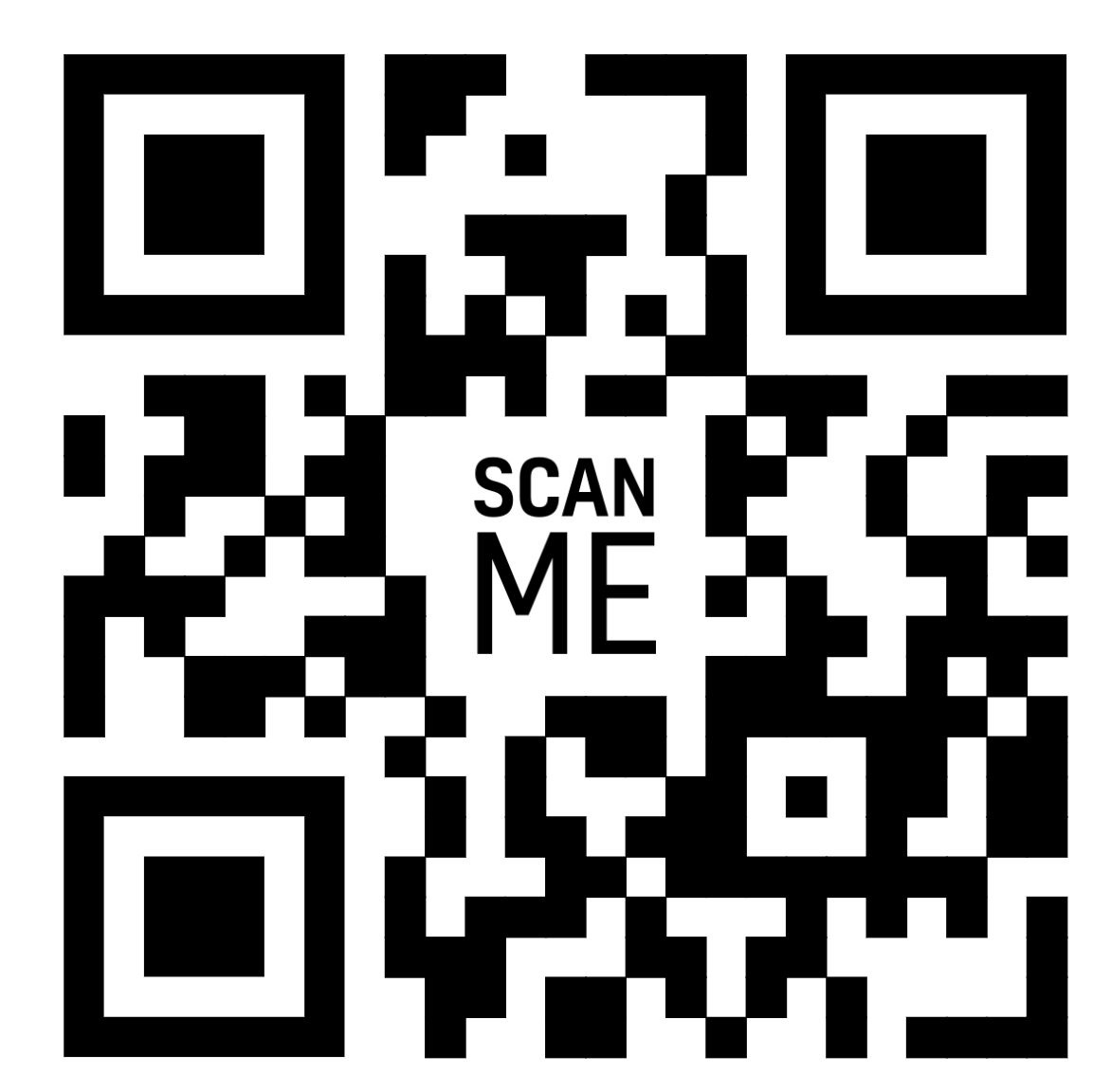}
    \caption*{\footnotesize\textit{``Scan me!''}}
    \end{subfigure}  
    \hfill
    \begin{subfigure}[b]{0.078\textwidth}
    \includegraphics[width=\textwidth]{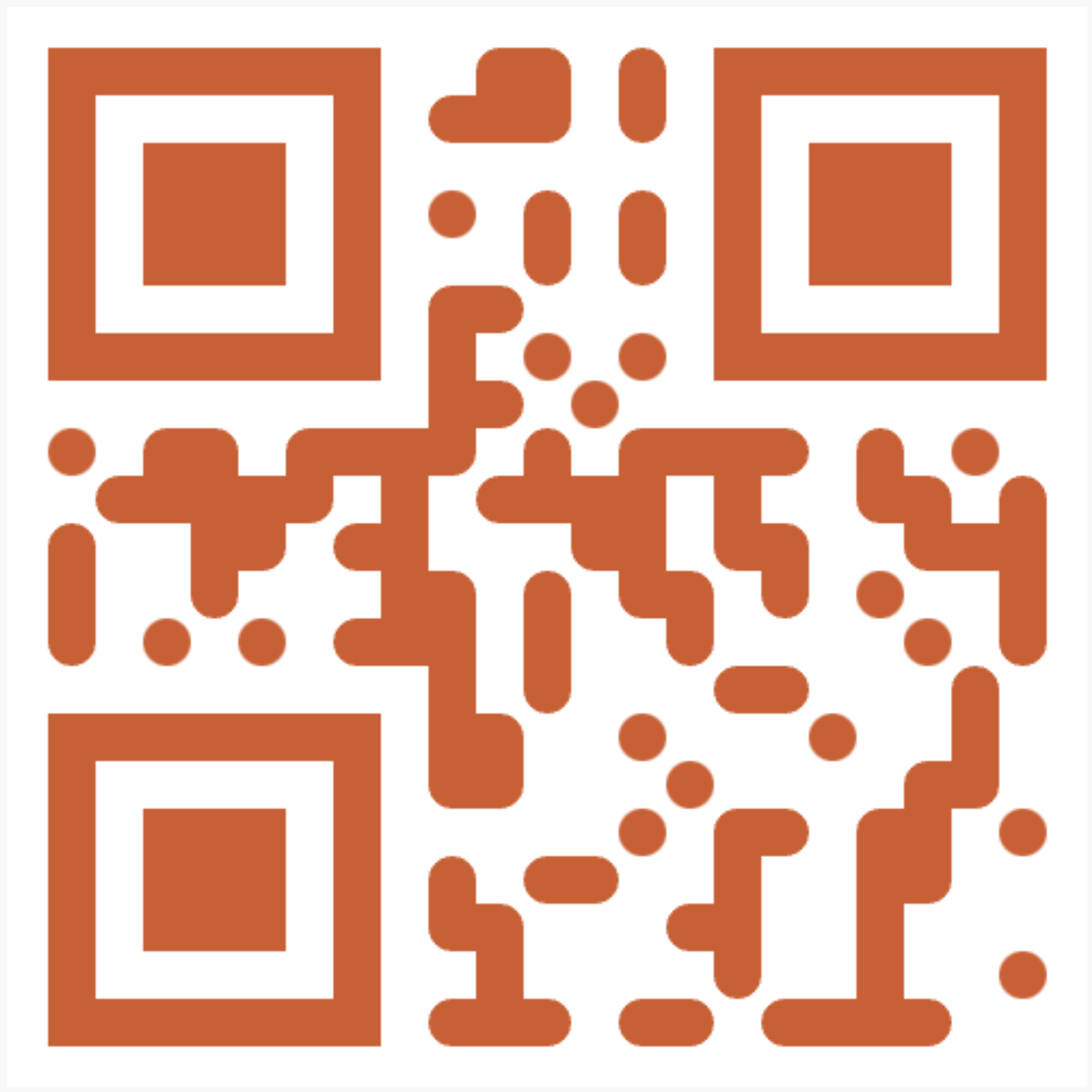}
    \caption*{\footnotesize\textit{``colors!''}}
    \end{subfigure}  
    \hfill
    \begin{subfigure}[b]{0.078\textwidth}
    \includegraphics[width=\textwidth]{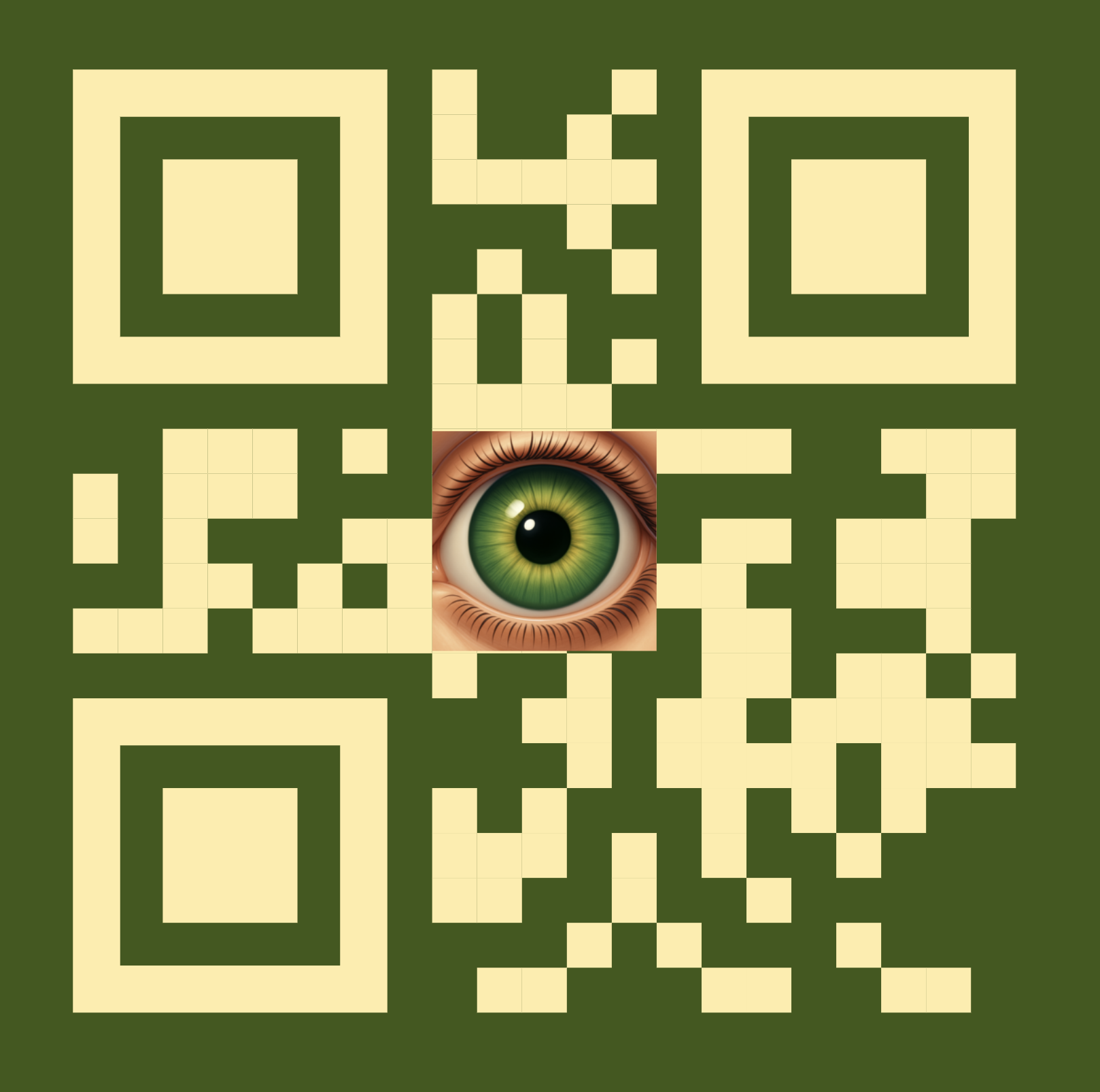}
    \caption*{\footnotesize\textit{``eye''}}
    \end{subfigure}  
    \hfill
    \begin{subfigure}[b]{0.078\textwidth}
    \includegraphics[width=\textwidth]{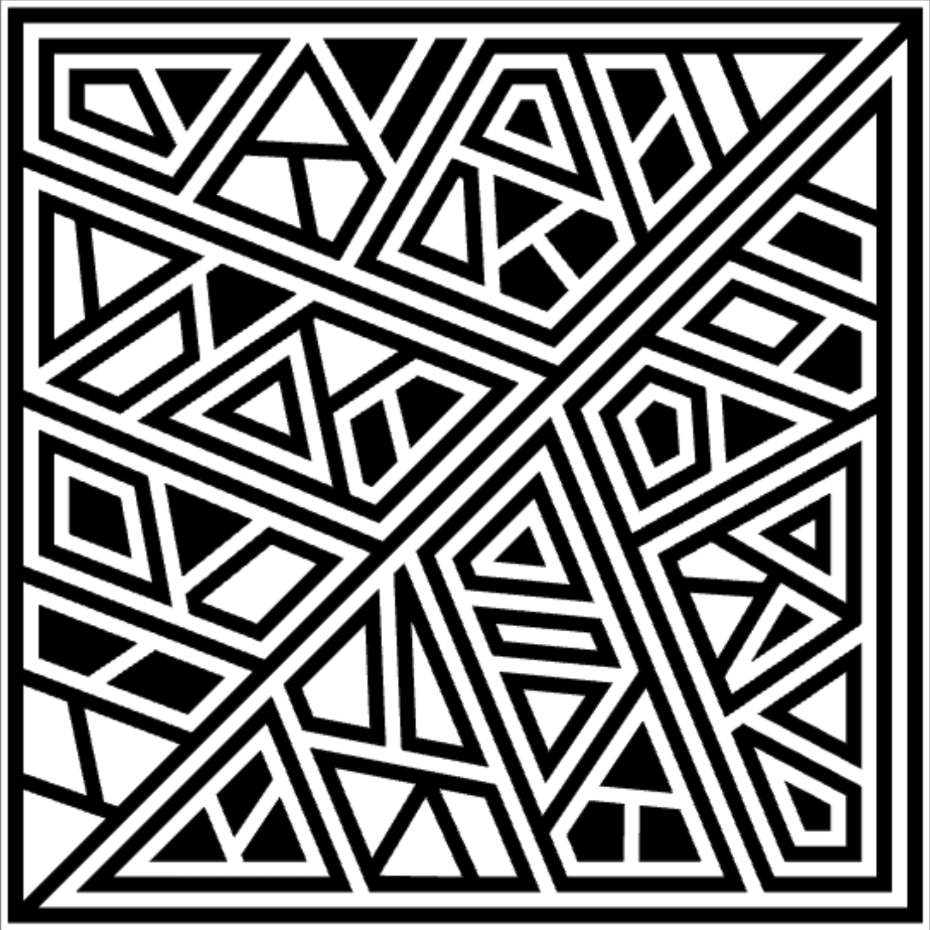}
    \caption*{\footnotesize\textit{``Claycode''}}
    \end{subfigure}  
    \hfill
    \begin{subfigure}[b]{0.078\textwidth}
    \includegraphics[width=\textwidth]{figures/banner/eye-clay.png}
    \caption*{\footnotesize\textit{``\emojieyes"}}
    \end{subfigure}  
    \hfill
    \begin{subfigure}[b]{0.078\textwidth}
    \includegraphics[width=\textwidth]{figures/banner/space-clay.png}
    \caption*{\footnotesize\textit{``space''}}
    \end{subfigure}  
    \hfill
    \begin{subfigure}[b]{0.078\textwidth}
    \includegraphics[width=\textwidth]{figures/pizza.jpeg}
    \caption*{\footnotesize\textit{``pizza''}}
    \end{subfigure}  
    \hfill
    \begin{subfigure}[b]{0.078\textwidth}
    \includegraphics[width=\textwidth]{figures/banner/magic-clay.png}
    \caption*{\footnotesize\textit{``magic''}}
    \end{subfigure}  
    \hfill
    \begin{subfigure}[b]{0.078\textwidth}
    \includegraphics[width=\textwidth]{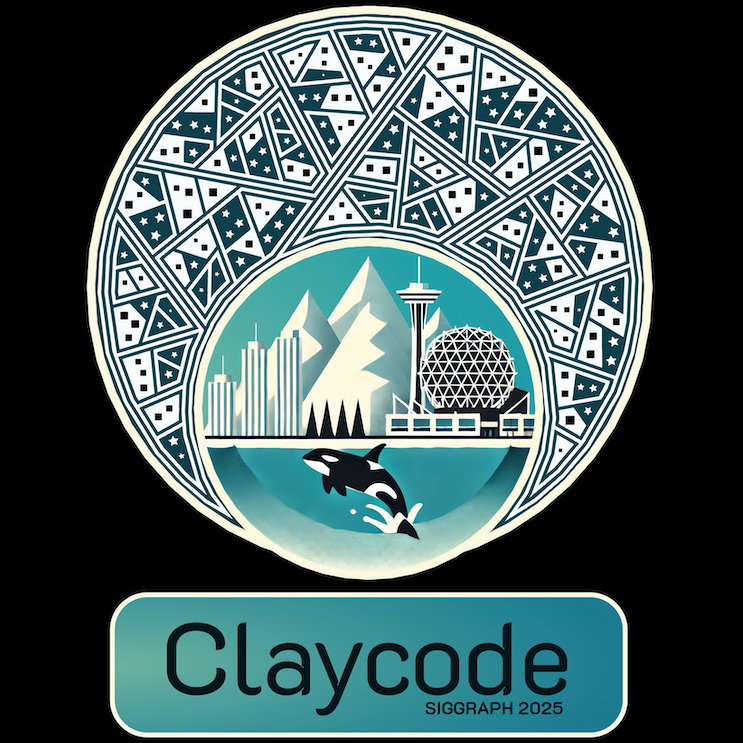}
    \caption*{\footnotesize\textit{``claycode.io''}}
    \end{subfigure}  
    \hfill
    \begin{subfigure}[b]{0.156\textwidth}
    \centering
    \includegraphics[width=0.8\textwidth]{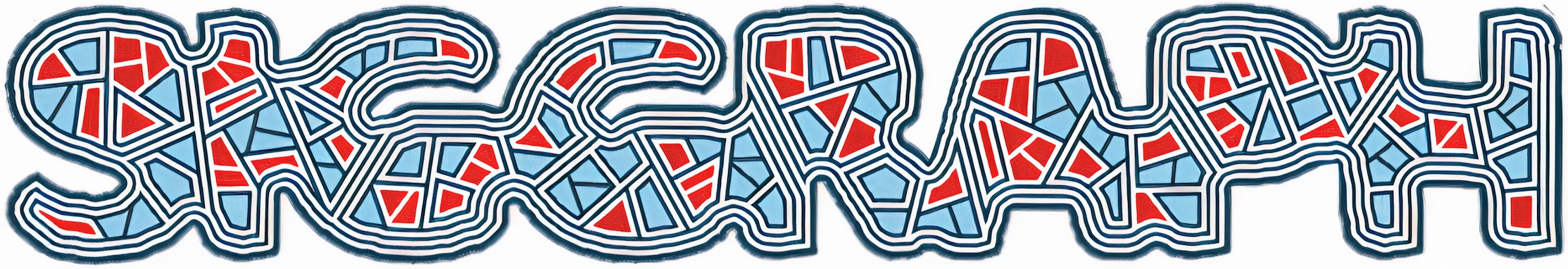}
    \includegraphics[width=0.8\textwidth]{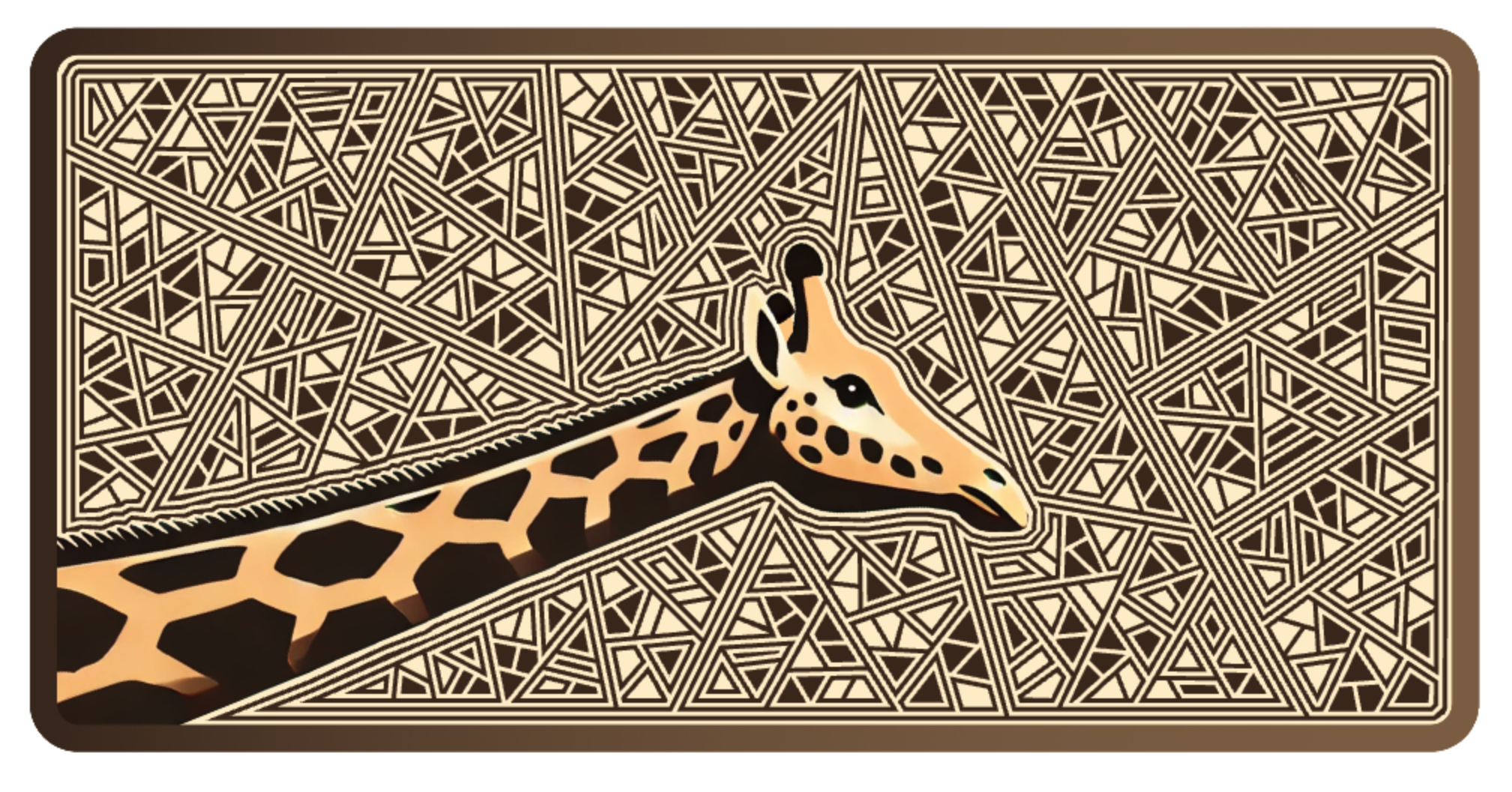}
    \caption*{\footnotesize\textit{``SIGGRAPH!'', ``Giraffe!''}}
    \end{subfigure}  
    \hfill

    \vspace{1mm}

    \begin{subfigure}[b]{0.163\textwidth}
    \includegraphics[width=\textwidth]{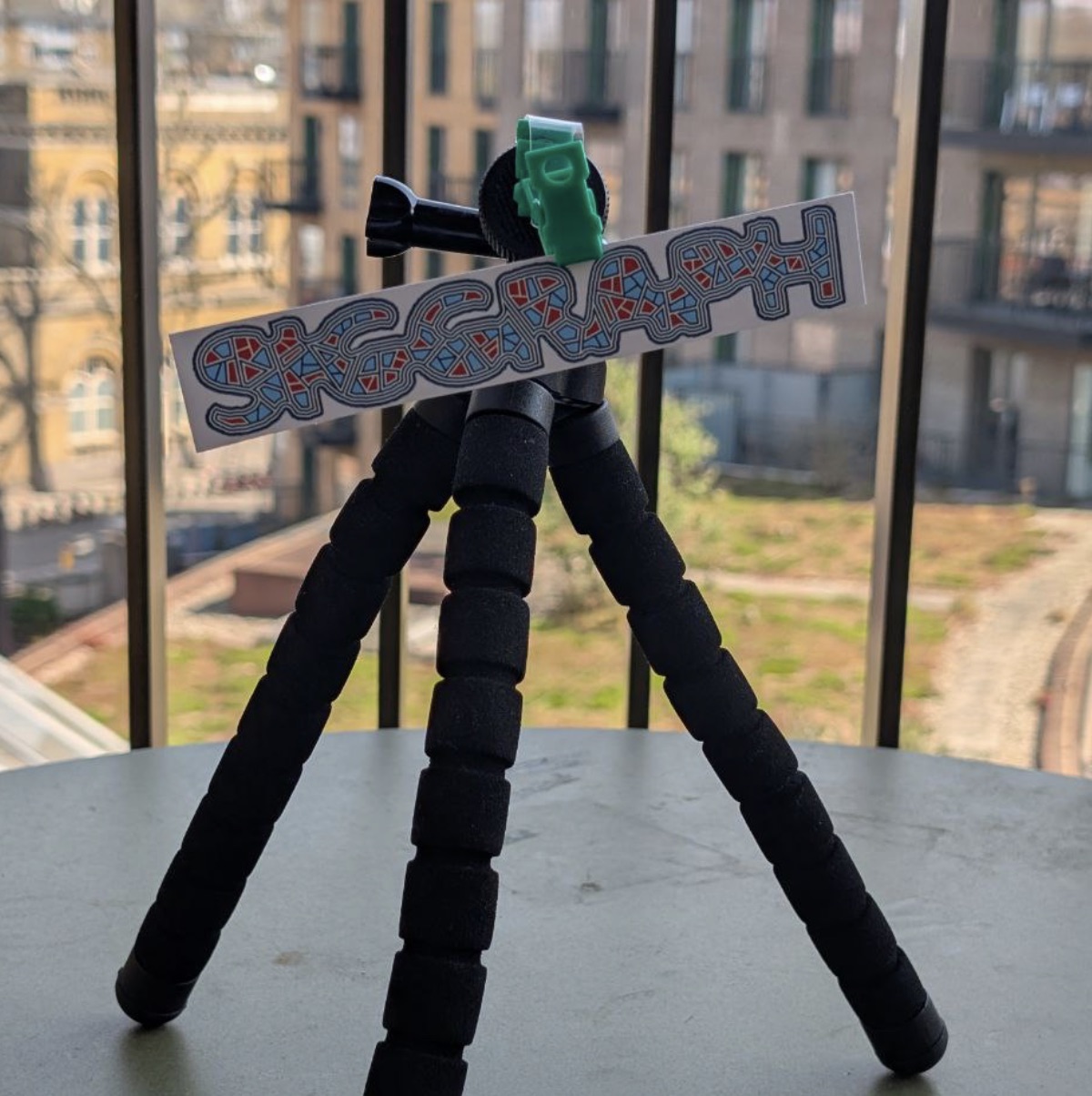}
    \caption{matte, outside}
    \label{fig:siggraph-outside}
    \end{subfigure}  
    \hfill
    \begin{subfigure}[b]{0.163\textwidth}
    \includegraphics[width=\textwidth]{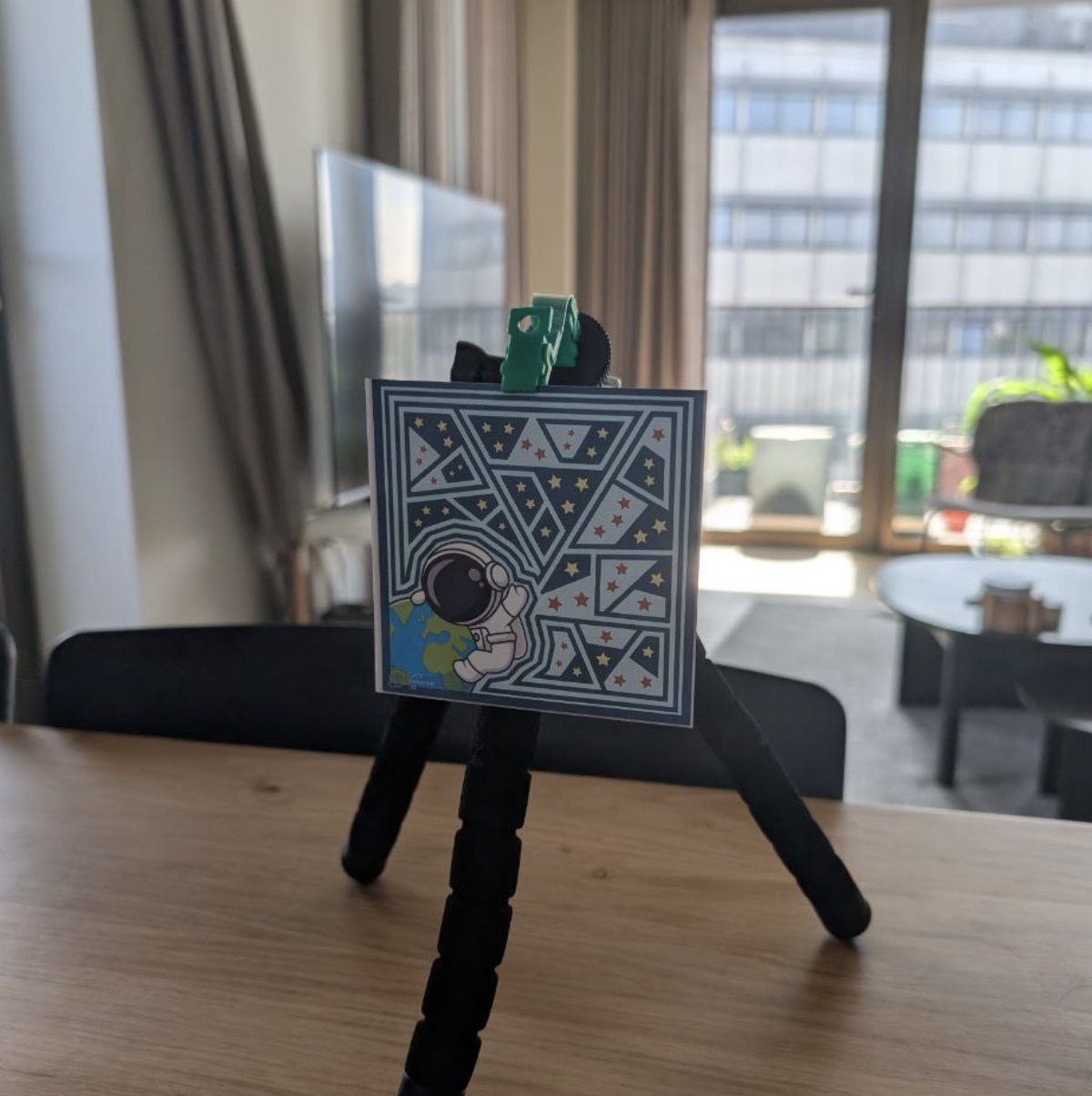}
    \caption{glossy, inside}
    \label{fig:space-matte-inside}
    \end{subfigure}
    \hfill
    \begin{subfigure}[b]{0.163\textwidth}
    \includegraphics[width=\textwidth]{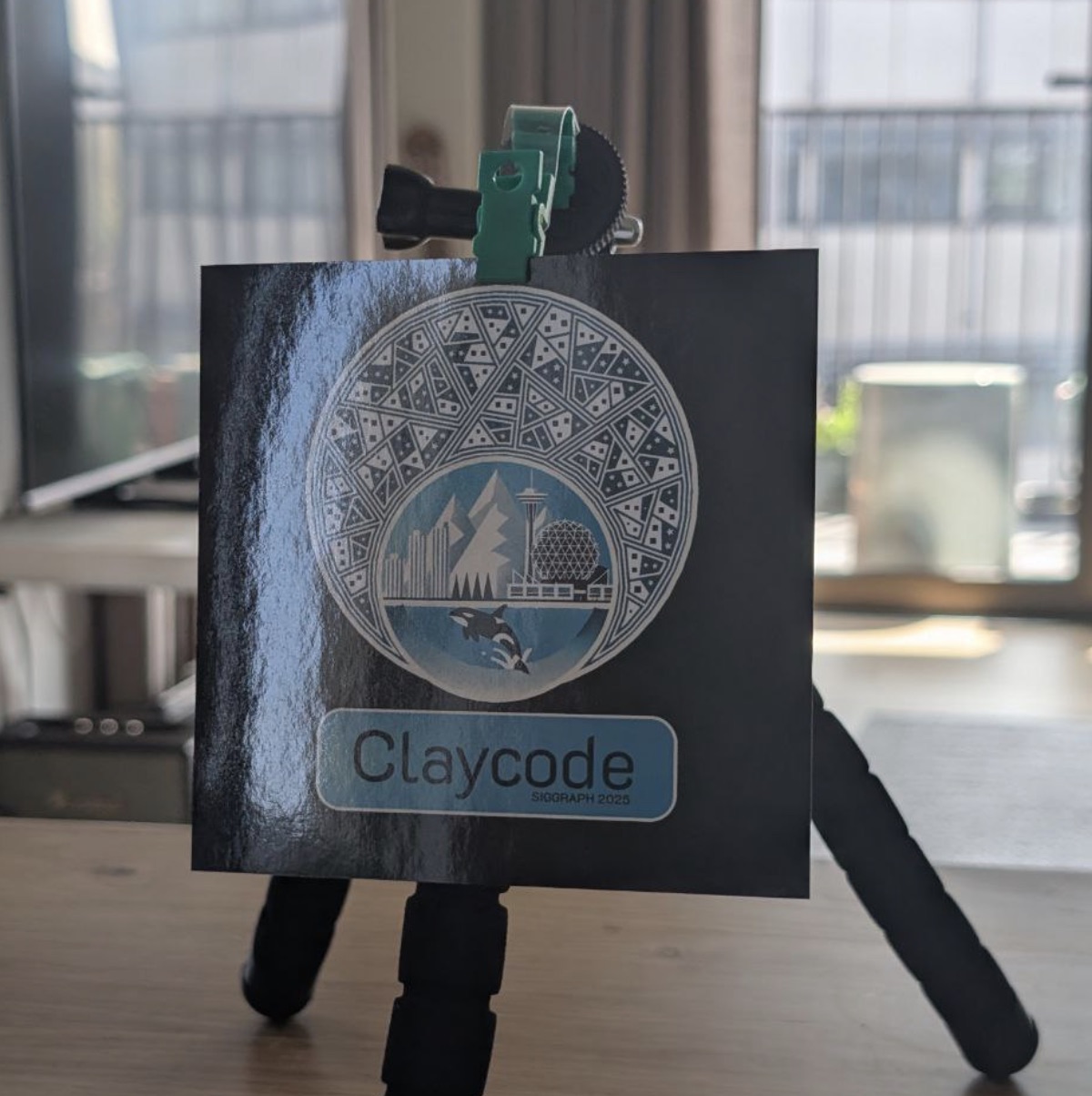}
    \caption{glossy, inside}
    \label{fig:claycodeio-glossy-inside}
    \end{subfigure}
    \hfill
    \begin{subfigure}[b]{0.163\textwidth}
    \includegraphics[width=\textwidth]{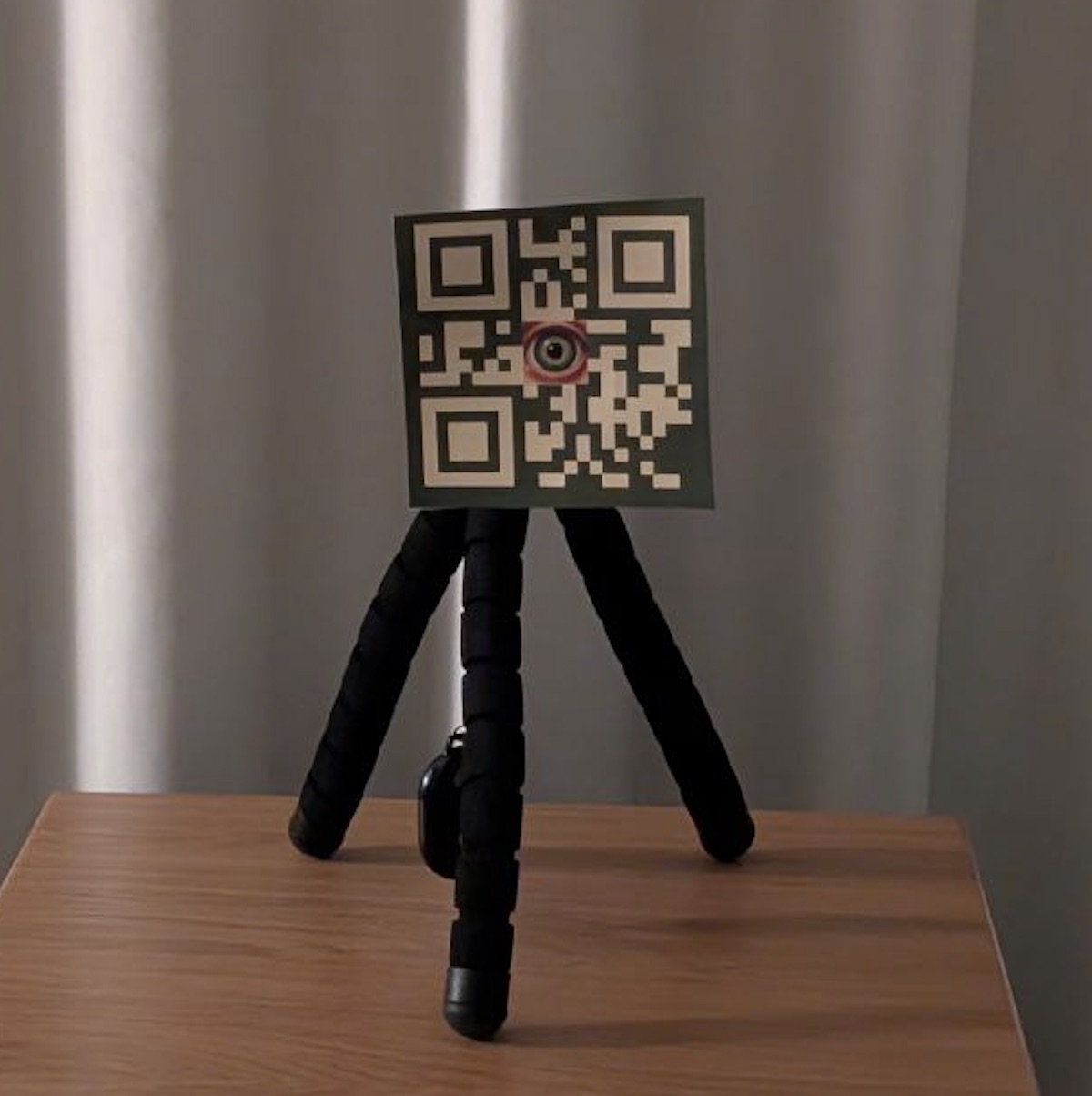}
    \caption{matte, uneven light}
    \label{fig:eye-qr-low-light}
    \end{subfigure}
    \hfill
    \begin{subfigure}[b]{0.163\textwidth}
    \includegraphics[width=\textwidth]{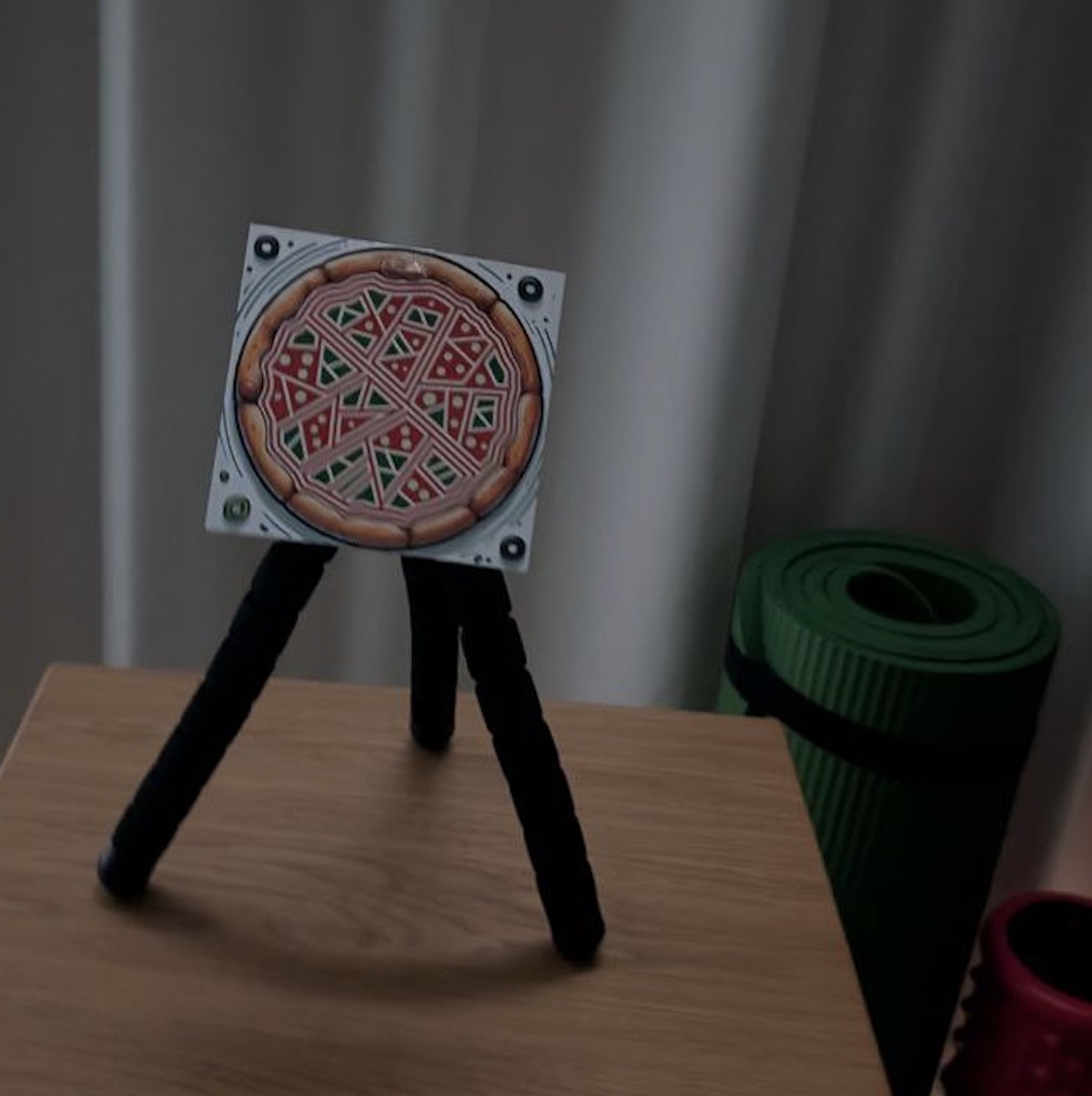}
    \caption{glossy, dark}
    \label{fig:pizza-dark}
    \end{subfigure}
    \hfill
    \begin{subfigure}[b]{0.163\textwidth}
    \includegraphics[width=\textwidth]{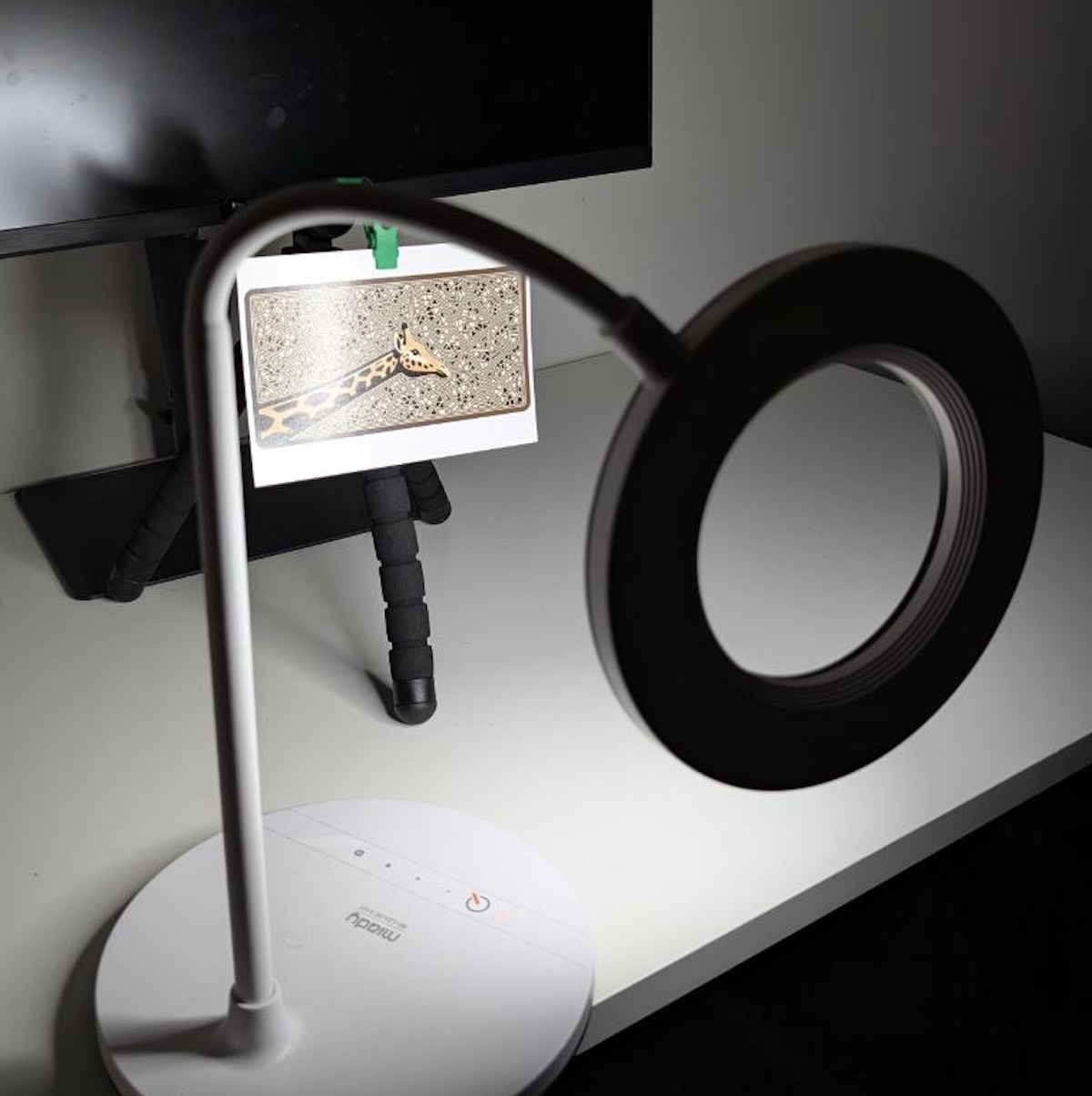}
    \caption{matte, reflections}
    \label{fig:giraffe-reflections}
    \end{subfigure}

    \begin{subfigure}[b]{0.163\textwidth}
    \includegraphics[width=\textwidth]{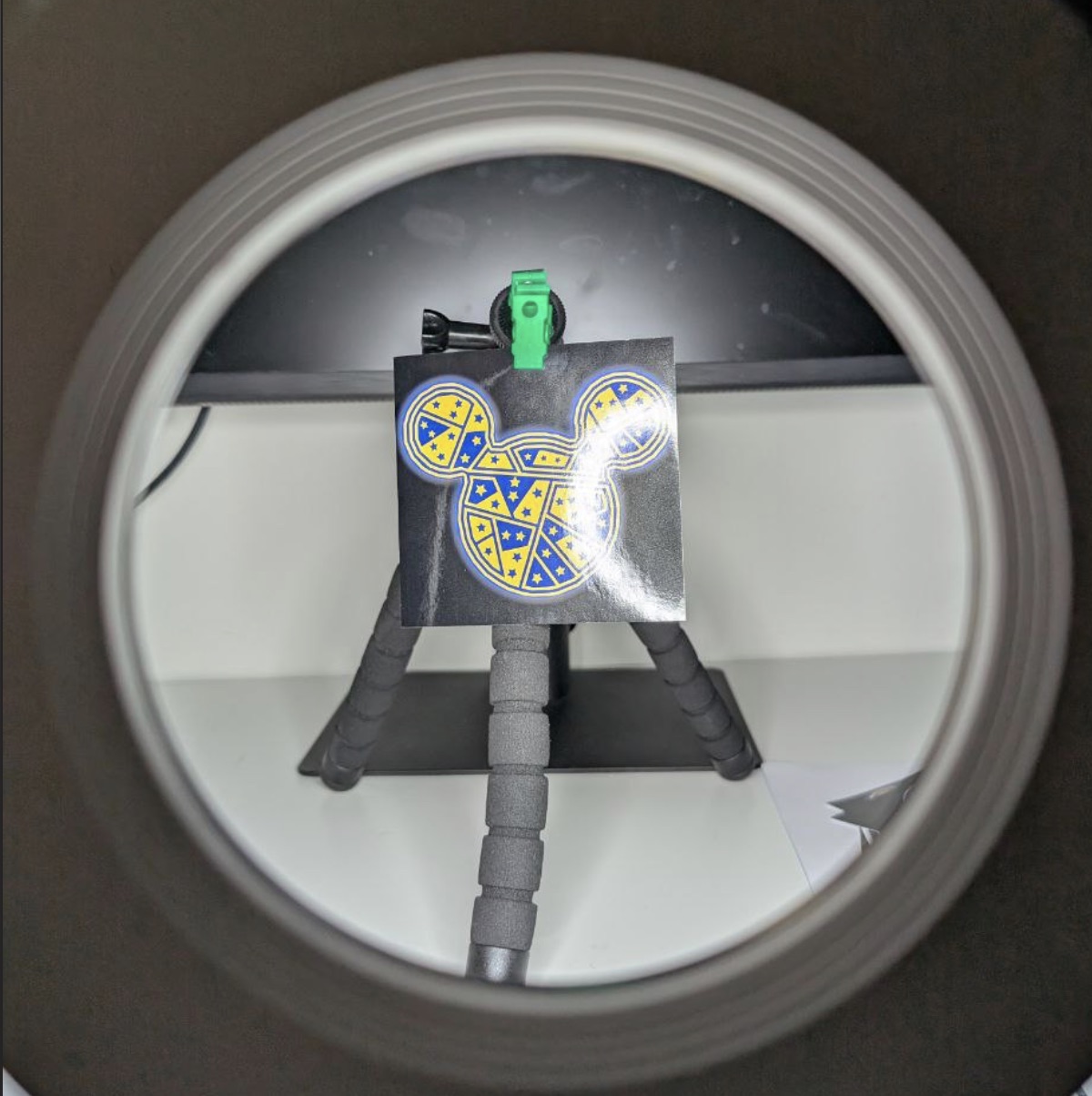}
    \caption{glossy, reflections}
    \label{fig:magic-glossy-reflections}
    \end{subfigure}
    \hfill
    \begin{subfigure}[b]{0.163\textwidth}
    \includegraphics[width=\textwidth]{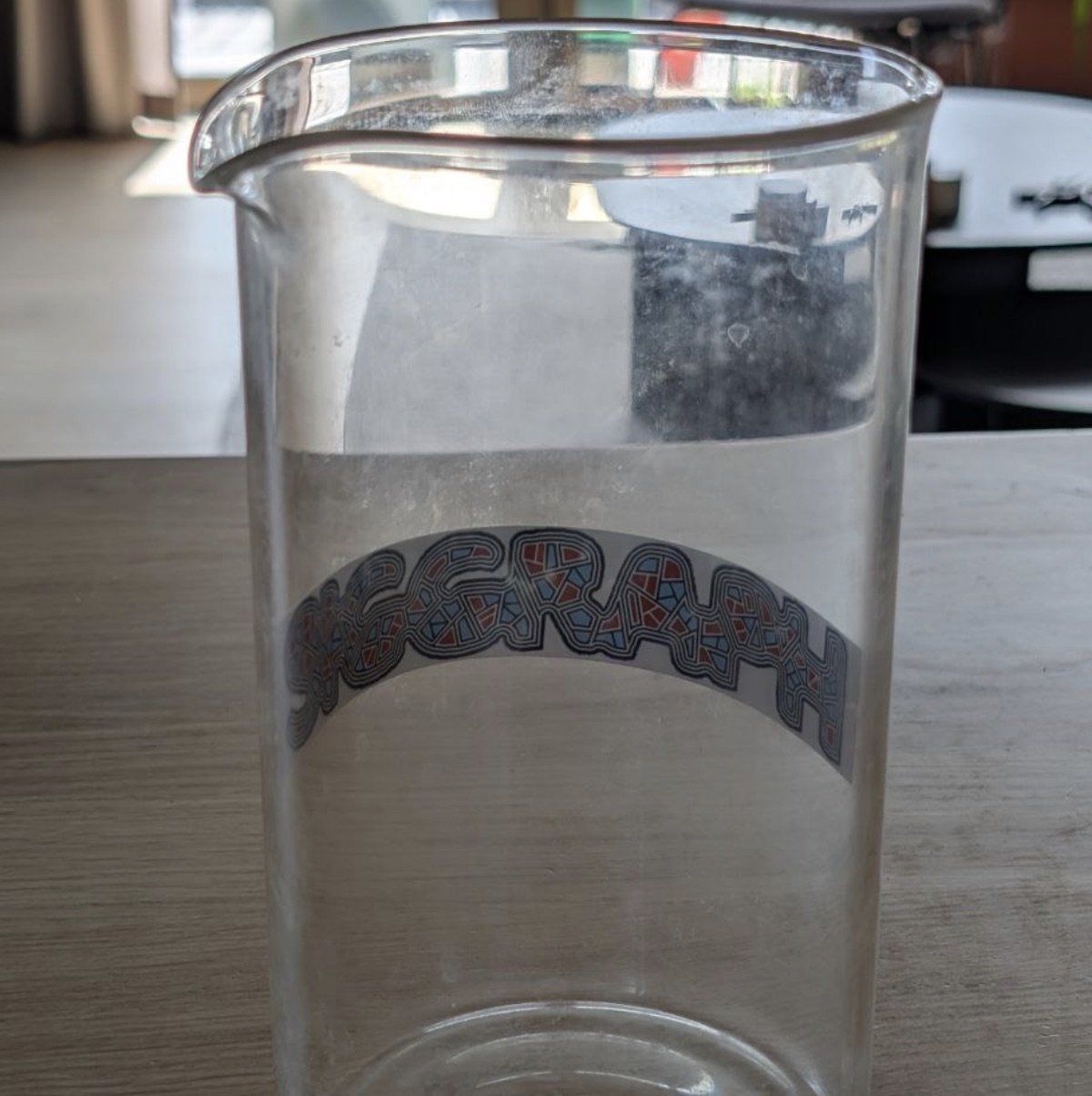}
    \caption{matte, carafe inw.}
    \label{fig:siggraph-carafe-inward}
    \end{subfigure}
    \hfill
    \begin{subfigure}[b]{0.163\textwidth}
    \includegraphics[width=\textwidth]{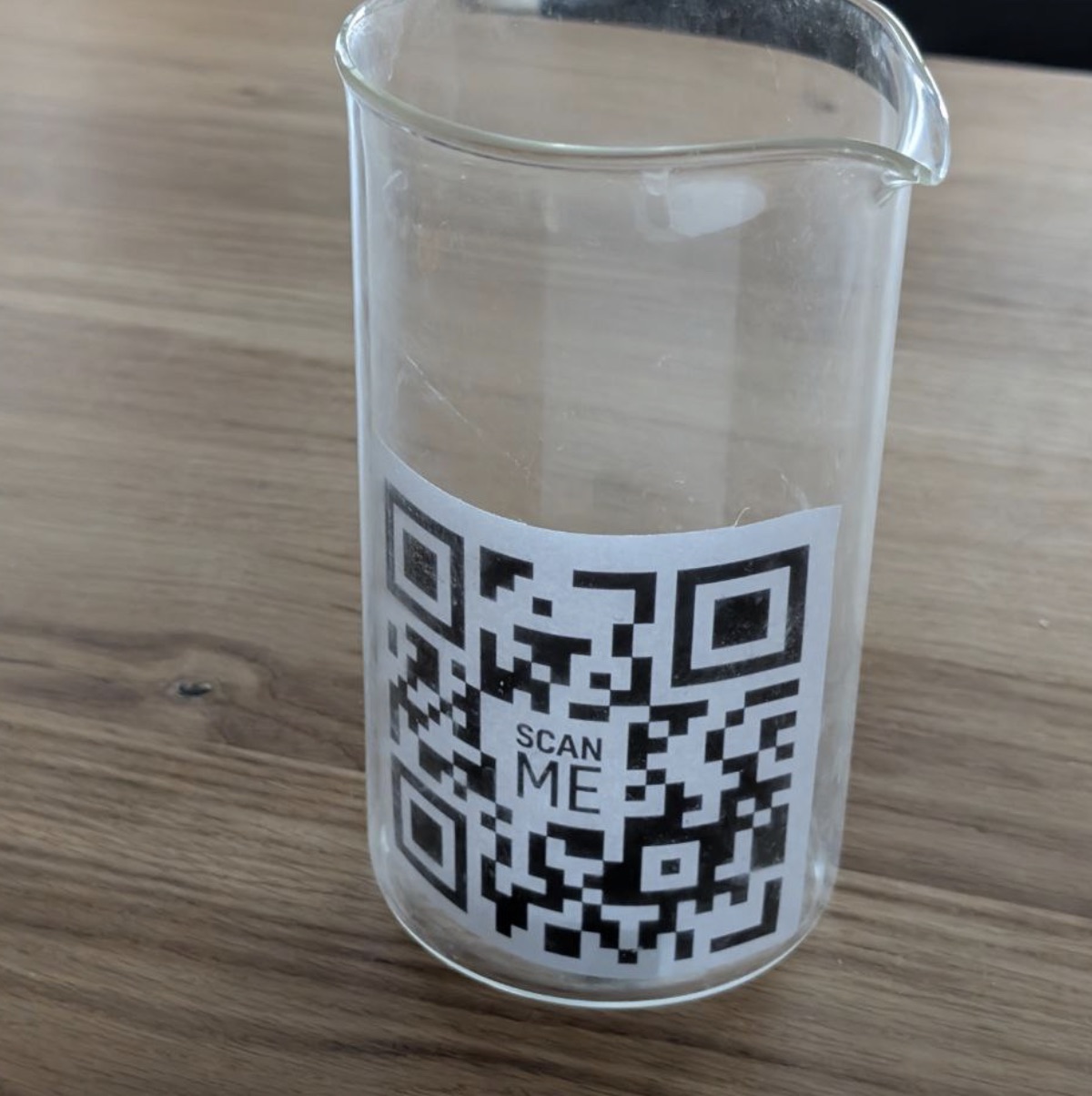}
    \caption{matte, carafe outw.}
    \label{fig:scanme-qr-outward}
    \end{subfigure}  
    \hfill
    \begin{subfigure}[b]{0.163\textwidth}
    \includegraphics[width=\textwidth]{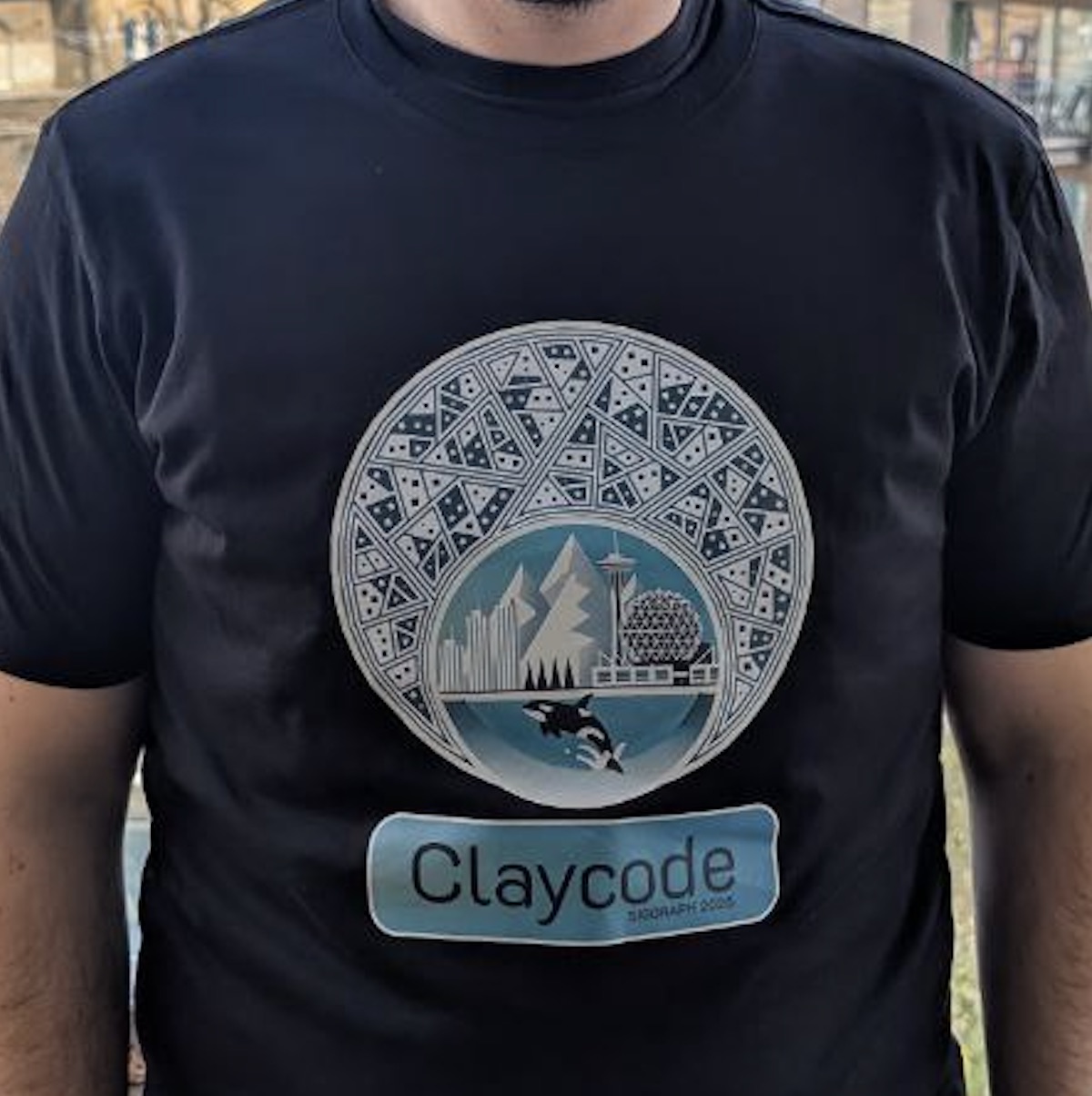}
    \caption{\textit{``claycode.io''}, T-shirt}
    \label{fig:claycodeio-tshirt}
    \end{subfigure}
    \hfill
    \begin{subfigure}[b]{0.163\textwidth}
    \includegraphics[width=\textwidth]{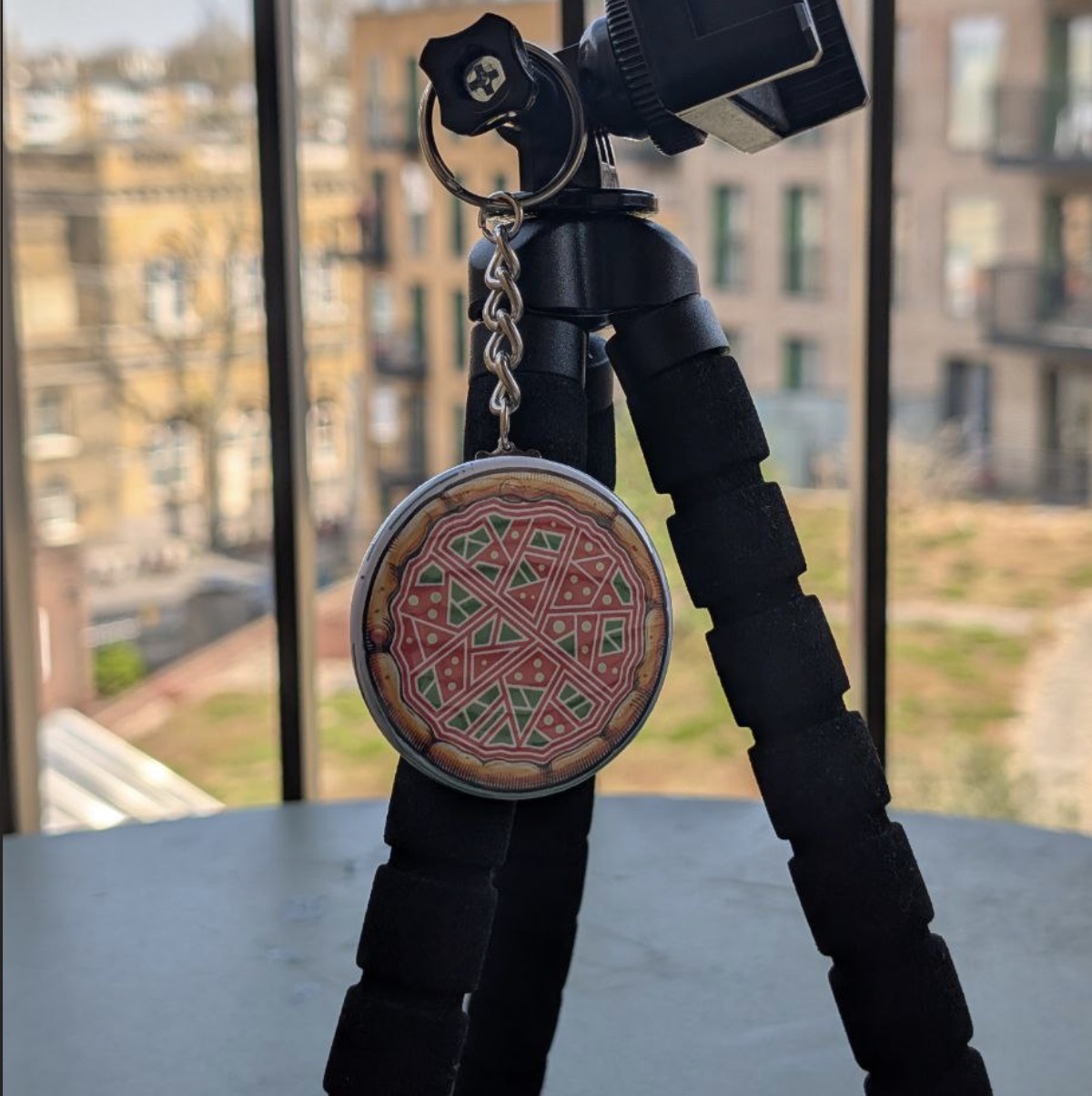}
    \caption{\textit{``pizza!''}, keychain}
    \label{fig:pizza-keychain}
    \end{subfigure}
    \hfill
    \begin{subfigure}[b]{0.163\textwidth}
    \includegraphics[width=\textwidth]{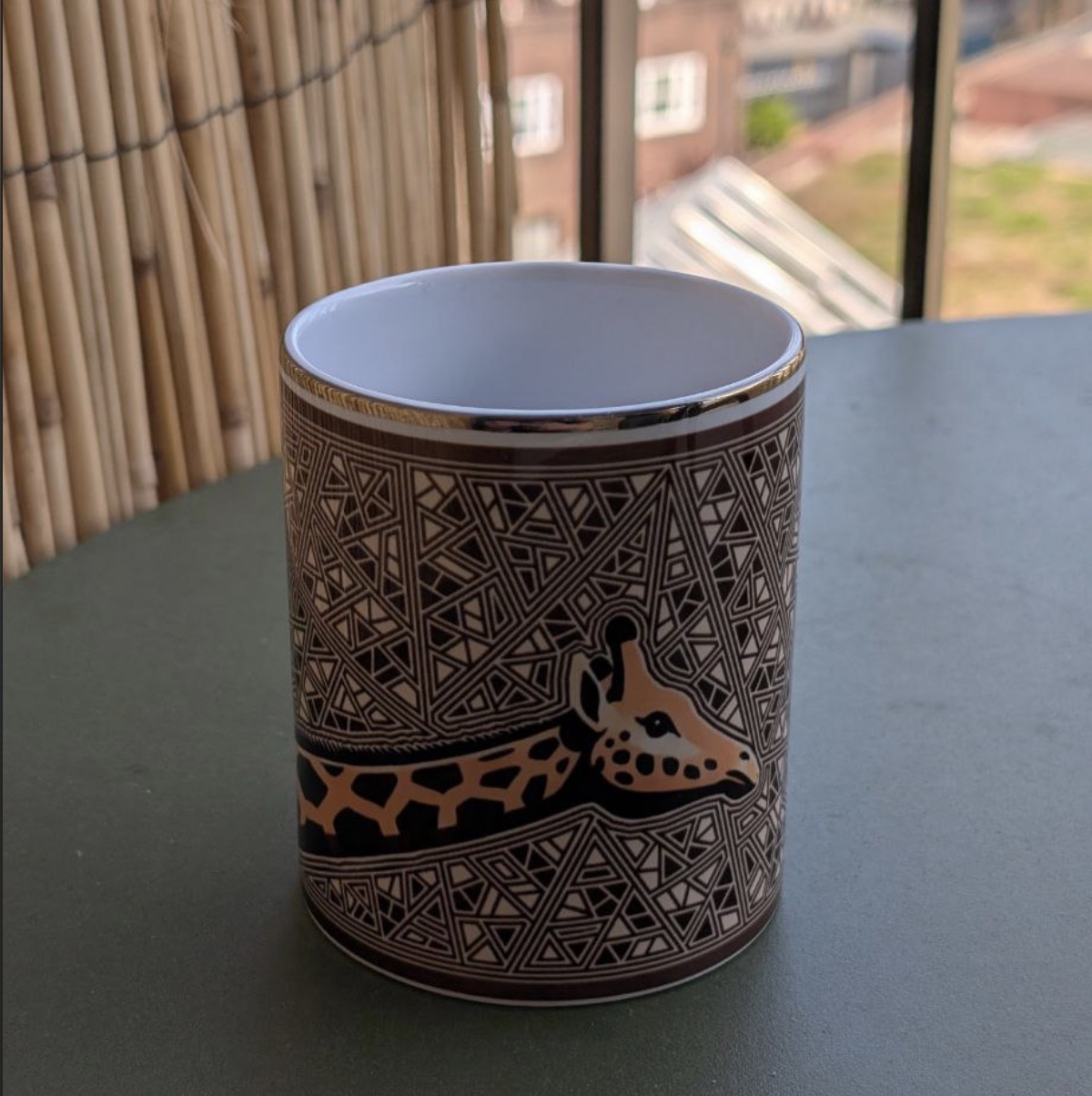}
    \caption{\textit{``Giraffe!''}, mug}
    \label{fig:giraffe-mug}
    \end{subfigure}
    \hfill

 \caption{The evaluation on printed media presented in \cref{sec:media}.}
  \label{fig:eval-media}
\end{figure*}

To further understand the practical limitations of Claycodes in real-world scenarios, we conducted a second evaluation that considered variations in physical media and environmental lighting.
We printed a set of $4$ QRCodes and $8$ Claycodes with varying designs, sizes, and colors, on both matte and glossy paper, and tested them across $7$ distinct physical scenarios, each featuring different backgrounds and lighting conditions. In contrast to the controlled setup in \cref{sec:results}, this evaluation recreated a natural user experience: a person attempted to scan each code as they normally would. \crmodsecond{Each scanning attempt was classified as a success (\gk) if the code was scanned in under three seconds without requiring any adjustment of the phone; a partial success (\yk) if scanning required repositioning or readjustment; or a failure (\rk) if the code could not be scanned within ten seconds. We scanned each code twice, and assigned the final outcome based on the lower-performing attempt: two successes were recorded as a success, a combination of success and partial success was recorded as a partial success, and any attempt that included a failure was recorded as a failure. }
The physical scenarios used in this evaluation are shown in \cref{fig:eval-media}:

\begin{itemize}
  \item \textbf{Outside} (\cref{fig:siggraph-outside}). A balcony in direct sunlight.
  \item \textbf{Inside} (\cref{fig:space-matte-inside}). A well-lit room with diffuse natural light.
  \item \textbf{Low, uneven light} (\cref{fig:eye-qr-low-light}). A partially illuminated room. Noticeable lighting variations on the surface of the code.
  \item \textbf{Dark} (\cref{fig:pizza-dark}). A room with low uniform illumination.
  \item \textbf{Strong reflections} (\cref{fig:giraffe-reflections}). A circular LED lamp is directed at the code, creating a harsh glare. The user must scan the code through the central opening of the lamp.
  \item \textbf{Carafe, inwards} (\cref{fig:siggraph-carafe-inward}). The code is bent inward and placed inside a transparent carafe, introducing surface distortion, visual noise, and partial occlusion.
  \item \textbf{Carafe, outwards} (\cref{fig:scanme-qr-outward}). As above, but with the code bent outward along the curvature of the carafe.
\end{itemize}

\begin{table*}
\centering
\caption{Experiment Evaluation Results}
\begin{tabular}{@{}l | l | c | cc | cc | cc | cc | cc | cc | cc@{}}
\toprule
& & \small\makecell{\textbf{Screen} \\ \textbf{(baseline)}}
& \multicolumn{2}{c|}{\small\makecell{\textbf{Outside} \\ \textbf{(nat. light)}}} 
& \multicolumn{2}{c|}{\small\makecell{\textbf{Inside} \\ \textbf{(nat. light)}}} 
& \multicolumn{2}{c|}{\small\makecell{\textbf{Low uneven} \\ \textbf{light}}} 
& \multicolumn{2}{c|}{\small\makecell{\textbf{Dark}}} 
& \multicolumn{2}{c|}{\small\makecell{\textbf{Strong} \\ \textbf{reflections}}} 
& \multicolumn{2}{c|}{\small\makecell{\textbf{Carafe} \\ \textbf{(Inwards)}}} 
& \multicolumn{2}{c}{\small\makecell{\textbf{Carafe} \\ \textbf{(Outwards)}}} \\ 

& & \footnotesize\textit{} 
& \scriptsize\textit{Matte} & \scriptsize\textit{Glossy} 
& \scriptsize\textit{Matte} & \scriptsize\textit{Glossy}  
& \scriptsize\textit{Matte} & \scriptsize\textit{Glossy}  
& \scriptsize\textit{Matte} & \scriptsize\textit{Glossy}  
& \scriptsize\textit{Matte} & \scriptsize\textit{Glossy}
& \scriptsize\textit{Matte} & \scriptsize\textit{Glossy}
& \scriptsize\textit{Matte} & \scriptsize\textit{Glossy} \\
\midrule

\multirow{3}{*}{\rotatebox{90}{\textbf{QRCode}}}       
& \textit{``baseline"}               & \gk & \gk & \gk & \gk & \gk & \gk & \gk & \gk & \gk & \gk & \gk & \yk & \yk & \yk & \yk \\
& \textit{``Scan Me"}                & \gk & \gk & \gk & \gk & \gk & \gk & \gk & \gk & \gk & \yk & \yk & \rk & \rk & \yk & \yk \\
& \textit{``colors!"} (\(coloured\)) & \gk & \gk & \gk & \gk & \gk & \gk & \gk & \gk & \gk & \gk & \gk & \rk & \yk & \yk & \yk \\
& \textit{``eye"} (\(stylised\))     & \gk & \gk & \gk & \gk & \gk & \gk & \gk & \yk & \yk & \yk & \rk & \yk & \yk & \yk & \yk \\
\midrule

\multirow{8}{*}{\rotatebox{90}{\textbf{Claycode}}} 
& \textit{``claycode"} (\(R=1\))    & \gk & \gk & \gk & \gk & \gk & \gk & \gk & \gk & \gk & \gk & \rk & \gk & \gk & \gk & \yk \\
& \textit{``\emojieyes"} (\(R=1\))         & \gk & \gk & \gk & \gk & \gk & \gk & \gk & \gk & \gk & \yk & \yk & \gk & \gk & \gk & \gk \\
& \textit{``space"} (\(R=1\))       & \gk & \gk & \gk & \gk & \gk & \gk & \gk & \gk & \gk & \yk & \rk & \rk & \rk & \gk & \gk \\
& \textit{``pizza!"} (\(R=1\))      & \gk & \gk & \gk & \yk & \yk & \gk & \gk & \gk & \gk & \rk & \rk & \gk & \rk & \gk & \gk \\
& \textit{``magic"} (\(R=1\))       & \gk & \gk & \gk & \gk & \gk & \gk & \yk & \yk & \rk & \rk & \rk & \rk & \rk & \rk & \rk \\
& \textit{``claycode.io"} (\(R=2\)) & \gk & \gk & \gk & \gk & \gk & \gk & \yk & \rk & \rk & \yk & \rk & \gk & \rk & \rk & \rk \\
& \textit{``SIGGRAPH!"} (\(R=2\))   & \gk & \yk & \yk & \yk & \yk & \gk & \gk & \yk & \rk & \yk & \gk & \yk & \yk & \rk & \rk \\
& \textit{``Giraffe!"} (\(R=7\))    & \gk & \gk & \gk & \gk & \gk & \yk & \yk & \rk & \yk & \yk & \yk & \yk & \yk & \gk & \yk \\
\bottomrule
\end{tabular}
\label{table:media-results}
\end{table*}

As can be seen in \cref{table:media-results}, all codes performed reliably under \textit{natural light}---both outdoors and indoors---as well as in \textit{low, uneven lighting} conditions. The few partial successes observed for Claycodes in these three scenarios were related to camera focus issues, which we consider orthogonal to the method itself and reflective of implementation limitations.

In the \textit{dark} scenario, several Claycodes failed to scan successfully. As expected, designs with less contrasted colors generally performed worse. However, Claycodes with greater redundancy (\textit{i.e.}, higher \(R\)) were particularly prone to failure in this condition: while redundancy can recover a partially occluded code, it also increases its visual complexity, making it more difficult to decode in low-light environments.

In the \textit{strong reflections} scenario, the difference between matte and glossy paper became more pronounced. Codes with higher redundancy generally performed better: as illustrated in \cref{fig:magic-glossy-reflections}, glare acts similarly to the occlusion damage modeled in \cref{sec:results}, and redundancy helped mitigate its effect.

The carafe experiments (\cref{fig:siggraph-carafe-inward,fig:scanme-qr-outward}) proved to be the most challenging overall for both QRCodes and Claycodes, due to a combination of distortion, partial occlusions, noise introduced by the semitransparent glass, and shadows. QRCodes were primarily hindered by distortion but would typically scan successfully once the user adjusted their position to minimize the perceived curvature. In contrast, Claycodes exhibited poorer performance with low-contrast codes but were generally scannable from multiple viewing angles.

Overall, the evaluation indicates that Claycodes are generally robust and versatile, although several areas for practical improvement remain, largely tied to the scanner's image preprocessing implementation: enhancements such as improved autofocus, adaptive zoom, and an ensemble of image pre-processing pipelines is likely to improve decoding performance in adverse conditions.

\subsection{Discussion on Additional Media}

To further examine the versatility of Claycodes as design artifacts, we applied a selection of designs to various physical media: a cotton T-shirt (\cref{fig:claycodeio-tshirt}), a ceramic mug (\cref{fig:giraffe-mug}), and a metallic keychain (\cref{fig:pizza-keychain}). We replicated the evaluation protocol described in \cref{table:media-results}, excluding the \textit{carafe} condition, and observed broadly consistent outcomes. All three objects were successfully scanned under natural lighting conditions, exhibited reduced performance under \textit{low, uneven light}, and failed to decode in the \textit{dark} scenario. The highly reflective metallic keychain proved particularly susceptible to scanning failure in the \textit{strong reflections} setting, while the T-shirt performed notably well. Textiles appear to be a promising medium for printed Claycodes, as the scanner is generally unaffected by natural wrinkles and creases. In the case of the T-shirt, no manual straightening was ever required to scan the code.

\section{Conclusion}

In this paper, we introduced Claycode, a novel type of 2D scannable code based on topological encoding. 
We have detailed the encoding/decoding pipeline and demonstrated that Claycodes are highly stylable and extremely tolerant to deformations. In all experiments, stylized Claycodes consistently performed equally or better than their non-stylized counterparts, showing that incorporating artistic elements does not negatively impact reliability. This stands in contrast with QR codes, where adding artistic features comes at the cost of damaging the code.
\crmod{}

In the future, we intend to improve the data capacity of Claycodes by abandoning UTF-8, in favor of a multimodal character encoding, devising more space-efficient bit-tree encodings, and implementing error correction schemes that go beyond the redundancy of the topology. Additionally, we plan to improve the reliability of the scan under challenging conditions, such as blur and poor lighting, by leveraging neural techniques similar to those applied to QR codes by ~\cite{chou-et-al}. In Claycodes, these techniques could further enable topology inference under partial occlusion, improving the robustness of Claycodes with no redundancy ($R=1$).

\begin{acks}
We thank Kimaya Bedarkar, Stratis Tsirtsis, and Valerio Di Donato for their valuable suggestions and discussions.
\end{acks}

\bibliographystyle{ACM-Reference-Format}
\bibliography{bibliography}


\begin{thebibliography}{57}

%%% the Web address.

\ifx \showCODEN    \undefined \def \showCODEN     #1{\unskip}     \fi
\ifx \showDOI      \undefined \def \showDOI       #1{#1}\fi
\ifx \showISBNx    \undefined \def \showISBNx     #1{\unskip}     \fi
\ifx \showISBNxiii \undefined \def \showISBNxiii  #1{\unskip}     \fi
\ifx \showISSN     \undefined \def \showISSN      #1{\unskip}     \fi
\ifx \showLCCN     \undefined \def \showLCCN      #1{\unskip}     \fi
\ifx \shownote     \undefined \def \shownote      #1{#1}          \fi
\ifx \showarticletitle \undefined \def \showarticletitle #1{#1}   \fi
\ifx \showURL      \undefined \def \showURL       {\relax}        \fi
% invisible to TeX
\providecommand\bibfield[2]{#2}
\providecommand\bibinfo[2]{#2}
\providecommand\natexlab[1]{#1}
\providecommand\showeprint[2][]{arXiv:#2}

\bibitem[Abe(1994)]%
        {abe1994tree}
\bibfield{author}{\bibinfo{person}{Y Abe}.} \bibinfo{year}{1994}\natexlab{}.
\newblock \showarticletitle{Tree representation of positive integers}.
\newblock \bibinfo{journal}{\emph{Applied Mathematics Letters}} \bibinfo{volume}{7}, \bibinfo{number}{1} (\bibinfo{year}{1994}), \bibinfo{pages}{57}.
\newblock


\bibitem[Apple(2020)]%
        {apple_hig_appclips}
\bibfield{author}{\bibinfo{person}{Apple}.} \bibinfo{year}{2020}\natexlab{}.
\newblock \bibinfo{title}{Human Interface Guidelines: App Clips}.
\newblock \bibinfo{howpublished}{\url{https://developer.apple.com/design/human-interface-guidelines/app-clips/}}.
\newblock
\newblock
\shownote{Accessed: 2024-12-30}.


\bibitem[Bencina and Kaltenbrunner(2005)]%
        {reactivision}
\bibfield{author}{\bibinfo{person}{Ross Bencina} {and} \bibinfo{person}{Martin Kaltenbrunner}.} \bibinfo{year}{2005}\natexlab{}.
\newblock \showarticletitle{Improved Topological Fiducial Tracking in the reacTIVision System}. In \bibinfo{booktitle}{\emph{2005 IEEE Computer Society Conference on Computer Vision and Pattern Recognition (CVPR'05) - Workshops}}. \bibinfo{pages}{99--99}.
\newblock
\urldef\tempurl%
\url{https://doi.org/10.1109/CVPR.2005.475}
\showDOI{\tempurl}


\bibitem[Beyer and Hedetniemi(1980)]%
        {beyer1980constant}
\bibfield{author}{\bibinfo{person}{Terry Beyer} {and} \bibinfo{person}{Sandra~Mitchell Hedetniemi}.} \bibinfo{year}{1980}\natexlab{}.
\newblock \showarticletitle{Constant time generation of rooted trees}.
\newblock \bibinfo{journal}{\emph{SIAM J. Comput.}} \bibinfo{volume}{9}, \bibinfo{number}{4} (\bibinfo{year}{1980}), \bibinfo{pages}{706--712}.
\newblock


\bibitem[{Bosch GmbH}(1991)]%
        {BoschCAN}
\bibfield{author}{\bibinfo{person}{{Bosch GmbH}}.} \bibinfo{year}{1991}\natexlab{}.
\newblock \bibinfo{booktitle}{\emph{CAN Specification Version 2.0}}.
\newblock \bibinfo{type}{{T}echnical {R}eport}. \bibinfo{institution}{Robert Bosch GmbH}, \bibinfo{address}{Stuttgart, Germany}.
\newblock
\urldef\tempurl%
\url{https://www.bosch-semiconductors.com/can}
\showURL{%
\tempurl}
\newblock
\shownote{Available from Bosch Automotive Electronics}.


\bibitem[Bradski(2000)]%
        {opencv_library}
\bibfield{author}{\bibinfo{person}{Gary Bradski}.} \bibinfo{year}{2000}\natexlab{}.
\newblock \showarticletitle{{The OpenCV Library}}.
\newblock \bibinfo{journal}{\emph{Dr. Dobb's Journal of Software Tools}} (\bibinfo{year}{2000}).
\newblock


\bibitem[Cappello(1988)]%
        {cappello1988new}
\bibfield{author}{\bibinfo{person}{Peter Cappello}.} \bibinfo{year}{1988}\natexlab{}.
\newblock \showarticletitle{A new bijection between natural numbers and rooted trees}. In \bibinfo{booktitle}{\emph{4th SIAM Conference on Discrete Mathematics}}. Citeseer.
\newblock


\bibitem[Chen et~al\mbox{.}(2016)]%
        {7479568}
\bibfield{author}{\bibinfo{person}{Changsheng Chen}, \bibinfo{person}{Wenjian Huang}, \bibinfo{person}{Baojian Zhou}, \bibinfo{person}{Chenchen Liu}, {and} \bibinfo{person}{Wai~Ho Mow}.} \bibinfo{year}{2016}\natexlab{}.
\newblock \showarticletitle{PiCode: A New Picture-Embedding 2D Barcode}.
\newblock \bibinfo{journal}{\emph{IEEE Transactions on Image Processing}} \bibinfo{volume}{25}, \bibinfo{number}{8} (\bibinfo{year}{2016}), \bibinfo{pages}{3444--3458}.
\newblock
\urldef\tempurl%
\url{https://doi.org/10.1109/TIP.2016.2573592}
\showDOI{\tempurl}


\bibitem[Chen and McMains(2005)]%
        {chen2005polygon}
\bibfield{author}{\bibinfo{person}{Xiaorui Chen} {and} \bibinfo{person}{Sara McMains}.} \bibinfo{year}{2005}\natexlab{}.
\newblock \showarticletitle{Polygon offsetting by computing winding numbers}. In \bibinfo{booktitle}{\emph{International Design Engineering Technical Conferences and Computers and Information in Engineering Conference}}, Vol.~\bibinfo{volume}{4739}. \bibinfo{pages}{565--575}.
\newblock


\bibitem[Chou et~al\mbox{.}(2015)]%
        {chou-et-al}
\bibfield{author}{\bibinfo{person}{Tzu-Han Chou}, \bibinfo{person}{Chuan-Sheng Ho}, {and} \bibinfo{person}{Yan-Fu Kuo}.} \bibinfo{year}{2015}\natexlab{}.
\newblock \showarticletitle{QR code detection using convolutional neural networks}. In \bibinfo{booktitle}{\emph{2015 International Conference on Advanced Robotics and Intelligent Systems (ARIS)}}. \bibinfo{pages}{1--5}.
\newblock
\urldef\tempurl%
\url{https://doi.org/10.1109/ARIS.2015.7158354}
\showDOI{\tempurl}


\bibitem[Chu et~al\mbox{.}(2013)]%
        {chu2013halftone}
\bibfield{author}{\bibinfo{person}{Hung-Kuo Chu}, \bibinfo{person}{Chia-Sheng Chang}, \bibinfo{person}{Ruen-Rone Lee}, {and} \bibinfo{person}{Niloy~J Mitra}.} \bibinfo{year}{2013}\natexlab{}.
\newblock \showarticletitle{Halftone QR codes}.
\newblock \bibinfo{journal}{\emph{ACM Transactions on Graphics (TOG)}} \bibinfo{volume}{32}, \bibinfo{number}{6} (\bibinfo{year}{2013}), \bibinfo{pages}{1--8}.
\newblock


\bibitem[Costanza and Huang(2009)]%
        {dtouch0}
\bibfield{author}{\bibinfo{person}{Enrico Costanza} {and} \bibinfo{person}{Jeffrey Huang}.} \bibinfo{year}{2009}\natexlab{}.
\newblock \showarticletitle{Designable visual markers}.
\newblock \bibinfo{journal}{\emph{Conference on Human Factors in Computing Systems - Proceedings}} (\bibinfo{date}{04} \bibinfo{year}{2009}).
\newblock
\urldef\tempurl%
\url{https://doi.org/10.1145/1518701.1518990}
\showDOI{\tempurl}


\bibitem[Costanza and Robinson(2003)]%
        {costanza2003region}
\bibfield{author}{\bibinfo{person}{Enrico Costanza} {and} \bibinfo{person}{John Robinson}.} \bibinfo{year}{2003}\natexlab{}.
\newblock \showarticletitle{A Region Adjacency Tree Approach to the Detection and Design of Fiducials.}
\newblock  (\bibinfo{year}{2003}).
\newblock


\bibitem[Costanza et~al\mbox{.}(2003)]%
        {dtouch1}
\bibfield{author}{\bibinfo{person}{Enrico Costanza}, \bibinfo{person}{Simon Shelley}, {and} \bibinfo{person}{John Robinson}.} \bibinfo{year}{2003}\natexlab{}.
\newblock \showarticletitle{D-touch: A Consumer-Grade Tangible Interface Module and Musical Applications}.
\newblock  (\bibinfo{date}{01} \bibinfo{year}{2003}).
\newblock


\bibitem[Cox(1927)]%
        {cox1927method}
\bibfield{author}{\bibinfo{person}{EP Cox}.} \bibinfo{year}{1927}\natexlab{}.
\newblock \showarticletitle{A method of assigning numerical and percentage values to the degree of roundness of sand grains}.
\newblock \bibinfo{journal}{\emph{Journal of paleontology}} \bibinfo{volume}{1}, \bibinfo{number}{3} (\bibinfo{year}{1927}), \bibinfo{pages}{179--183}.
\newblock


\bibitem[Cox(2012)]%
        {cox2012qartcodes}
\bibfield{author}{\bibinfo{person}{Russ Cox}.} \bibinfo{year}{2012}\natexlab{}.
\newblock \bibinfo{title}{Qartcodes}.
\newblock \bibinfo{howpublished}{\url{http://research.swtch.com/qart}}.
\newblock
\newblock
\shownote{Accessed: 2025-01-02}.


\bibitem[de~Alicante(2016)]%
        {navilens}
\bibfield{author}{\bibinfo{person}{Universidad de Alicante}.} \bibinfo{year}{Patent ES2616146A1, 2016}\natexlab{}.
\newblock \bibinfo{title}{Method of detection and recognition of visual markers of long reach and high density}.
\newblock
\newblock


\bibitem[Dogan et~al\mbox{.}(2023a)]%
        {Dogan2023StructCode}
\bibfield{author}{\bibinfo{person}{Mustafa~Doga Dogan}, \bibinfo{person}{Vivian~Hsinyueh Chan}, \bibinfo{person}{Richard Qi}, \bibinfo{person}{Grace Tang}, \bibinfo{person}{Thijs Roumen}, {and} \bibinfo{person}{Stefanie Mueller}.} \bibinfo{year}{2023}\natexlab{a}.
\newblock \showarticletitle{StructCode: Leveraging Fabrication Artifacts to Store Data in Laser-Cut Objects}. In \bibinfo{booktitle}{\emph{Proceedings of the 8th ACM Symposium on Computational Fabrication (SCF '23)}}. \bibinfo{publisher}{Association for Computing Machinery}, \bibinfo{pages}{1--13}.
\newblock
\urldef\tempurl%
\url{https://doi.org/10.1145/3623263.3623353}
\showDOI{\tempurl}


\bibitem[Dogan et~al\mbox{.}(2023b)]%
        {Dogan2023BrightMarker}
\bibfield{author}{\bibinfo{person}{Mustafa~Doga Dogan}, \bibinfo{person}{Raul Garcia-Martin}, \bibinfo{person}{Patrick~William Haertel}, \bibinfo{person}{Jamison~John O'Keefe}, \bibinfo{person}{Ahmad Taka}, \bibinfo{person}{Akarsh Aurora}, \bibinfo{person}{Raul Sanchez-Reillo}, {and} \bibinfo{person}{Stefanie Mueller}.} \bibinfo{year}{2023}\natexlab{b}.
\newblock \showarticletitle{BrightMarker: 3D Printed Fluorescent Markers for Object Tracking}. In \bibinfo{booktitle}{\emph{Proceedings of the 36th Annual ACM Symposium on User Interface Software and Technology (UIST '23)}}. \bibinfo{publisher}{Association for Computing Machinery}, \bibinfo{pages}{1--13}.
\newblock
\urldef\tempurl%
\url{https://doi.org/10.1145/3586183.3606758}
\showDOI{\tempurl}


\bibitem[Dogan et~al\mbox{.}(2022)]%
        {Dogan2022InfraredTags}
\bibfield{author}{\bibinfo{person}{Mustafa~Doga Dogan}, \bibinfo{person}{Ahmad Taka}, \bibinfo{person}{Michael Lu}, \bibinfo{person}{Yunyi Zhu}, \bibinfo{person}{Akshat Kumar}, \bibinfo{person}{Aakar Gupta}, {and} \bibinfo{person}{Stefanie Mueller}.} \bibinfo{year}{2022}\natexlab{}.
\newblock \showarticletitle{InfraredTags: Embedding Invisible AR Markers and Barcodes Using Low-Cost, Infrared-Based 3D Printing and Imaging Tools}. In \bibinfo{booktitle}{\emph{Proceedings of the CHI Conference on Human Factors in Computing Systems (CHI)}}. \bibinfo{publisher}{Association for Computing Machinery}, \bibinfo{pages}{1--12}.
\newblock
\urldef\tempurl%
\url{https://doi.org/10.1145/3491102.3501951}
\showDOI{\tempurl}


\bibitem[Effantin(2004)]%
        {effantin2004generation}
\bibfield{author}{\bibinfo{person}{Brice Effantin}.} \bibinfo{year}{2004}\natexlab{}.
\newblock \showarticletitle{Generation of Unordered Binary Trees}. In \bibinfo{booktitle}{\emph{Computational Science and Its Applications--ICCSA 2004: International Conference, Assisi, Italy, May 14-17, 2004, Proceedings, Part III 4}}. Springer, \bibinfo{pages}{648--655}.
\newblock


\bibitem[Feick et~al\mbox{.}(2025)]%
        {Feick2025Imprinto}
\bibfield{author}{\bibinfo{person}{Martin Feick}, \bibinfo{person}{Xuxin Tang}, \bibinfo{person}{Raul Garcia-Martin}, \bibinfo{person}{Alexandru Luchianov}, \bibinfo{person}{Roderick Wei~Xiao Huang}, \bibinfo{person}{Chang Xiao}, \bibinfo{person}{Alexa Siu}, {and} \bibinfo{person}{Mustafa~Doga Dogan}.} \bibinfo{year}{2025}\natexlab{}.
\newblock \showarticletitle{Imprinto: Enhancing Infrared Inkjet Watermarking for Human and Machine Perception}. In \bibinfo{booktitle}{\emph{Proceedings of the CHI Conference on Human Factors in Computing Systems (CHI)}}. \bibinfo{publisher}{Association for Computing Machinery}, \bibinfo{pages}{to appear}.
\newblock
\urldef\tempurl%
\url{https://doi.org/10.1145/3491102.3501951}
\showDOI{\tempurl}


\bibitem[Fiala(2004)]%
        {ARTag}
\bibfield{author}{\bibinfo{person}{Mark Fiala}.} \bibinfo{year}{2004}\natexlab{}.
\newblock \showarticletitle{ARTag Revision 1, A Fiducial Marker System Using Digital Techniques}.
\newblock  (\bibinfo{date}{01} \bibinfo{year}{2004}).
\newblock


\bibitem[Flowcode(2024)]%
        {flowcode}
\bibfield{author}{\bibinfo{person}{Flowcode}.} \bibinfo{year}{2024}\natexlab{}.
\newblock \bibinfo{title}{Flowcode: Create and Share QR Codes Instantly}.
\newblock \bibinfo{howpublished}{\url{https://www.flowcode.com/}}.
\newblock
\newblock
\shownote{Accessed: 2024-12-30}.


\bibitem[{Gamma Play}(2024)]%
        {qr_code_scanner_gamma_play}
\bibfield{author}{\bibinfo{person}{{Gamma Play}}.} \bibinfo{year}{2024}\natexlab{}.
\newblock \bibinfo{title}{QR Code Scanner}.
\newblock \bibinfo{howpublished}{\url{https://play.google.com/store/apps/details?id=com.gamma.scan&hl=en_GB}}.
\newblock
\newblock
\shownote{Accessed: 2025-01-20}.


\bibitem[Getschmann and Echtler(2021)]%
        {seedmarkers}
\bibfield{author}{\bibinfo{person}{Christopher Getschmann} {and} \bibinfo{person}{Florian Echtler}.} \bibinfo{year}{2021}\natexlab{}.
\newblock \showarticletitle{Seedmarkers: Embeddable Markers for Physical Objects}. \bibinfo{pages}{1--11}.
\newblock
\urldef\tempurl%
\url{https://doi.org/10.1145/3430524.3440645}
\showDOI{\tempurl}


\bibitem[Higashino et~al\mbox{.}(2016)]%
        {higashino2016arttag}
\bibfield{author}{\bibinfo{person}{Shinichi Higashino}, \bibinfo{person}{Sakiko Nishi}, {and} \bibinfo{person}{Ryuuki Sakamoto}.} \bibinfo{year}{2016}\natexlab{}.
\newblock \showarticletitle{ARTTag: aesthetic fiducial markers based on circle pairs}.
\newblock In \bibinfo{booktitle}{\emph{ACM SIGGRAPH 2016 Posters}}. \bibinfo{pages}{1--2}.
\newblock


\bibitem[{ISO}(2007)]%
        {ISO15417-2007}
\bibfield{author}{\bibinfo{person}{{ISO}}.} \bibinfo{year}{2007}\natexlab{}.
\newblock \bibinfo{title}{{ISO/IEC 15417:2007 Information technology -- Automatic identification and data capture techniques -- Bar code symbology specification -- Code 128}}.
\newblock
\newblock
\urldef\tempurl%
\url{https://www.iso.org/standard/43896.html}
\showURL{%
\tempurl}


\bibitem[Jung et~al\mbox{.}(2019)]%
        {Jung2019Automating}
\bibfield{author}{\bibinfo{person}{Joshua D.~A. Jung}, \bibinfo{person}{Rahul~N. Iyer}, {and} \bibinfo{person}{Daniel Vogel}.} \bibinfo{year}{2019}\natexlab{}.
\newblock \showarticletitle{Automating the Intentional Encoding of Human-Designable Markers}. In \bibinfo{booktitle}{\emph{Proceedings of the 2019 CHI Conference on Human Factors in Computing Systems (CHI)}}. \bibinfo{publisher}{Association for Computing Machinery}, \bibinfo{pages}{187:1--187:12}.
\newblock
\urldef\tempurl%
\url{https://doi.org/10.1145/3290605.3300417}
\showDOI{\tempurl}


\bibitem[Jung and Vogel(2018)]%
        {Jung2018Methods}
\bibfield{author}{\bibinfo{person}{Joshua D.~A. Jung} {and} \bibinfo{person}{Daniel Vogel}.} \bibinfo{year}{2018}\natexlab{}.
\newblock \showarticletitle{Methods for Intentional Encoding of High Capacity Human-Designable Visual Markers}. In \bibinfo{booktitle}{\emph{Proceedings of the 2018 CHI Conference on Human Factors in Computing Systems (CHI '18)}}. \bibinfo{publisher}{Association for Computing Machinery}, \bibinfo{pages}{313:1--313:12}.
\newblock
\urldef\tempurl%
\url{https://doi.org/10.1145/3173574.3173887}
\showDOI{\tempurl}


\bibitem[Kornprobst et~al\mbox{.}(2009)]%
        {bilateral}
\bibfield{author}{\bibinfo{person}{Pierre Kornprobst}, \bibinfo{person}{Jack Tumblin}, {and} \bibinfo{person}{Frédo Durand}.} \bibinfo{year}{2009}\natexlab{}.
\newblock \showarticletitle{Bilateral Filtering: Theory and Applications}.
\newblock \bibinfo{journal}{\emph{Foundations and Trends in Computer Graphics and Vision}}  \bibinfo{volume}{4} (\bibinfo{date}{01} \bibinfo{year}{2009}), \bibinfo{pages}{1--74}.
\newblock
\urldef\tempurl%
\url{https://doi.org/10.1561/0600000020}
\showDOI{\tempurl}


\bibitem[Kurniawan et~al\mbox{.}(2019)]%
        {qrcodes2019}
\bibfield{author}{\bibinfo{person}{Wendy Kurniawan}, \bibinfo{person}{Hiroshi Okumura}, \bibinfo{person}{Muladi Muladi}, {and} \bibinfo{person}{Anik Handayani}.} \bibinfo{year}{2019}\natexlab{}.
\newblock \showarticletitle{An Improvement on QR Code Limit Angle Detection using Convolution Neural Network}. In \bibinfo{booktitle}{\emph{nternational Conference on Electrical, Electronics and Information Engineering (ICEEIE)}}. \bibinfo{pages}{234--238}.
\newblock
\urldef\tempurl%
\url{https://doi.org/10.1109/ICEEIE47180.2019.8981449}
\showDOI{\tempurl}


\bibitem[Li et~al\mbox{.}(2017)]%
        {Li2017AirCode}
\bibfield{author}{\bibinfo{person}{Dingzeyu Li}, \bibinfo{person}{Avinash~S. Nair}, \bibinfo{person}{Shree~K. Nayar}, {and} \bibinfo{person}{Changxi Zheng}.} \bibinfo{year}{2017}\natexlab{}.
\newblock \showarticletitle{AirCode: Unobtrusive Physical Tags for Digital Fabrication}. In \bibinfo{booktitle}{\emph{Proceedings of the 30th Annual ACM Symposium on User Interface Software and Technology (UIST)}}. \bibinfo{publisher}{Association for Computing Machinery}, \bibinfo{pages}{449--460}.
\newblock
\urldef\tempurl%
\url{https://doi.org/10.1145/3126594.3126641}
\showDOI{\tempurl}


\bibitem[Li(1997)]%
        {li1997generation}
\bibfield{author}{\bibinfo{person}{Gang Li}.} \bibinfo{year}{1997}\natexlab{}.
\newblock \bibinfo{booktitle}{\emph{Generation of rooted trees and free trees.}}
\newblock \bibinfo{publisher}{University of Victoria}.
\newblock


\bibitem[Lin et~al\mbox{.}(2015)]%
        {7112509}
\bibfield{author}{\bibinfo{person}{Shih-Syun Lin}, \bibinfo{person}{Min-Chun Hu}, \bibinfo{person}{Chien-Han Lee}, {and} \bibinfo{person}{Tong-Yee Lee}.} \bibinfo{year}{2015}\natexlab{}.
\newblock \showarticletitle{Efficient QR Code Beautification With High Quality Visual Content}.
\newblock \bibinfo{journal}{\emph{IEEE Transactions on Multimedia}} \bibinfo{volume}{17}, \bibinfo{number}{9} (\bibinfo{year}{2015}), \bibinfo{pages}{1515--1524}.
\newblock
\urldef\tempurl%
\url{https://doi.org/10.1109/TMM.2015.2437711}
\showDOI{\tempurl}


\bibitem[Lin et~al\mbox{.}(2013)]%
        {lin2013appearance}
\bibfield{author}{\bibinfo{person}{Yu-Hsun Lin}, \bibinfo{person}{Yu-Pei Chang}, {and} \bibinfo{person}{Ja-Ling Wu}.} \bibinfo{year}{2013}\natexlab{}.
\newblock \showarticletitle{Appearance-based QR code beautifier}.
\newblock \bibinfo{journal}{\emph{IEEE Transactions on Multimedia}} \bibinfo{volume}{15}, \bibinfo{number}{8} (\bibinfo{year}{2013}), \bibinfo{pages}{2198--2207}.
\newblock


\bibitem[Maia et~al\mbox{.}(2019)]%
        {Maia2019LayerCode}
\bibfield{author}{\bibinfo{person}{Henrique~Teles Maia}, \bibinfo{person}{Dingzeyu Li}, \bibinfo{person}{Yuan Yang}, {and} \bibinfo{person}{Changxi Zheng}.} \bibinfo{year}{2019}\natexlab{}.
\newblock \showarticletitle{LayerCode: Optical Barcodes for 3D Printed Shapes}.
\newblock \bibinfo{journal}{\emph{ACM Transactions on Graphics (TOG)}} \bibinfo{volume}{38}, \bibinfo{number}{4} (\bibinfo{year}{2019}), \bibinfo{pages}{90:1--90:13}.
\newblock
\urldef\tempurl%
\url{https://doi.org/10.1145/3306346.3322960}
\showDOI{\tempurl}


\bibitem[{Mark Keil}(2000)]%
        {MARKKEIL2000491}
\bibfield{author}{\bibinfo{person}{J. {Mark Keil}}.} \bibinfo{year}{2000}\natexlab{}.
\newblock \showarticletitle{Chapter 11 - Polygon Decomposition}.
\newblock In \bibinfo{booktitle}{\emph{Handbook of Computational Geometry}}, \bibfield{editor}{\bibinfo{person}{J.-R. Sack} {and} \bibinfo{person}{Jorge Urrutia}} (Eds.). \bibinfo{publisher}{North-Holland}, \bibinfo{address}{Amsterdam}, \bibinfo{pages}{491--518}.
\newblock
\showISBNx{978-0-444-82537-7}
\urldef\tempurl%
\url{https://doi.org/10.1016/B978-044482537-7/50012-7}
\showDOI{\tempurl}


\bibitem[Ozkaya et~al\mbox{.}(2015)]%
        {ozkaya2015factors}
\bibfield{author}{\bibinfo{person}{Elif Ozkaya}, \bibinfo{person}{H~Erkan Ozkaya}, \bibinfo{person}{Juanita Roxas}, \bibinfo{person}{Frank Bryant}, {and} \bibinfo{person}{Debbora Whitson}.} \bibinfo{year}{2015}\natexlab{}.
\newblock \showarticletitle{Factors affecting consumer usage of QR codes}.
\newblock \bibinfo{journal}{\emph{Journal of Direct, Data and Digital Marketing Practice}}  \bibinfo{volume}{16} (\bibinfo{year}{2015}), \bibinfo{pages}{209--224}.
\newblock


\bibitem[Pollack and Trevi{\~n}o(2018)]%
        {pollack2018finding}
\bibfield{author}{\bibinfo{person}{Paul Pollack} {and} \bibinfo{person}{Enrique Trevi{\~n}o}.} \bibinfo{year}{2018}\natexlab{}.
\newblock \showarticletitle{Finding the Four Squares in Lagrange's Theorem.}
\newblock \bibinfo{journal}{\emph{Integers}} \bibinfo{volume}{18}, \bibinfo{number}{A15} (\bibinfo{year}{2018}), \bibinfo{pages}{7--17}.
\newblock


\bibitem[{QRCode AI}(2024)]%
        {qrcode_ai}
\bibfield{author}{\bibinfo{person}{{QRCode AI}}.} \bibinfo{year}{2024}\natexlab{}.
\newblock \bibinfo{title}{QRCode AI: Advanced QR Code Generator}.
\newblock \bibinfo{howpublished}{\url{https://qrcode-ai.com/}}.
\newblock
\newblock
\shownote{Accessed: 2024-12-30}.


\bibitem[Reed and Solomon(1960)]%
        {reed1960polynomial}
\bibfield{author}{\bibinfo{person}{Irving~S Reed} {and} \bibinfo{person}{Gustave Solomon}.} \bibinfo{year}{1960}\natexlab{}.
\newblock \showarticletitle{Polynomial codes over certain finite fields}.
\newblock \bibinfo{journal}{\emph{Journal of the society for industrial and applied mathematics}} \bibinfo{volume}{8}, \bibinfo{number}{2} (\bibinfo{year}{1960}), \bibinfo{pages}{300--304}.
\newblock


\bibitem[Shin-ichi Nakano(2003)]%
        {NAKANORooted}
\bibfield{author}{\bibinfo{person}{Takeaki~Uno Shin-ichi Nakano}.} \bibinfo{year}{2003}\natexlab{}.
\newblock \showarticletitle{Efficient generation of rooted trees}.
\newblock  (\bibinfo{year}{2003}).
\newblock


\bibitem[Skliar et~al\mbox{.}(2020)]%
        {skliar2020one}
\bibfield{author}{\bibinfo{person}{Osvaldo Skliar}, \bibinfo{person}{Sherry Gapper}, {and} \bibinfo{person}{Ricardo E.~Monge Monge}.} \bibinfo{year}{2020}\natexlab{}.
\newblock \showarticletitle{A One-to-One Correspondence between Natural Numbers and Binary Trees}.
\newblock \bibinfo{journal}{\emph{arXiv preprint arXiv:2002.04477}} (\bibinfo{year}{2020}).
\newblock


\bibitem[Spotify(2017)]%
        {spotifycodes}
\bibfield{author}{\bibinfo{person}{Spotify}.} \bibinfo{year}{2017}\natexlab{}.
\newblock \bibinfo{title}{Spotify Codes - Share music the easy way}.
\newblock \bibinfo{howpublished}{\url{https://www.spotifycodes.com/}}.
\newblock
\newblock
\shownote{Accessed: 2024-12-30}.


\bibitem[Su et~al\mbox{.}(2021)]%
        {su2021artcoder}
\bibfield{author}{\bibinfo{person}{Hao Su}, \bibinfo{person}{Jianwei Niu}, \bibinfo{person}{Xuefeng Liu}, \bibinfo{person}{Qingfeng Li}, \bibinfo{person}{Ji Wan}, \bibinfo{person}{Mingliang Xu}, {and} \bibinfo{person}{Tao Ren}.} \bibinfo{year}{2021}\natexlab{}.
\newblock \showarticletitle{Artcoder: an end-to-end method for generating scanning-robust stylized qr codes}. In \bibinfo{booktitle}{\emph{Proceedings of the IEEE/CVF Conference on Computer Vision and Pattern Recognition}}. \bibinfo{pages}{2277--2286}.
\newblock


\bibitem[Suzuki and Abe(1985)]%
        {Suzuki1985TopologicalSA}
\bibfield{author}{\bibinfo{person}{Satoshi Suzuki} {and} \bibinfo{person}{Keiichi Abe}.} \bibinfo{year}{1985}\natexlab{}.
\newblock \showarticletitle{Topological structural analysis of digitized binary images by border following}.
\newblock \bibinfo{journal}{\emph{Comput. Vis. Graph. Image Process.}}  \bibinfo{volume}{30} (\bibinfo{year}{1985}), \bibinfo{pages}{32--46}.
\newblock
\urldef\tempurl%
\url{https://api.semanticscholar.org/CorpusID:205113350}
\showURL{%
\tempurl}


\bibitem[Tancik et~al\mbox{.}(2020)]%
        {Tancik2020StegaStamp}
\bibfield{author}{\bibinfo{person}{Matthew Tancik}, \bibinfo{person}{Ben Mildenhall}, {and} \bibinfo{person}{Ren Ng}.} \bibinfo{year}{2020}\natexlab{}.
\newblock \showarticletitle{StegaStamp: Invisible Hyperlinks in Physical Photographs}. In \bibinfo{booktitle}{\emph{Proceedings of the IEEE/CVF Conference on Computer Vision and Pattern Recognition (CVPR)}}. \bibinfo{publisher}{IEEE}, \bibinfo{pages}{2117--2126}.
\newblock
\urldef\tempurl%
\url{https://doi.org/10.1109/CVPR42600.2020.00219}
\showDOI{\tempurl}


\bibitem[Tiwari(2016)]%
        {tiwari2016introduction}
\bibfield{author}{\bibinfo{person}{Sumit Tiwari}.} \bibinfo{year}{2016}\natexlab{}.
\newblock \showarticletitle{An introduction to QR code technology}. In \bibinfo{booktitle}{\emph{2016 international conference on information technology (ICIT)}}. IEEE, \bibinfo{pages}{39--44}.
\newblock


\bibitem[Wan et~al\mbox{.}(2023)]%
        {wan2023implantable}
\bibfield{author}{\bibinfo{person}{Nan Wan}, \bibinfo{person}{Pengcheng Zhang}, \bibinfo{person}{Zuheng Liu}, \bibinfo{person}{Zhe Li}, \bibinfo{person}{Wei Niu}, \bibinfo{person}{Xiuye Rui}, \bibinfo{person}{Shibo Wang}, \bibinfo{person}{Myeongsu Seong}, \bibinfo{person}{Pengbo He}, \bibinfo{person}{Siqi Liang}, {et~al\mbox{.}}} \bibinfo{year}{2023}\natexlab{}.
\newblock \showarticletitle{Implantable QR code subcutaneous microchip using photoacoustic and ultrasound microscopy for secure and convenient individual identification and authentication}.
\newblock \bibinfo{journal}{\emph{Photoacoustics}}  \bibinfo{volume}{31} (\bibinfo{year}{2023}), \bibinfo{pages}{100504}.
\newblock


\bibitem[Xiao et~al\mbox{.}(2018)]%
        {Xiao2018FontCode}
\bibfield{author}{\bibinfo{person}{Chang Xiao}, \bibinfo{person}{Cheng Zhang}, {and} \bibinfo{person}{Changxi Zheng}.} \bibinfo{year}{2018}\natexlab{}.
\newblock \showarticletitle{FontCode: Embedding Information in Text Documents Using Glyph Perturbation}.
\newblock \bibinfo{journal}{\emph{ACM Transactions on Graphics (TOG)}} \bibinfo{volume}{37}, \bibinfo{number}{2} (\bibinfo{year}{2018}), \bibinfo{pages}{15:1--15:16}.
\newblock
\urldef\tempurl%
\url{https://doi.org/10.1145/3152823}
\showDOI{\tempurl}


\bibitem[Xie et~al\mbox{.}(2024)]%
        {Xie2024GladCoderSQ}
\bibfield{author}{\bibinfo{person}{Yuqiu Xie}, \bibinfo{person}{Bolin Jiang}, \bibinfo{person}{Jiawei Li}, \bibinfo{person}{Naiqi Li}, \bibinfo{person}{Bin Chen}, \bibinfo{person}{Tao Dai}, \bibinfo{person}{Yuang Peng}, {and} \bibinfo{person}{Shu-Tao Xia}.} \bibinfo{year}{2024}\natexlab{}.
\newblock \showarticletitle{GladCoder: Stylized QR Code Generation with Grayscale-Aware Denoising Process}. In \bibinfo{booktitle}{\emph{International Joint Conference on Artificial Intelligence}}.
\newblock
\urldef\tempurl%
\url{https://api.semanticscholar.org/CorpusID:271504025}
\showURL{%
\tempurl}


\bibitem[Xu et~al\mbox{.}(2021)]%
        {xu2021art}
\bibfield{author}{\bibinfo{person}{Mingliang Xu}, \bibinfo{person}{Qingfeng Li}, \bibinfo{person}{Jianwei Niu}, \bibinfo{person}{Hao Su}, \bibinfo{person}{Xiting Liu}, \bibinfo{person}{Weiwei Xu}, \bibinfo{person}{Pei Lv}, \bibinfo{person}{Bing Zhou}, {and} \bibinfo{person}{Yi Yang}.} \bibinfo{year}{2021}\natexlab{}.
\newblock \showarticletitle{ART-UP: A novel method for generating scanning-robust aesthetic QR codes}.
\newblock \bibinfo{journal}{\emph{ACM Transactions on Multimedia Computing, Communications, and Applications (TOMM)}} \bibinfo{volume}{17}, \bibinfo{number}{1} (\bibinfo{year}{2021}), \bibinfo{pages}{1--23}.
\newblock


\bibitem[Xu et~al\mbox{.}(2019)]%
        {8604076}
\bibfield{author}{\bibinfo{person}{Mingliang Xu}, \bibinfo{person}{Hao Su}, \bibinfo{person}{Yafei Li}, \bibinfo{person}{Xi Li}, \bibinfo{person}{Jing Liao}, \bibinfo{person}{Jianwei Niu}, \bibinfo{person}{Pei Lv}, {and} \bibinfo{person}{Bing Zhou}.} \bibinfo{year}{2019}\natexlab{}.
\newblock \showarticletitle{Stylized Aesthetic QR Code}.
\newblock \bibinfo{journal}{\emph{IEEE Transactions on Multimedia}} \bibinfo{volume}{21}, \bibinfo{number}{8} (\bibinfo{year}{2019}), \bibinfo{pages}{1960--1970}.
\newblock
\urldef\tempurl%
\url{https://doi.org/10.1109/TMM.2019.2891420}
\showDOI{\tempurl}


\bibitem[Yang et~al\mbox{.}(2016)]%
        {artcodes}
\bibfield{author}{\bibinfo{person}{Zhe Yang}, \bibinfo{person}{Yuting Bao}, \bibinfo{person}{Chuhao Luo}, \bibinfo{person}{Xingya Zhao}, \bibinfo{person}{Siyu Zhu}, \bibinfo{person}{Chunyi Peng}, \bibinfo{person}{Yunxin Liu}, {and} \bibinfo{person}{Xinbing Wang}.} \bibinfo{year}{2016}\natexlab{}.
\newblock \showarticletitle{ARTcode: preserve art and code in any image}. In \bibinfo{booktitle}{\emph{Proceedings of the 2016 ACM International Joint Conference on Pervasive and Ubiquitous Computing}} (Heidelberg, Germany) \emph{(\bibinfo{series}{UbiComp '16})}. \bibinfo{publisher}{Association for Computing Machinery}, \bibinfo{address}{New York, NY, USA}, \bibinfo{pages}{904–915}.
\newblock
\showISBNx{9781450344616}
\urldef\tempurl%
\url{https://doi.org/10.1145/2971648.2971733}
\showDOI{\tempurl}


\bibitem[Yang et~al\mbox{.}(2018)]%
        {Yang_2018}
\bibfield{author}{\bibinfo{person}{Zhibo Yang}, \bibinfo{person}{Huanle Xu}, \bibinfo{person}{Jianyuan Deng}, \bibinfo{person}{Chen~Change Loy}, {and} \bibinfo{person}{Wing~Cheong Lau}.} \bibinfo{year}{2018}\natexlab{}.
\newblock \showarticletitle{Robust and Fast Decoding of High-Capacity Color QR Codes for Mobile Applications}.
\newblock \bibinfo{journal}{\emph{IEEE Transactions on Image Processing}} \bibinfo{volume}{27}, \bibinfo{number}{12} (\bibinfo{date}{Dec.} \bibinfo{year}{2018}), \bibinfo{pages}{6093–6108}.
\newblock
\showISSN{1941-0042}
\urldef\tempurl%
\url{https://doi.org/10.1109/tip.2018.2855419}
\showDOI{\tempurl}


\bibitem[Yu et~al\mbox{.}(2020)]%
        {yu2020topotag}
\bibfield{author}{\bibinfo{person}{Guoxing Yu}, \bibinfo{person}{Yongtao Hu}, {and} \bibinfo{person}{Jingwen Dai}.} \bibinfo{year}{2020}\natexlab{}.
\newblock \showarticletitle{TopoTag: A robust and scalable topological fiducial marker system}.
\newblock \bibinfo{journal}{\emph{IEEE Transactions on Visualization and Computer Graphics}} \bibinfo{volume}{27}, \bibinfo{number}{9} (\bibinfo{year}{2020}), \bibinfo{pages}{3769--3780}.
\newblock


\end{thebibliography}

\clearpage
\appendix
\section{A Bijection Between Natural Numbers and Bit Strings}\label{sec:appendix-bijection}

While explaining the bit-tree encoding, we have modified the problem to work on natural numbers instead of bit strings, defining $f'$ and $g'$~so~that:

\begin{align}
    &f': \mathbb{N} \to \mathcal{T}, \quad g': \mathcal{T} \to \mathbb{N}, \notag \\
    &f(b) = f'(nat(b)) \quad\quad g(T) = bits(g'(T)) 
\end{align}

and assumed that the $nat: \{0, 1\}^* \to \mathbb{N}$ is a mapping from bit strings to natural numbers, while $bits: \mathbb{N} \to \{0, 1\}^*$ is its inverse. In the following, we present the bijection used in our implementation.

A simple mapping that converts a bit string $b = b_k b_{k-1} \dots b_1$ to a natural number $n$ by treating it as the binary representation of $n$, \ie, $nat(b) = \sum_{i=0}^{k-1} b_{k-i} \cdot 2^i$, is not enough. In fact, if $b_k=0$, then the function would map $b'=b_{k-1} \dots b_1$ to the same number as $b$, breaking the bijection. This problem can be avoided by appending a $1$ to the input string, resulting in:
\begin{equation}
    nat(b) = 2^{k+1}+\sum_{i=0}^{k-1} b_{k-i} \cdot 2^i
\end{equation}
and its inverse $bits(n) = n_{\mathcal{B}}[1:]$ where $n_{\mathcal{B}}$ is the binary string representation of $n$, and $[1:]$ discards the first element of $n_{\mathcal{B}}$.

\section{Reduction of the Partitioning Minimization Problem to the Binary Case}
\label{sec:appendix-packer-proof}

The packer's partitioning minimization problem weighs circularity and area proportionality to produce a partitioning that increases scannability:   
\begin{equation}
\min_{[P_1, \ldots, P_k]\in \mathcal{P}} \sum_{i=1}^k \alpha \left| A(P_i) - A^*(P_i) \right| + (1-\alpha)\left(1-R(P_i)\right).
\label{eq:packermin-suppl}
\end{equation}

In our packer implementation, we reduce this to the binary case, and then implement an algorithm that finds a solution of \cref{eq:packermin-suppl} by recursively applying the binary case. In the following, we prove the reduction to the binary case. First, we instantiate \cref{eq:packermin-suppl} with only two polygons $P1$ and $P2$:
\begin{equation}
\begin{split}
\min_{[P_1, P_2] \in \mathcal{P}} 
&\ \alpha \left| A(P_1) - A^*(P_1) \right| + (1-\alpha)\left(1 - R(P_1)\right) \\
&+ \alpha \left| A(P_2) - A^*(P_2) \right| + (1-\alpha)\left(1 - R(P_2)\right).
\end{split}
\end{equation}

Factoring $\alpha$ and $(1-\alpha)$, we obtain:
\begin{equation}
\begin{split}
\min_{[P_1, P_2] \in \mathcal{P}} 
&\ \alpha \left(\left| A(P_1) - A^*(P_1) \right| + \left| A(P_2) - A^*(P_2) \right| \right) + \\
&+(1-\alpha)\left(2 - (R(P_1) + R(P_2))\right).
\end{split}
\label{eq:packermin-suppl-binary-factored}
\end{equation}

Since $P = P_1 \cup P_2$ and $P_1 \cap P_2 = \emptyset$, then it holds that $A(P_2)=A(P) - A(P_1)$. Moreover, Since in the binary case, 

\begin{align*}
A^*(P_2) &= A(P) \cdot \frac{F(T_2)}{F(T_1) + F(T_2)} \\
&\implies A^*(P_2) = A(P) \cdot \left(1 - \frac{F(T_1)}{F(T_1) + F(T_2)}\right) \\
&\implies A^*(P_2) = A(P) - A(P) \cdot \frac{F(T_1)}{F(T_1) + F(T_2)}.
\end{align*}

Then, $A^*(P_2) = A(P)-A^*(P_1)$. Therefore, we rewrite \cref{eq:packermin-suppl-binary-factored}:

\begin{equation}
\begin{split}
\min_{[P_1, P_2] \in \mathcal{P}} 
&\ \alpha \left(\left| A(P_1) - A^*(P_1) \right| + \left| (A(P) - A(P_1)) - (A(P)-A^*(P_1)) \right| \right) + \\
&+(1-\alpha)\left(2 - (R(P_1) + R(P_2))\right).
\end{split}
\end{equation}

Which simplifies into: 

\begin{equation}
\begin{split}
\min_{[P_1, P_2] \in \mathcal{P}} 
&\ 2\alpha \left(\left| A(P_1) - A^*(P_1) \right| \right) + (1-\alpha)\left(2 - (R(P_1) + R(P_2))\right).
\end{split}
\end{equation}

Finally, we divide by $2$, obtaining the formulation:

\begin{equation}
\min_{[P_1, P_2]\in \mathcal{P}} \alpha\left|A(P_1) - w_1A(P)\right| + (1-\alpha)\left(1 - \frac{R(P_1)+R(P_2)}{2}\right)
\end{equation}

where $w_1$ is $\frac{F(T_1)}{F(T_1)+F(T_2)}$.

\section{Contours Detection Algorithm}\label{sec:appendix-contours}

The algorithm scans the binary image row by row. When an unvisited foreground pixel $(x, y)$ is encountered (\ie, $B(x, y) = 1$), it initiates the boundary tracing. The boundary is traced by iteratively exploring the 8-connected neighborhood:
\begin{equation}
N(x, y) = \{(x+i, y+j) \mid i, j \in \{-1, 0, 1\}, (i, j) \neq (0, 0)\}.
\end{equation}
The boundary tracing process continues until the starting pixel is reached, forming a closed contour.

Each detected contour is stored as a sequence of points $\pi = \{p_1, p_2, \dots, p_k\}$, where $p_i = (x_i, y_i)$ represents the coordinates of the $i$-th pixel on the contour.

Contours are detected using a border-following technique on the binary image. 
Suzuki and Abe algorithm is used \cite{Suzuki1985TopologicalSA,opencv_library} which efficiently detects contours and processes their hierarchy.
The algorithm also organizes contours into a hierarchical structure, representing their nesting relationships. For each contour $\pi_i$, the hierarchy is described by a 4-tuple:
\begin{equation}
H[i] = [\text{\texttt{Next}}, \text{\texttt{Previous}}, \text{\texttt{FirstChild}}, \text{\texttt{Parent}}],
\end{equation}
where \texttt{Next} is the index of the next contour at the same hierarchical level; \texttt{Previous} is the index of the previous contour at the same level; \texttt{FirstChild} is the index of the first child contour (nested within $\pi_i$); \texttt{Parent} is the index of the parent contour (enclosing $\pi_i$).
The hierarchy is constructed by examining the spatial relationships between contours. A contour $\pi_j$ is considered a child of $\pi_i$ if all points of $\pi_j$ lie within the boundary of $\pi_i$. This can be expressed mathematically as:
\begin{equation}
\forall (x, y) \in \pi_j, \, (x, y) \in \text{Interior}(\pi_i).
\end{equation}
Here, $\text{Interior}(\pi_i)$ denotes the region enclosed by the contour $\pi_i$.

This hierarchical representation enables the scanner to reconstruct a comprehensive topological tree for the input image.

\end{document}